\documentclass{aa}

\usepackage{graphicx}
\usepackage{txfonts}
\usepackage{amsmath}
\usepackage{threeparttable}
\usepackage{pifont}
\usepackage{float}
\usepackage[colorlinks=true,linkcolor=blue,citecolor=blue,urlcolor=blue]{hyperref}

\newcommand{\angstrom}{\text{\AA}}
\newcommand{\kms}{\mbox{$\mathrm{km\,s^{-1}}$}}
\newcommand{\cmq}{\mbox{$\mathrm{cm^{-3}}$}}
\newcommand{\Zsun}{\mbox{Z$_{\odot}$}}
\newcommand{\Msun}{\mbox{M$_{\odot}$}}
\newcommand{\Msunyr}{\mbox{M$_{\odot}\,\mathrm{yr^{-1}}$}}
\newcommand{\Mvir}{\mbox{$M_\mathrm{vir}$}}
\newcommand{\Rvir}{\mbox{$R_\mathrm{vir}$}}
\newcommand{\dxmin}{\mbox{$\Delta x_\mathrm{min}$}}
\newcommand{\dxminsq}{\mbox{$\Delta x^2_\mathrm{min}$}}
\newcommand{\epsff}{\mbox{$\varepsilon_\mathrm{ff}$}}
\newcommand{\tff}{\mbox{$t_\mathrm{ff}$}}
\newcommand{\gttco}{\texttt{GTT-noFB}}
\newcommand{\gtt}{\texttt{GTT}}
\newcommand{\sink}{\texttt{SINK}}
\newcommand{\sinkhr}{\texttt{SINK-HR}}
\newcommand{\fesc}{\mbox{$f_\mathrm{esc}$}}

\begin{document}

\title{Mitigating the overcooling problem with sink-based bursty star formation in a high-$z$ dwarf galaxy}
\titlerunning{Radiation-driven regulation via bursty star formation}

\author{Cheonsu Kang\inst{1}\thanks{astro.ckang@gmail.com}
\and Taysun Kimm\inst{1}\corrauth{tkimm@yonsei.ac.kr}
\and Daniel Han\inst{1}
\and Maxime Rey\inst{1}
\and Fred Thompson\inst{2}
\and Martin P. Rey\inst{3}
\and Harley Katz\inst{4}}

\institute{Department of Astronomy, Yonsei University, 50 Yonsei-ro, Seodaemun-gu, Seoul 03722, Republic of Korea
\and Sub-department of Astrophysics, University of Oxford, Denys Wilkinson Building, Keble Road, Oxford OX1 3RH, UK
\and Department of Physics, University of Bath, Claverton Down, Bath, BA2 7AY, UK
\and Department of Astronomy \& Astrophysics, University of Chicago, 5640 S Ellis Avenue, Chicago, IL 60637, USA}

\date{Received XXX; accepted XXX}

\abstract
{
Star formation is a fundamental driver of galaxy evolution, yet many galaxy formation models still fail to regulate it realistically, allowing gas to collapse too efficiently and overproduce stars.
To investigate a possible solution to this overcooling problem, we perform cosmological zoom-in radiation-hydrodynamics simulations of a dark matter halo reaching $10^{10}\,\Msun$ at $z=6$, adopting two distinct star formation models: a Schmidt-type model, in which the star formation criteria and efficiency per free-fall time are tied to local gravo-thermo-turbulent conditions, and a sink-based model, in which star formation is governed by gas inflows onto sink particles.
The sink-based model naturally produces bursty star formation through rapid accretion onto young sink particles embedded in strongly convergent gas flows. The resulting intense radiation ionizes and disperses star-forming clumps through photoionization heating before the first supernova explodes. Consequently, supernovae occur in lower-density environments, imparting greater terminal momentum and driving stronger galactic outflows. In contrast, star formation within individual gas clumps is less efficient in the Schmidt-type model because, despite ongoing clump-scale collapse, individual star formation events locally modify cell conditions, temporarily suppressing subsequent star formation and lowering the degree of burstiness. Relative to the Schmidt-type model, the sink-based model yields a total stellar mass lower by a factor of $\sim3$ and a Lyman continuum escape fraction higher by a factor of $\sim10$ by $z=6$. The bursty model drives stronger metal-enriched outflows and suppresses excess central star formation, exhibiting better agreement with constraints from JWST observations in gas-phase metallicity and galaxy size.
Our results suggest that bursty star formation is a key mechanism for enhancing feedback and alleviating the overcooling problem in galaxy formation simulations.
}

\keywords{galaxies: high-redshift -- galaxies: formation -- galaxies: evolution -- galaxies: star formation}

\maketitle
\nolinenumbers

\section{Introduction}

Galaxy formation models must capture the complex interplay of multiple physical processes operating across a wide range of spatial and temporal scales, which makes constructing accurate models in numerical simulations a formidable challenge \citep{Somerville2015, Naab2017, Vogelsberger2020, Crain2023}. Among these, star formation within the interstellar medium (ISM) and its associated feedback are of particular importance, as they not only govern the local conversion of cold gas into stars but also the regulation of the thermal and dynamical states of galaxies on considerably larger scales \citep{Krumholz2014, Girichidis2020}.

The fraction of baryons converted into stars is among the most stringent tests of baryonic physics in galaxy formation models because it directly reflects the integrated effect of star formation and feedback. Observational constraints on the stellar-to-halo mass relation--derived from abundance matching techniques \citep{Moster2013, Behroozi2013, Behroozi2019, Girelli2020, Moster2021}, weak gravitational lensing \citep{Mandelbaum2006, vanUitert2016}, halo occupation modeling \citep{Leauthaud2012}, satellite kinematics \citep{More2011, Karachentsev2021}, or \ion{H}{i} rotation curves \citep{Read2017} indicate that galaxy formation is fundamentally inefficient. The stellar-to-halo mass ratio peaks at only a few percent for dark matter halos (DMHs) with masses of $M_\mathrm{halo}\sim10^{12}\,\Msun$ and declines toward both lower and higher halo masses, albeit by different physical processes \citep[see][for a comprehensive review]{Wechsler2018}.

Reproducing these constraints has remained a persistent challenge in numerical simulations largely because of the overcooling problem \citep[e.g.,][]{White1991, Katz1992, Balogh2001, Ceverino2009}, in which excessive radiative cooling leads to unrealistically efficient star formation.
This issue is often attributed in part to limited numerical resolution when shock fronts or multiphase gas structures are unresolved and artificially mixed \citep{Creasey2011, KimCG2015, Rey2024}. Developing a galaxy formation model that remains robust across a wide dynamic range is non-trivial because such numerical artifacts are inherently resolution-dependent. Consequently, substantial effort has been devoted to mitigating overcooling through the empirical calibration of feedback prescriptions. For example, the SPHINX simulation \citep{Rosdahl2018} calibrated its stellar feedback to reproduce the observed UV luminosity function at $z=6$ by boosting the supernova (SN) rate by a factor of four relative to that expected for the adopted initial mass function (IMF). A suite of THESAN-zoom simulations \citep{Kannan2025} introduced additional early feedback beyond stellar winds, radiation, or SNe, reproducing the predicted stellar-to-halo mass relation at $z\gtrsim3$. At lower redshifts, EAGLE \citep{Schaye2015} and IllustrisTNG \citep{Pillepich2018} explicitly calibrated feedback parameters associated with black holes and SNe to reproduce key observables such as the stellar mass function and gas mass fraction at $z=0$. More recently, the COLIBRE galaxy formation model \citep{Schaye2025} employed machine learning to perform resolution-dependent calibration. Although such approaches effectively counterbalance numerical overcooling and regulate star formation, they rely on ad hoc parameterizations of baryonic physics, limiting the physical interpretation that can be drawn from simulations.

Stellar feedback is inherently driven by star formation, and therefore, an alternative strategy to improve galaxy formation models is to refine the star formation prescriptions themselves. By running cosmological zoom-in simulations with a stochastic Schmidt-law-based star formation model \citep{Schmidt1959, Kennicutt1998}, \citet{Agertz2015, Agertz2016} showed that adopting a higher local star formation efficiency (SFE) per free-fall time ($\epsff=10\%$) can reduce the final stellar mass by up to an order of magnitude compared to models with $\epsff=1\%$ in a $\sim10^{12}\,\Msun$ DMH at $z=0$. Such self-regulation is achieved through strongly correlated feedback, and this can only be realized when the local SFE is sufficiently high to trigger clustered star formation. Adopting an SFE of $\epsff=100\%$ in gravitationally bound regions, FIRE simulations \citep{Hopkins2014, Hopkins2018} also achieved the robust regulation of galaxy growth across a wide range of masses and redshifts. These examples underscore that the efficiency of stellar feedback is tightly coupled to the mode of star formation, with burstier star formation producing stronger feedback. Indeed, \citet{Muratov2015} and \citet{Pandya2021} showed that the galactic outflow for a given SFR (i.e., mass loading factor) in FIRE-1 and FIRE-2 galaxies, respectively, is stronger at higher redshifts and in lower-mass galaxies, where star formation histories (SFHs) are characterized by episodic bursts rather than continuous activity. Analytic modeling by \citet{Faucher2018} further suggested that burstier star formation at high redshift enhances the ability of clustered SNe to launch powerful galactic winds, while numerical studies directly demonstrated the importance of SN clustering in driving galactic outflows and regulating star formation \citep{Hu2019, Smith2019, Smith2021}.
The physical mechanism by which SN clustering enhances feedback efficiency has been investigated in idealized experiments. For instance, one-dimensional hydrodynamic simulations assuming spherical symmetry by \citet{Gentry2017} showed that the momentum injection per SN can be boosted by a factor of $\sim4$--10 when multiple explosions occur within a common superbubble \citep[see also][]{Keller2014, Fielding2018}. In three-dimensional simulations, \citet{KimCG2017} found that the mass and thermal energy of hot gas per SN increase with increasing SN cadence, particularly when the time interval between explosions is shorter than the shell formation timescale, beyond which dense gas efficiently loses energy via radiative cooling.

Recent studies have also suggested that such bursty and clustered star formation may be essential for reproducing the UV luminosity function at high redshifts. Using semi-empirical and numerical models, \citet{Shen2023} and \citet{Kravtsov2024} argued that substantial UV variability ($\sigma_\mathrm{UV}$) is required to explain the abundance of UV-bright galaxies observed by the JWST at $z\ge10$ \citep[e.g.,][]{Naidu2022, Finkelstein2023, Harikane2024}. Simulating the same galaxy with different values of $\epsff$, \citet{Semenov2025bursty} showed that high $\sigma_\mathrm{UV}$ at high redshifts was reproduced only when \epsff\ is sufficiently high ($\ge10\%$), as clustered star formation and feedback drive large temporal fluctuations in SFR and the corresponding UV luminosity \citep[see also][]{Sun2023}. This suggests that capturing clustered star formation in galaxy formation simulations may play a central role not only in regulating feedback and galaxy growth, but also in shaping the UV variability and luminosity functions of galaxies in the early universe.

Despite the success of simulations that employ locally efficient star formation, many models still rely on simple density-based prescriptions with a constant efficiency. Although these models are computationally efficient and straightforward to implement, they often neglect key physical processes that regulate gravitational collapse, such as turbulent pressure and the degree of gravitational boundedness \citep{MacLow2004}. To address this limitation, more sophisticated ``multi-freefall'' models have been developed in which $\epsff$ varies according to local ISM conditions \citep{KM05, PN11, HC11, FK12, Semenov2016}. However, despite their ability to capture aspects of local gas dynamics, the practical implementation of a variable $\epsff$ introduces several sources of uncertainty. First, the contribution from unresolved turbulence to supporting gas against gravitational collapse must be estimated \citep[cf. see][for subgrid modeling of the unresolved turbulence]{Kretschmer2020, Girma2024, Semenov2025turb}. Second, since the high-density tail of the probability distribution function (PDF) of gas density arises from supersonic shocks that compress the gas \citep[e.g.,][]{Padoan1997, Passot1998, Federrath2008}, the evaluation of $\epsff$ may become unreliable in subsonic regimes \citep[e.g.,][]{FK12, Molina2012}. Furthermore, while multi-freefall models assume a log-normal PDF \citep{Vazquez1994, Nordlund1999, Padoan2002}, it has been widely suggested that the high-density tail, where gravity dominates over turbulence, is better described by a power-law distribution \citep{Kainulainen2009, Ballesteros-Paredes2011}. More recently, \citet{Hennebelle2024} and \citet{Brucy2024} demonstrated that deviations from log-normality become increasingly pronounced with increasing Mach number, as strong shocks induce sharper density contrasts \citep[see also][]{Castaing1996, Hopkins2013PDF}. Consequently, the inferred mass fraction of gas available for gravitational collapse may vary significantly \citep{Burkhart2018}, introducing additional uncertainty into the determination of $\epsff$.

Instead of relying on prescriptions calibrated from small-scale, high-resolution simulations, an alternative approach is to explicitly model star formation using sink particles \citep{Bate1995, Bromm2002, Krumholz2004, Federrath2010, Gong2013, Bleuler2014, Grudic2021}. In this framework, high-density regions undergoing gravitational collapse are replaced by point masses that grow through accretion of surrounding gas. Because the gas accretion rate, and thus the local SFR, is determined directly by the simulated gas density and velocity fields, the sink particle method naturally links local gas dynamics to star formation. Owing to its effectiveness in tracking the growth of individual stars in high-resolution simulations, this method has been widely used to study the formation of star clusters \citep{Myers2014, Guszejnov2022cluster}, the IMF \citep{Bonnell2011, Guszejnov2022IMF}, and protostellar outflows \citep{Federrath2014, Guszejnov2021}. However, because resolving gravitational collapse requires a sufficiently high spatial resolution, sink particle methods have not been extensively applied in cosmological simulations. In our previous work \citep{Kang2025}, we demonstrated that sink-based star formation is viable and numerically convergent in cosmological simulations at resolutions $\lesssim10\,\mathrm{pc}$ and that it promotes burstier SFHs. Nevertheless, we found that this approach is not inherently more effective at reducing the stellar mass than a multi-freefall model unless the feedback energy is artificially enhanced. This limitation arises in part because the target DMH was relatively small ($10^9\,\Msun$) such that feedback could efficiently expel gas even in the absence of spatially correlated feedback. To further investigate these effects, we extend the work of \citet{Kang2025} to a system an order of magnitude more massive ($10^{10}\,\Msun$) while introducing several modifications to the sink-based star formation model. The primary objective of this study is to demonstrate that clustered star formation on a short timescale can mitigate overcooling without requiring additional calibration.

The remainder of this paper is organized as follows. In Sect.~\ref{sec:simulations}, we describe the simulations and numerical methods, including the two star formation models employed: a Schmidt-type multi-freefall model and a sink-particle model. Sect.~\ref{sec:results} presents our main results, demonstrating how global galaxy properties depend on the adopted star formation model. We further analyze star formation activity on clump scale and its implications for star cluster formation and clump dispersal. In Sect.~\ref{sec:discussion}, we examine the numerical convergence of the sink-based model and discuss the physical origin of bursty star formation, as well as the impact of accretion efficiency on the growth of stellar mass. Finally, we summarize our conclusions in Sect.~\ref{sec:summary}.

\section{Simulations}
\label{sec:simulations}
We perform cosmological radiation hydrodynamic simulations using the adaptive mesh refinement code \textsc{ramses-rt} \citep{Teyssier2002, Rosdahl2013, Rosdahl2015} to investigate the effect of bursty star formation on galaxy properties. Euler and Poisson equations are solved using the Harten-Lax-van Leer-contact scheme \citep{Toro1994} and a multigrid method \citep{Guillet2011}, respectively, with a Courant factor of 0.8. For radiative transfer, we adopt the M1 closure relation \citep{Levermore1984} with the GLF flux solver. We adopt a reduced speed of light equal to 1\% of its true value to accelerate the simulations while keeping it sufficiently larger than gas velocities to accurately capture radiation feedback in star-forming regions. Photons are grouped into six energy bins (5.6--11.2, 11.2--13.6, 13.6--15.2, 15.2--24.59, 24.59--54.42, 54.42--$\infty$ eV); these groups are selected to model momentum transfer, photoionization, non-equilibrium chemistry, and cooling rates self-consistently for $\mathrm{H_2}$, \ion{H}{i}, \ion{H}{ii}, \ion{He}{i}, \ion{He}{ii}, \ion{He}{iii}, and $e^{-}$ \citep{Katz2017, Kimm2017}. In addition to primordial species, we include cooling from metal species, where the cooling rates at $T>10^4\,\mathrm{K}$ and $T\le10^4\,\mathrm{K}$ are estimated based on \citet{Ferland1998} and \citet{Rosen1995}, respectively. Finally, we include a redshift-dependent uniform UV background radiation \citep{Haardt2012} when computing photoionization and heating rates, with a self-shielding factor of $\exp(-n_\mathrm{H}/0.01\,\cmq)$ applied to each cell.

\subsection{Cosmological initial conditions}
We generate the initial condition for our cosmological zoom-in simulations using {\sc genetIC} \citep{Stopyra2021} and assume a cosmology compatible with \citet{Planck2020} ($\Omega_\mathrm{m}=0.3158$, $\Omega_\Lambda=0.6842$, $H_0=67.32\,\kms\,\mathrm{Mpc^{-1}}$, $\Omega_\mathrm{b}=0.045$). First, we run a simulation with $256^3$ dark matter particles in a $(30\,\mathrm{cMpc})^3$ box and select a halo with a virial mass of $\Mvir\sim10^{10}\,\Msun$ at $z=6$. We select a region of three virial radii around this halo, in which we increase the mass resolution of dark matter particles to $m_\mathrm{dm}\sim1600\,\Msun$. No coarser dark matter particles exist within this region at the end of our simulations. Moreover, we use the procedure from \citet{Pontzen2021} to position this zoomed region at rest with respect to the large-scale structure of the initial conditions. This ensures that the bulk velocity of our main halo at $z=6$ is $\sim16\,\kms$, which is less than a quarter of the virial velocity ($\sim70\,\kms$), thereby minimizing advection errors when integrating with a grid code \citep[see discussion in][]{Pontzen2021}.


We identify DMHs using an {\sc AdaptaHop} algorithm \citep{Tweed2009}. $\Rvir$ is defined as the radius of a sphere whose mean density is equal to $\Delta_\mathrm{crit}\rho_\mathrm{crit}$, where $\Delta_\mathrm{crit}=18\pi^2+82x-39x^2$ represents the virial overdensity \citep{Bryan1998}, $x\equiv\Omega_\mathrm{m}/(\Omega_\mathrm{m}+a^3\Omega_\Lambda) - 1$, and $\rho_\mathrm{crit}(z)$ represents the critical density of the universe as a function of redshift. Although the halo is initially centered at the density peak of dark matter particles, we redefine the halo center as the center of mass of star particles within $0.2\Rvir$ once the halo hosts more than 100 star particles ($10^5\,\Msun$). The corresponding $\Rvir$ is recomputed from the redefined center, and the virial mass is defined as $\Mvir=4\pi R^3_\mathrm{vir}\Delta_\mathrm{crit}\rho_\mathrm{crit}(z)/3$.

The simulation box is initially divided into $256^3$ cells and further refined within the zoom-in region if at least one of the following two criteria is met. First, a cell is refined when the thermal Jeans length ($\lambda_\mathrm{J}=\sqrt{\pi c^2_\mathrm{s}/G\rho_\mathrm{gas}}$) becomes shorter than 16 times the cell size, where $c_\mathrm{s}$ and $\rho_\mathrm{gas}$ represent the isothermal sound speed and gas density of a cell, respectively. Second, refinement is triggered when the mass of dark matter particles plus baryons divided by the baryon fraction ($\Omega_\mathrm{b}/\Omega_\mathrm{m}$) within a cell exceeds $8m_\mathrm{dm}\sim12400\,\Msun$. In addition to these criteria, cells located within the accretion zone of each sink particle are always refined to the maximum level for capturing the accretion process with the highest accuracy in simulations with sink particles (Sect.~\ref{sec:SINK}).

A higher level of refinement is unlocked at $z=19$ and 9 ($a=0.05$ and 0.1) to maintain the physical resolution, $\dxmin$, nearly constant throughout simulations. Given this refinement strategy, $\dxmin$ in our fiducial runs varies within a fixed range of 5.7 and 11.4 pc from $z=39$ to 6. Although simulations with a star formation model based on a simple density threshold may suffer from an artificial spike in the SFR after a new level of refinement is released \citep{Snaith2018}, such a hold-back effect is not observed in our simulations because the star formation criteria in our two models require a correspondingly higher density threshold for the smaller, denser cells that emerge at a higher level of refinement.

\subsection{Star formation models}
\label{sec:SFmodel}

In our simulations, gas metallicity is initially set to zero to model the formation of Population (Pop) III stars. The physical prescription for Pop~III star formation is identical throughout this work and is adopted from \citet{Kimm2017}. The formation criteria and computation of $\epsff$ are the same as those described in Sect.~\ref{sec:GTT}. The primary difference from Pop~II formation is that we stochastically sample individual Pop~III masses between $10$ and $10^3\,\Msun$ from an IMF given by $P(\log{m}) \propto m^{-1.3}\,\exp{[-(m/100\,\Msun)^{-1.6}]}$ instead of forming a stellar particle representing a simple stellar population. This functional form corresponds to a Salpeter-like IMF \citep{Salpeter1955} with an exponential suppression below a characteristic mass of $100\,\Msun$ \citep{Wise2012}. Pop~III star formation is restricted to gas with metallicity $Z_\mathrm{gas}\le 10^{-6}\,\Zsun$, where $\Zsun=0.0134$ denotes the solar metallicity \citep{Asplund2009}.

For Pop~II stars, we adopt two distinct formation models: one based on a Schmidt law with a variable \epsff\ and the other based on a sink particle algorithm. The former incorporates local ISM conditions to provide a more physically motivated description of star formation and has been adopted in several recent cosmological simulations, such as SPHINX \citep{Rosdahl2018}, NewHorizon2 \citep{Yi2024}, and NewCluster \citep{HanS2026}. The latter is motivated by star formation studies that explicitly model the growth of individual star particles through gas accretion. We summarize the main features below and refer the reader to \citet{Kang2025} for a detailed description of our methodology.

\subsubsection{Multi-freefall model with gravo-thermo-turbulent conditions}
\label{sec:GTT}

The first model is based on the traditional Schmidt law but adopts a variable \epsff\ to account for local gravo-thermo-turbulent (GTT) conditions. To assess whether gravitational collapse can proceed against thermal and turbulent support, the model compares the local cell size with the turbulent Jeans length \citep{Bonazzola1986} at the highest refinement level.
\begin{equation}
\label{eq:GTT}
    \lambda_\mathrm{J,turb}=\frac{\pi \sigma^2_\mathrm{gas}+\sqrt{\pi^2 \sigma^4_\mathrm{gas} + 36\pi G \rho_\mathrm{tot} c^2_\mathrm{s} \dxminsq}}{6G \rho_\mathrm{tot} \dxmin} \le \dxmin,
\end{equation}
where $\sigma_\mathrm{gas}$ and $\rho_\mathrm{tot}$ represent the velocity dispersion and total mass density of the cell, respectively. We approximate $\sigma_\mathrm{gas}$ as the Frobenius norm of the density-weighted velocity gradient tensor $\vec{\nabla} \mathbf{v}$ scaled by the cell size: $\sigma_\mathrm{gas}=\dxmin\,\|\vec{\nabla} \mathbf{v}\|$. The rank-2 tensor $\vec{\nabla} \mathbf{v}$ is constructed from the velocity differences across the faces of six neighboring cells. This calculation is performed after subtracting symmetric divergence along the three axes and rotation velocity along the three planes. We restrict star formation to cells that exceed a density threshold ($n_\mathrm{H} \ge 100\,\cmq$), are undergoing converging flows (i.e., have negative velocity divergence), and are denser than their six neighboring cells. The specific choice of the density threshold does not change our results because stars are born within cells with considerably higher densities in this model because of Eq.~\eqref{eq:GTT}.

Once potential star-forming cells are identified, the conversion efficiency in each cell is computed based on the multi-freefall theory, where $\epsff$ is governed by the relative strength of gravity, thermal pressure, and turbulent motions \citep{PN11, FK12}. $\epsff$ is obtained by integrating the log-normal PDF of turbulent density fluctuations $p(s\equiv\ln(\rho/\langle\rho\rangle))$ weighted by $\rho/\tff(\rho)$ from $s=s_\mathrm{crit}$ to $\infty$.
\begin{equation}
\label{eq:epsff}
    \epsff=\frac{\varepsilon_\mathrm{acc}}{2\phi_\mathrm{t}} \, \exp \left(\frac{3}{8}\sigma^2_\mathrm{s} \right) \left[1+\mathrm{erf} \left(\frac{\sigma^2_\mathrm{s}-s_\mathrm{crit}}{\sqrt{2\sigma^2_\mathrm{s}}}\right)\right].
\end{equation}
The lower limit of the integral $s_\mathrm{crit}$ is the critical density above which gas is assumed to collapse, and $\sigma_\mathrm{s}$ represents the standard deviation of $p(s)$. The parameters $\varepsilon_\mathrm{acc}=0.5$ and $1/\phi_\mathrm{t}=0.57$ are the best-fit values calibrated to high-resolution simulations \citep{FK12}, where the former corresponds to the fraction of collapsing gas that is accreted onto a star, and the latter accounts for the uncertainty in the free-fall time factor $\tff(\langle\rho\rangle)/\tff(\rho)$.
The mass of a star particle is determined following Poisson statistics, $P(N)=\frac{\lambda^N}{N!}\exp(-\lambda)$, and the expectation value $\lambda$ is derived based on a Schmidt law.
\begin{equation}
\label{eq:poiss}
    \lambda=\frac{dM_*}{dt}\frac{\Delta t_\mathrm{step}}{m_\mathrm{PopII,min}}=\epsff\frac{M_\mathrm{gas}}{\tff}\frac{\Delta t_\mathrm{step}}{m_\mathrm{PopII,min}},
\end{equation}
where $\Delta t_\mathrm{step}$, $m_\mathrm{PopII,min}=1000\,\Msun$, and $M_\mathrm{gas}$ represent the timestep of the simulation, the minimum mass of a Pop~II star particle, and gas mass of the cell, respectively. The mass of each star particle is given by an integer multiple $N$ of $m_\mathrm{PopII,min}$.
With these criteria, we find that the majority of star particles ($\approx 80\%$ in mass) are formed at the minimum allowed stellar mass in the fiducial run.

\subsubsection{Sink-based model with a flux accretion scheme}
\label{sec:SINK}
The second model adopts a sink particle approach \citep{Bleuler2014}. Gas clumps are identified using the \textsc{phew} algorithm \citep{Bleuler2015} when the hydrogen number density at the density peak exceeds $n_\mathrm{H} \ge 100\,\cmq$. A sink particle with an initial seed mass of $0.1\,\Msun$ is allowed to form at the density peak of a collapsing virialized clump if the cell density exceeds $8.86 c^2_\mathrm{s}/\pi G \dxminsq$, which is the density of a collapsing isothermal sphere at a distance of $0.5\dxmin$ from the center \citep{Larson1969, Penston1969, Gong2013}. Sink particles grow in mass through gas accretion, where the accretion rate is given by the net mass flux across the boundary of the accretion zone.
\begin{equation}
\label{eq:acc}
    \dot M_\mathrm{acc} = \oint \rho_\mathrm{gas} (\vec{v}_\mathrm{gas}-\vec{v}_\mathrm{sink}) \cdot \vec{dA},
\end{equation}
where $\vec{v}_\mathrm{gas}$ and $\vec{v}_\mathrm{sink}$ are the velocities of the gas cells and sink particle, respectively\footnote{In \citet{Kang2025}, accretion onto sink particles was modeled using both a flux-based scheme and the Bondi-Hoyle-Lyttleton rate \citep{Hoyle1939, Bondi1952}. In this study, we do not adopt the latter because the ambient gas properties at infinity, such as velocity and density, are not well defined in turbulent media \citep{Krumholz2004}. Therefore, we adopt a purely flux-based approach that provides a more self-consistent description by directly capturing local gas flows.}.
The accretion zone is defined as a sphere with a radius of $r_\mathrm{acc}=\dxmin$.
Within this volume, 33 cloud particles of equal mass are uniformly distributed along the three axes with a spacing of $0.5\dxmin$ to sample the local flow. This corresponds to the integral in Eq.~\eqref{eq:acc} over the 11 nearest cells that contain at least one cloud particle.

Once a sink particle accretes more than $1000\,\Msun$, a star particle is spawned, and the corresponding mass is removed from the sink particle. To prevent numerical instabilities and excessive gas depletion, we limit the timestep such that no more than 75\% of the gas mass from any cell is removed by accretion within a single step \citep{Bleuler2014}. Finally, two sink particles are allowed to merge when their separation becomes smaller than $2r_\mathrm{acc}$.


\subsection{Stellar feedback}
Our simulations include multiple feedback processes from massive stars, namely stellar radiation, Type II SNe, and mass-dependent explosions of Pop III stars. The feedback prescriptions adopted in this study are summarized below. Further details are provided in \citet{Rosdahl2013, Kimm2017}.

Radiation feedback is modeled by injecting photons from each star particle, where the photon flux per unit mass at a specific wavelength varies with age and metallicity. We use the predefined tables of \citet{Schaerer2002} and \citet[][Binary Population and Spectral Synthesis v2.2.1]{Stanway2018} to compute radiation from Pop~III and Pop~II stars, respectively. Each Pop~II star particle represents a simple stellar population with a Kroupa IMF \citep{Kroupa2002}, assuming an upper (lower) mass limit of $300\,\Msun$ ($0.1\,\Msun$). The probability that a photon interacts with a given species is proportional to the product of its number density and wavelength-dependent photoionization cross-section \citep{Hui1997}. Lyman continuum (LyC) photons ionize \ion{H}{i}, \ion{He}{i}, \ion{He}{ii}, and $\mathrm{H_2}$ depending on their photon energies \citep[see Table~1 of][]{Katz2017}. We also model the photodissociation of H$_2$ by Lyman-Werner radiation with energies $11.2 < E < 13.6\,\mathrm{eV}$. Any excess photon energy remaining after ionization or dissociation is transferred to free electrons, thereby heating the gas. In addition to photoionization heating, both ionizing and non-ionizing photons exert direct radiation pressure by transferring momentum to the gas upon absorption.
Non-thermal pressure due to trapped infrared photons or Lyman-$\alpha$ is not considered in this study.

For Pop~II stars, core-collapse SNe are assumed to originate from massive stars with masses $\ge 8\,\Msun$. We adopt a rate of one SN per $100\,\Msun$ of stars formed, which is consistent with the Kroupa IMF used in this study. Progenitor lifetimes spanning $\approx4$ to 50 Myr are randomly sampled based on \citet{Leitherer1999}. SN explosions are implemented using the mechanical feedback scheme of \citet{Kimm2014, Kimm2015}, in which the terminal radial momentum is deposited into cells neighboring the SN sites. The terminal radial momentum depends on the local gas properties as
\begin{equation}
\label{eq:SN}
    P_\mathrm{SN}=2.5\times10^5\,\Msun\,\kms\, E_\mathrm{51}^{16/17} n_\mathrm{H}^{-2/17} Z'^{-0.14},
\end{equation}
where $E_\mathrm{51}$, $n_\mathrm{H}$, and $Z'=\max[0.01,\,Z/\Zsun]$ are the SN energy normalized to $10^{51}\,\mathrm{erg}$, hydrogen number density in units of \cmq, and gas-phase metallicity normalized to the solar value, respectively. The chemical yield from each explosion is set to $\eta_Z=0.075$.

The explosion of Pop~III stars releases different amounts of energy and metals depending on the progenitor mass. We adopt core-collapse SNe for $11-20\,\Msun$ \citep{Nomoto2006}, hypernovae for $20-40\,\Msun$ \citep{Nomoto2006}, and pair-instability SNe for $140-260\,\Msun$ \citep{Heger2002}. The corresponding energy and metal yields are given by
\begin{equation}
\begin{cases}
\begin{array}{l}
E_\mathrm{SN} = 10^{51}\,\mathrm{erg} \\
M_{Z,\mathrm{ejec}} = 0.1077+0.3383(m_\mathrm{PopIII}-11)\,\Msun \hfill (11 \le m_\mathrm{PopIII} < 20) \\[3pt]

E_\mathrm{SN} = (1.08571m_\mathrm{PopIII}-13.7143)\times10^{51}\,\mathrm{erg} \\
M_{Z,\mathrm{ejec}} = 0.27935m_\mathrm{PopIII}-2.76499\,\Msun \hfill (20 \le m_\mathrm{PopIII} < 40) \\[3pt]

E_\mathrm{SN} = [5+1.304(\frac{13}{24}(m_\mathrm{PopIII}-20)-64)]\times10^{51}\,\mathrm{erg} \\
M_{Z,\mathrm{ejec}} = \frac{13}{24}(m_\mathrm{PopIII}-20)\,\Msun \hfill (140 \le m_\mathrm{PopIII} < 260)
\end{array}
\end{cases}
\end{equation}
where $m_\mathrm{PopIII}$ is the mass of Pop~III star particles in units of $\Msun$. Pop~III stars outside these mass ranges are assumed to collapse directly into black holes without releasing any energy or metals.

Table~\ref{table1} summarizes the four simulations presented in this study. We perform one simulation without stellar feedback (\gttco) to serve as a reference case in which SN explosions and radiative feedback are disabled while metal enrichment is still included. By comparing our two fiducial runs (\gtt\ and \sink), we examine the relative strength of stellar feedback and its connection to the burstiness of star formation. We also carry out a higher-resolution simulation (\sinkhr) to assess resolution effects and the convergence of both the accretion process and the total stellar mass. Our analysis focuses on the two fiducial runs in Sect.~\ref{sec:results}, whereas the \sinkhr\ run is compared with the \sink\ run in Sect.~\ref{sec:res}.

\begin{table}
\caption{Summary of the simulation setup.}
\label{table1}
\centering
\begin{tabular*}{\linewidth}{@{\extracolsep{\fill}} l c c c c}
\hline\hline
Simulation &  $\dxmin$ & SF & RT & SN \\
Label      & [pc]     &     &    & \\
\hline
\gttco     & 5.7--11.4 & Gravo-thermo-turbulent & \ding{55} & \ding{55} \\
\gtt       & 5.7--11.4 & Gravo-thermo-turbulent &\ding{51} & \ding{51} \\
\sink      & 5.7--11.4 & Flux accretion &\ding{51} & \ding{51} \\
\sinkhr    & 2.8--5.7  & Flux accretion & \ding{51} & \ding{51} \\
\hline
\end{tabular*}
\tablefoot{From left to right: simulation label,  maximum spatial resolution, star formation model used, inclusion of radiation feedback, and inclusion of SN explosions. All simulations adopt a dark matter particle mass of $1616\,M_\odot$ and an initial star particle mass of $1000\,M_\odot$.}
\end{table}

\section{Results}
\label{sec:results}
We demonstrate overall structures and galaxy-integrated properties sensitive to star formation activities and feedback strength, providing a general picture of how two fiducial simulations differ. We perform a more detailed analysis by constructing linking trees of star clusters and gas clumps to illustrate how bursty star formation enhances the strength of feedback channels and stops star formation within individual star-forming sites.

\subsection{Global trends in galactic structure and mass assembly}
\begin{figure*}
    \hfill
    \includegraphics[height=0.95\textheight]{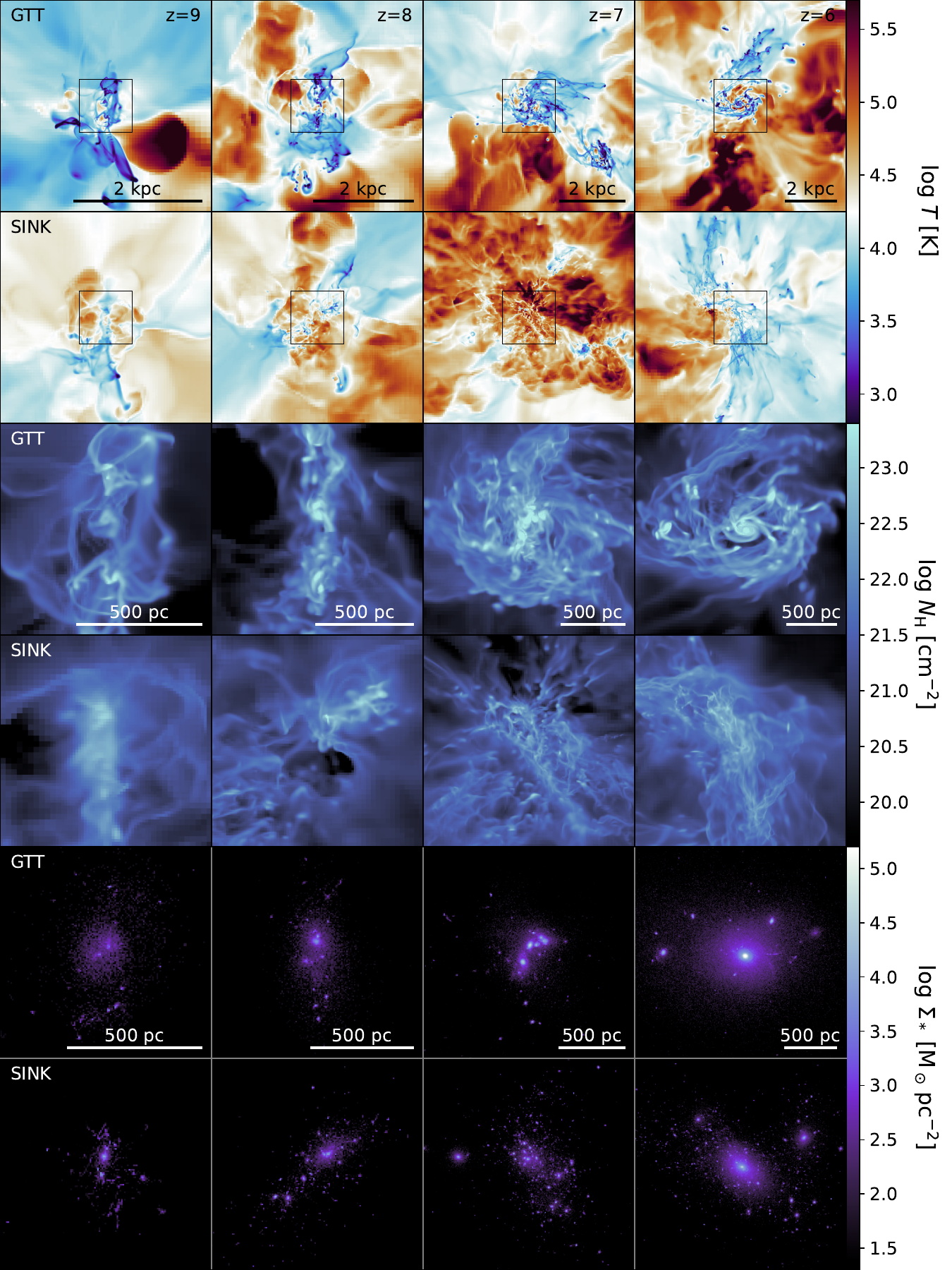}
    \caption{Redshift evolution of the main galaxy from $z=9$ (left) to $z=6$ (right). The rows display the mass-weighted temperature (top), hydrogen column density (middle), and stellar mass surface density (bottom). In the top rows, the central square indicates a region with a side length of $0.2\Rvir$. The middle and bottom rows show zoomed-in views within the area enclosed by the square.}
    \label{fig:map}
\end{figure*}

Fig.~\ref{fig:map} shows the the mass-weighted temperature in the central region of the main halo at four different redshifts, along with the hydrogen column density and stellar mass surface density for a smaller volume.
A clear difference emerges immediately: at every epoch shown, the galaxy center in the \sink\ run is significantly warmer and less clumpy than that in the \gtt\ run. The circumgalactic medium (CGM) is more extended in the \sink\ run because the gas is more effectively blown away from the center. In the \gtt\ run, massive clumps ($\ge 10^7\,\Msun$) develop at $z\sim7$ in the galaxy center as the main halo continues to grow. Significant star formation occurs and the stellar mass surface density reaches $>10^{4.5}\,\Msun\,\mathrm{pc^{-2}}$ at this galaxy center. Meanwhile, a satellite halo whose evolution diverges between the two simulations enters the virial sphere at $z\sim7.4$ and approaches $0.2\,\Rvir$. Gas clumps in this satellite galaxy are again considerably denser and colder in the \gtt\ run, and they survive until the end of the simulation. Within these clumps, star formation continues for more than 400 Myr (see Sect.~\ref{sec:clump&cluster} for a detailed clump analysis). They eventually settle at the center of the main halo at $z=6$, forming a dense and large core with a prominent rotational feature. This occurs because the core is directly connected to dense cold gas streams falling in from larger distances, which effectively increases the angular momentum of the central region of the galaxy and fuels star formation \citep[e.g.,][]{Kimm2011, Danovich2015}. This yields a peak stellar surface density of $\ge 10^{5.5}\,\Msun\,\mathrm{pc^{-2}}$ at the final snapshot. In contrast, both the cold inflows and rotational feature are significantly weaker in the \sink\ run. This is because gas clumps within the same satellite are destroyed completely by $z=6.5$ due to stronger feedback. Although they collapse again to some extent and migrate to the galaxy center at $z=6$ as in the \gtt\ run, their contribution to star formation is drastically lower, which results in a central stellar surface density of $\sim10^4\,\Msun\,\mathrm{pc^{-2}}$. In the \sink\ run, star formation from smaller clumps results in a larger number of local peaks with smaller sizes and lower surface densities located off-center.

\begin{figure}
    \includegraphics[width=\linewidth]{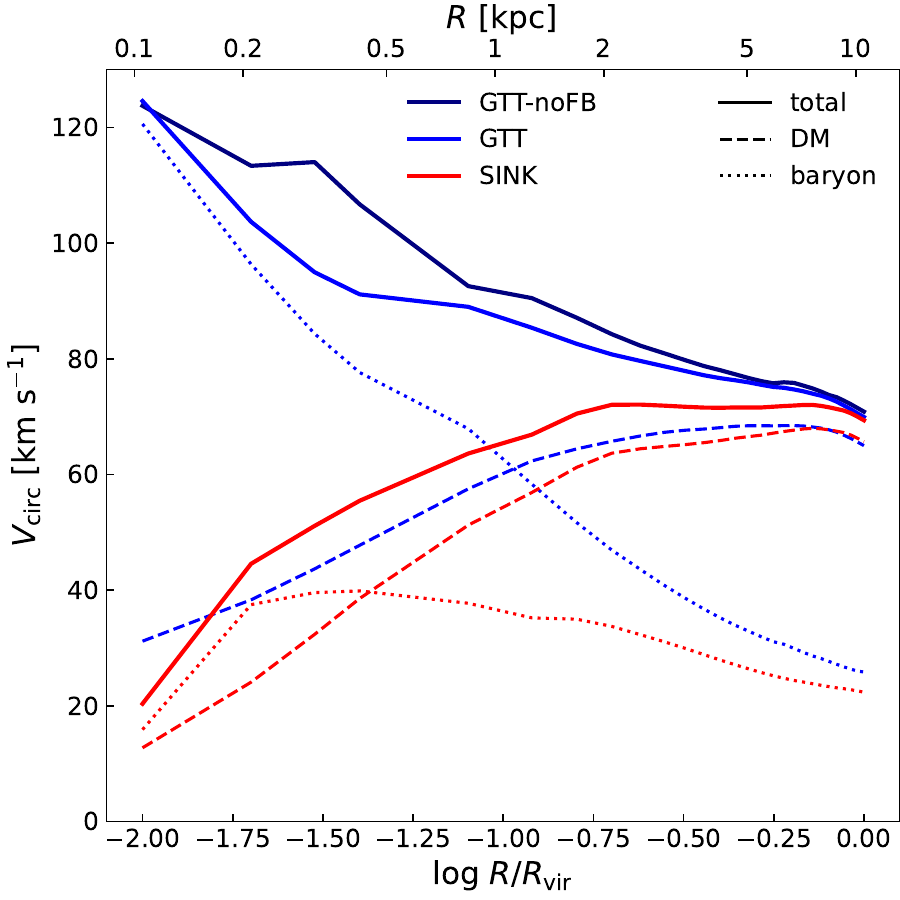}
    \vspace{-5mm}
    \caption{Circular velocities of total (solid), dark matter (dashed), and baryons (gas + stars, dashed) as a function of radius at $z=6$ for the \gtt\ (blue) and \sink\ (red) simulations. The total circular velocity of the \gttco\ run is shown as a solid navy line for comparison. The simulation employing sink-particle-based star formation exhibits a smoothly rising rotation curve.}
    \label{fig:Vcirc}
\end{figure}

We present the circular velocity profiles ($V_\mathrm{circ}$) of the dark matter and baryonic components at $z=6$ in Fig.~\ref{fig:Vcirc} to illustrate the mass distribution within the main halo more quantitatively. We also show $V_\mathrm{circ}$ for \gttco\ for comparison because this run serves as an extreme case in which mass cannot be transported outwards by stellar feedback. Although the total enclosed mass (baryon + dark matter) within the virial sphere is the same in all three runs, the mass distributions at smaller radii differ between simulations. Compared with the \gttco\ run, both feedback runs (\gtt\ and \sink) exhibit lower total $V_\mathrm{circ}$ as stellar feedback prevents gas collapse toward the galaxy center.
However, this effect is considerably stronger in the \sink\ run than in the \gtt\ run. The velocity profile of the \gtt\ galaxy decreases monotonically with an increasing radius, and the circular velocity at the galactic center ($R=0.01\Rvir\sim100\,\mathrm{pc}$) is nearly indistinguishable from that of the \gttco\ run, indicating inefficient feedback in the \gtt\ run. In contrast, the \sink\ run exhibits an increasing profile for the total $V_\mathrm{circ}$ toward larger radii, which suggests an effective mass removal from the center by stronger feedback. This interpretation is further supported by the circular velocity originating from the baryonic component, which remains consistently lower at all radii in the \sink\ run.

Although dark matter particles are not directly affected by baryonic processes, we find that their mass density is lower in the \sink\ run with the discrepancy from the \gtt\ run widening toward the center. This occurs because the gravitational potential in the inner region is dominated by baryons, and the redistribution of baryonic mass via stellar feedback alters the orbits of dark matter particles \citep{Pontzen2012, Chan2015}.
The baryonic component shows an even more pronounced difference between the two models. Unlike the dark matter mass, the circular velocity contributed by baryons continues to show a marked offset that persists to the virial radius. Although the total stellar mass in the \sink\ run is smaller by a factor of $\sim3.2$ at the virial radius, the mass contrast becomes even more extreme in the inner regions, where the \gtt\ run produces a compact stellar core. Together, the \gtt\ run forms a highly concentrated DMH and galaxy, whereas the \sink\ run produces more spatially extended features.

\begin{figure}
    \includegraphics[width=\linewidth]{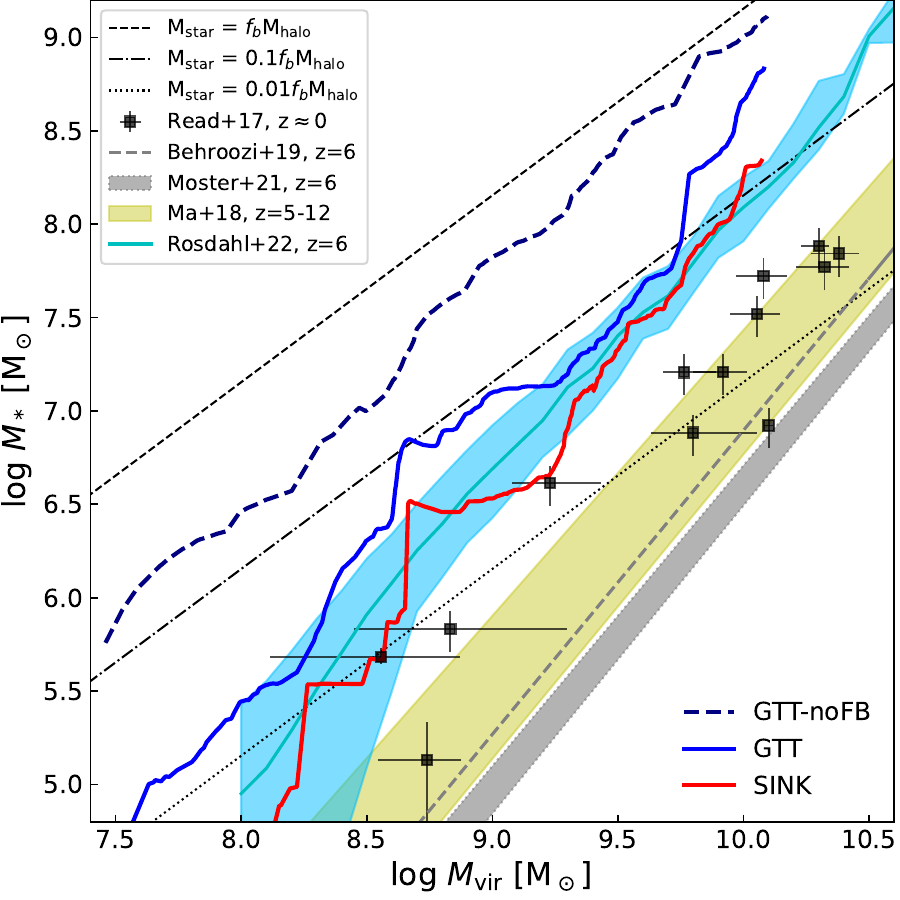}
    \vspace{-5mm}
    \caption{Stellar-to-halo mass relations of the \gttco\ (dashed navy), \gtt\ (blue), and \sink\ (red) simulations. The three gray diagonal lines correspond to the global SFE of 100, 10, and 1\%. Local dwarf observation results \citep{Read2017}, theoretical predictions extrapolated to low masses \citep{Behroozi2019, Moster2021}, and the best-fit relation from FIRE-2 simulation \citep{Ma2018} are overplotted for comparison. The median and 16--84\% relation from $\mathrm{SPHINX^{20}}$ \citep{Rosdahl2022}, which employs the same GTT prescription but with a boosted SN feedback, is also presented.}
    \label{fig:SMHM}
\end{figure}

\begin{table}
\caption{Summary of simulation results at $z=6$.}
\label{table2}
\centering
\begin{tabular*}{\linewidth}{@{\extracolsep{\fill}} l c c c c c}
\hline\hline
Simulation & $\Mvir$            & $M_*$           & $Z_*$     & SFR         & $M_\mathrm{UV}$ \\
Label      & [$10^{10}\,\Msun$] & [$10^8\,\Msun$] & [$\Zsun$] & [$\Msunyr$] &        \\
\hline
\gttco     & 1.25 & 13.0 & 1.14 & 2.10 & -19.51 \\
\gtt       & 1.21 & 6.80 & 0.72 & 2.53 & -19.48 \\
\sink      & 1.18 & 2.15 & 0.17 & 0.86 & -18.53 \\
\sinkhr    & 1.18 & 2.42 & 0.19 & 0.94 & -19.15 \\
\hline
\end{tabular*}
\tablefoot{From left to right: simulation label, virial mass, stellar mass, mass-weighted stellar metallicity, mean SFR for the last 100 Myr, and intrinsic UV magnitude at 1500$\,\AA$.}
\end{table}

Fig.~\ref{fig:SMHM} shows the stellar-to-halo mass relation in the three simulations (see also Table~\ref{table2} for a summary of the global galaxy properties at $z=6$). In the absence of stellar feedback (\gttco), the stellar mass accounts for $\approx73\%$ of the total baryon mass at $z=6$. As expected, this fraction is reduced by introducing stellar feedback into the simulations. However, even with feedback (\gtt), $\sim40\%$ of gas is converted into stars, producing a galaxy that is overly massive for its halo mass. The conversion efficiency increases steeply at the high mass end ($\Mvir\gtrsim10^{9.7}\,\Msun$), which implies that feedback becomes less effective in a deeper gravitational potential. In contrast, star formation is more strongly suppressed in the \sink\ run throughout the simulation, thereby yielding a smaller total stellar mass by a factor of $\sim3.2$ at $z=6$ compared to the \gtt\ run. The global stellar-to-halo mass ratio rises gradually with halo mass and reaches 10\% at $\sim10^{10}\,\Msun$.

However, local observations of dwarf galaxies \citep{Read2017} indicate a substantially lower global conversion efficiency of $\sim1\%$ at $\Mvir\sim10^{10}\,\Msun$, although a direct comparison is limited by the significantly large redshift difference. Extrapolations of empirical abundance-matching models \citep{Behroozi2019, Moster2021} to $z=6$ predict efficiencies of $\lesssim1\%$ at comparable halo masses. Similarly, the best-fit relation from the FIRE-2 zoom-in simulations \citep{Ma2018}, which assumes a constant SFE per free-fall time of $\epsff=100\%$ in gravitationally bound regions, lies $\sim1\,\mathrm{dex}$ below our results.

It is also useful to compare our results with $\mathrm{SPHINX^{20}}$ \citep{Rosdahl2022}, which was performed with the same \textsc{ramses-rt} code, a comparable spatial resolution, and the same star formation model as in the \gtt\ run, but with calibrated feedback implemented by increasing the SN frequency by a factor of 4. As expected, our \gtt\ galaxy mostly lies above the $1\sigma$ relation of $\mathrm{SPHINX^{20}}$.
The \sink\ run follows an evolutionary track similar to the median of the $\mathrm{SPHINX^{20}}$ simulation. This indicates that, although the stellar mass may remain higher than empirical or theoretical expectations, feedback strength can be naturally enhanced by changing the star formation model, thereby suppressing the formation of an overly massive nuclear cluster at the galactic center.

\subsection{Impact of bursty star formation on galaxy properties}
\begin{figure}
    \includegraphics[width=\linewidth]{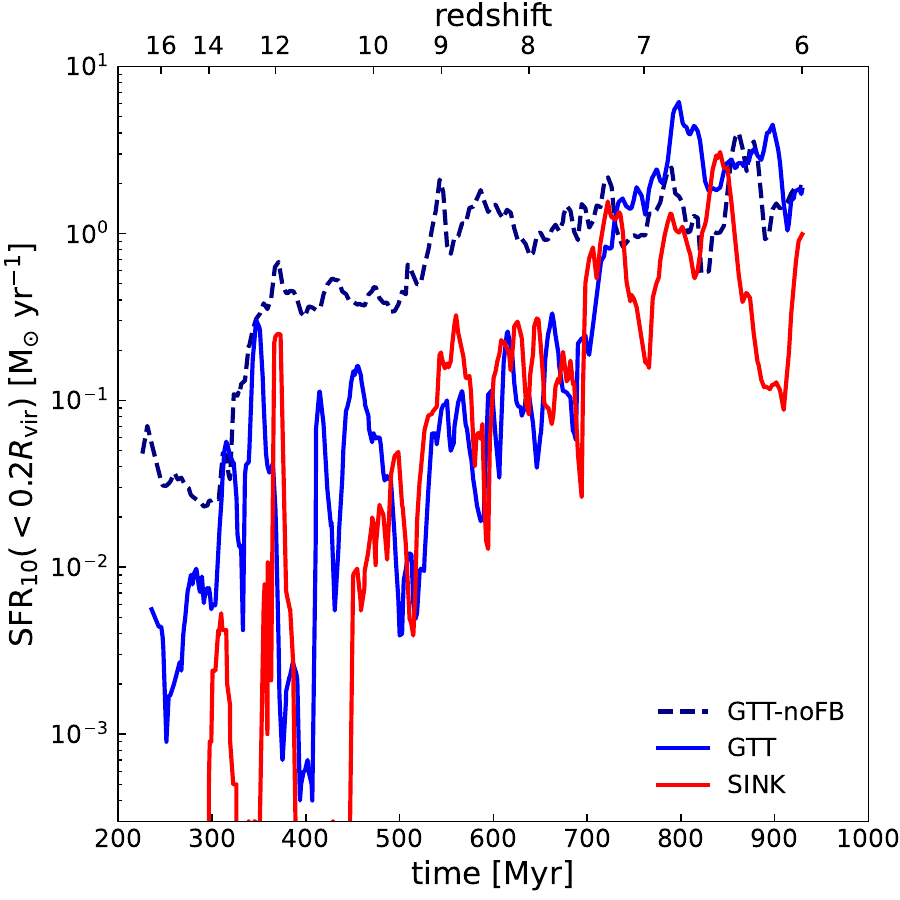}
    \vspace{-5mm}
    \caption{SFR within $0.2\Rvir$ averaged over 10 Myr for the \gttco\ (dashed navy), \gtt\ (blue), and \sink\ (red) runs.}
    \label{fig:SFR}
\end{figure}

In Fig.~\ref{fig:SFR}, we present the SFR within $0.2\Rvir$ averaged over 10 Myr as a function of time. The SFR in the \gttco\ run remains higher than those of the other two simulations except at later times ($z<7$), when it becomes comparable to that in the \gtt\ run. This is because most baryons have already been converted into stars, thereby leaving less gas ($3.5\times10^8$ and $4.2\times10^8\,\Msun$ in the \gttco\ and \gtt\ run, respectively) available for further star formation (see the increasing efficiency with halo mass of the \gttco\ run in Fig.~\ref{fig:SMHM}). In the \gtt\ and \sink\ runs, the SFR exhibits larger temporal variations because stellar feedback suppresses subsequent star formation. These fluctuations are more pronounced in the \sink\ run, where the larger amplitude originates from deeper SFR troughs. For example, star formation in the \sink\ run is completely quenched between $400<t/\mathrm{Myr}<450$ after a short starburst at $z\sim12$. In contrast, the SFR declines less and recovers within a shorter time over the same epoch in the \gtt\ run. Another notable difference emerges at $t>750\,\mathrm{Myr}$ when the halo mass approaches $10^{10}\,\Msun$. At later times, the SFR in the \gtt\ run remains nearly flat and exceeds that of the \gttco\ run, whereas the \sink\ run still experiences larger fluctuations with an overall lower SFR. This indicates that the SFHs in the \sink\ run are better characterized by short starbursts followed by the efficient quenching of star formation, whereas the smoother SFR in the \gtt\ run indicates ineffective quenching.

\begin{figure}
    \includegraphics[width=\linewidth]{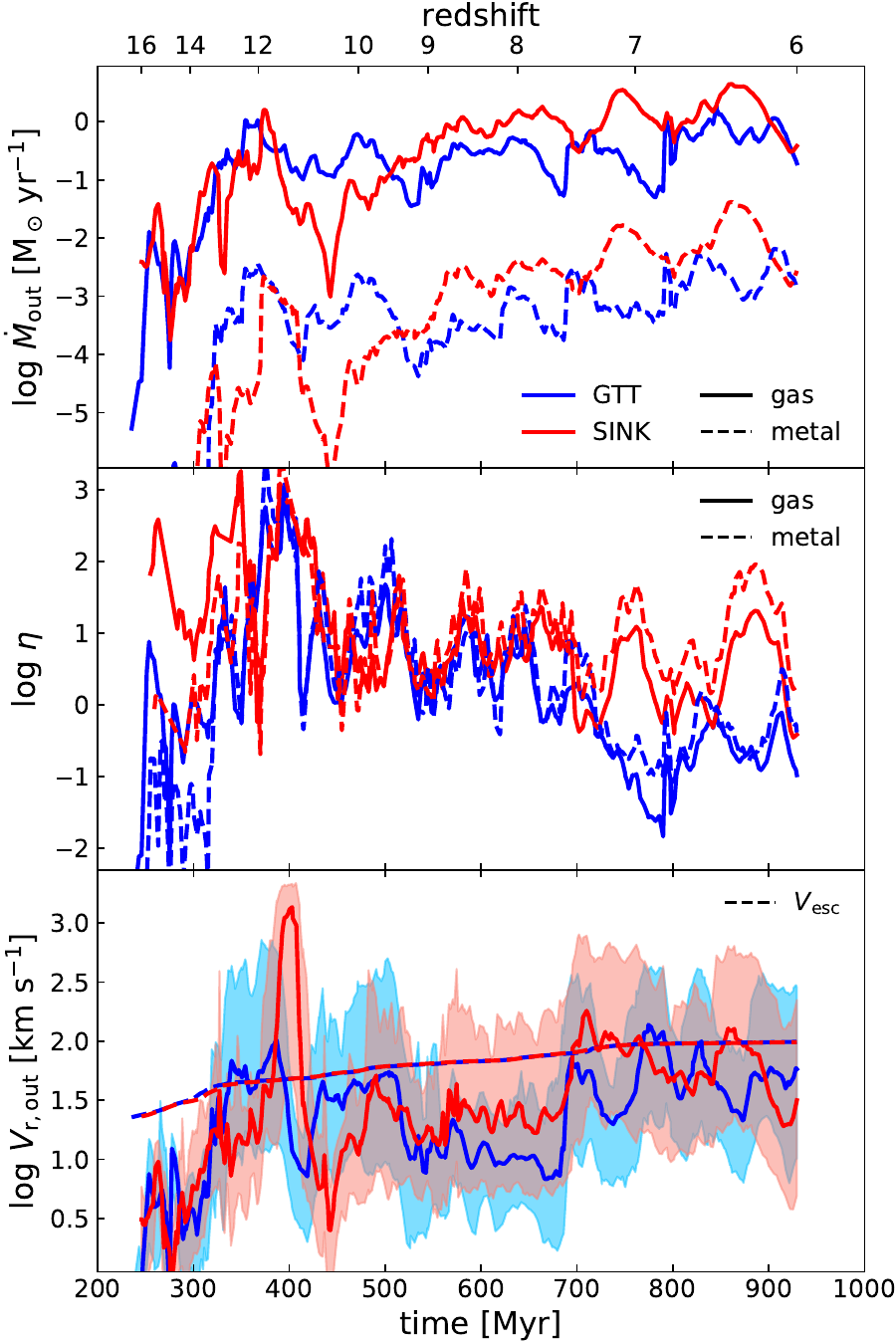}
    \vspace{-5mm}
    \caption{Top: outflow rates of gas (solid lines) and metals (dashed lines) for the two simulations measured at $0.15<R/\Rvir<0.25$. Middle: mass loading factors of gas (solid lines) and metals (dashed lines). Bottom: Median radial velocity of outflowing gas cells with the central 16--84\% distribution shown as the shaded region. Two dashed lines mark the escape velocity of the DMH ($V_\mathrm{esc}$) at a given time.}
    \label{fig:outflow}
\end{figure}

We measure the galactic outflow rate within a spherical shell in the range of $0.15<R/\Rvir<0.25$ as $\dot M_\mathrm{out}=\sum_\mathrm{i} M_\mathrm{i}\,V_\mathrm{r,i} / \Delta R$ \citep[e.g.,][]{Muratov2015}, where $M_\mathrm{i}$ and $V_\mathrm{r,i}$ represent the mass and radial velocity of each cell within the shell, respectively, and $\Delta R=0.1\Rvir$ represents the shell width. We also measure the metal outflow rate by replacing $M_\mathrm{i}$ with $Z_\mathrm{i}M_\mathrm{i}$, where $Z_\mathrm{i}$ is the cell metallicity. Only cells with positive radial velocities are included in the summation, whereas those within the virial radius of satellite halos are excluded. The resulting gas and metal outflow rates are plotted in the top panel of Fig.~\ref{fig:outflow}. During early times when the halo is less massive than $10^9\,\Msun$ ($z>9.5$), the \sink\ run maintains a lower SFR. However, the outflow rates between the two runs are comparable, which indicates that the \sink\ run is more effective at driving large-scale winds. Both the gas and metal outflow rates are consistently higher in the \sink\ run as the halo grows further and the gravitational potential deepens. Cumulatively, the masses of gas and metals escaping the $R=0.2\Rvir$ sphere in the \sink\ run are $M_\mathrm{g,out}\sim5.9\times10^8\,\Msun$ and $M_\mathrm{Z,out}\sim3.0\times10^6\,\Msun$, respectively, which are greater by factors of 2.6 and 4 than those in the \gtt\ run. As shown in Fig.~\ref{fig:Vcirc}, the stronger outflows in the \sink\ run expel more gas from the ISM, thereby better suppressing star formation.

The SFR is on average higher in the \gtt\ run, and therefore, the differences in the mass loading factors ($\eta=\dot M_\mathrm{out}/\mathrm{SFR}$) between the two runs are even more pronounced at later times (middle panel). After the first starburst, both galaxies reach a maximum $\eta$ of 1,000 at $t\sim400\,\mathrm{Myr}$. $\eta$ then decreases gradually as the halos grow, remaining below 10 (1) from $t=660\,\mathrm{Myr}$ (710 Myr) in the \gtt\ run. The metal loading factor is slightly higher, but falls below 3 after $t\sim700\,\mathrm{Myr}$. Conversely, in the \sink\ run, $\eta$ rarely falls below 1, and the two local peaks of $\sim15$ occur at $t=770$ and 890 Myr, which correspond to two SFR troughs. The metal loading factors are higher, reaching $\sim70$ at the same epochs.

The bottom panel of Fig.~\ref{fig:outflow}, which shows the radial velocity of outflowing cells within the shell, further demonstrates that outflows in the \sink\ run achieve higher radial velocities from $z\approx9.5$ and are thus more efficient in transporting gas to larger distances. Higher radial velocities slow down recollapse, preventing rapid rejuvenation of star formation. The radial velocities are different between the two runs even when the SFHs are similar ($600<t/\mathrm{Myr}<700$). During this period, $\sim12$--41\% of the outflowing cells in the \sink\ run maintain sufficiently high radial velocities to escape the halo. In contrast, this fraction drops significantly to only a few percent in the \gtt\ run, leaving most of the gas inside the halo for future star formation.

\begin{figure}
    \includegraphics[width=\linewidth]{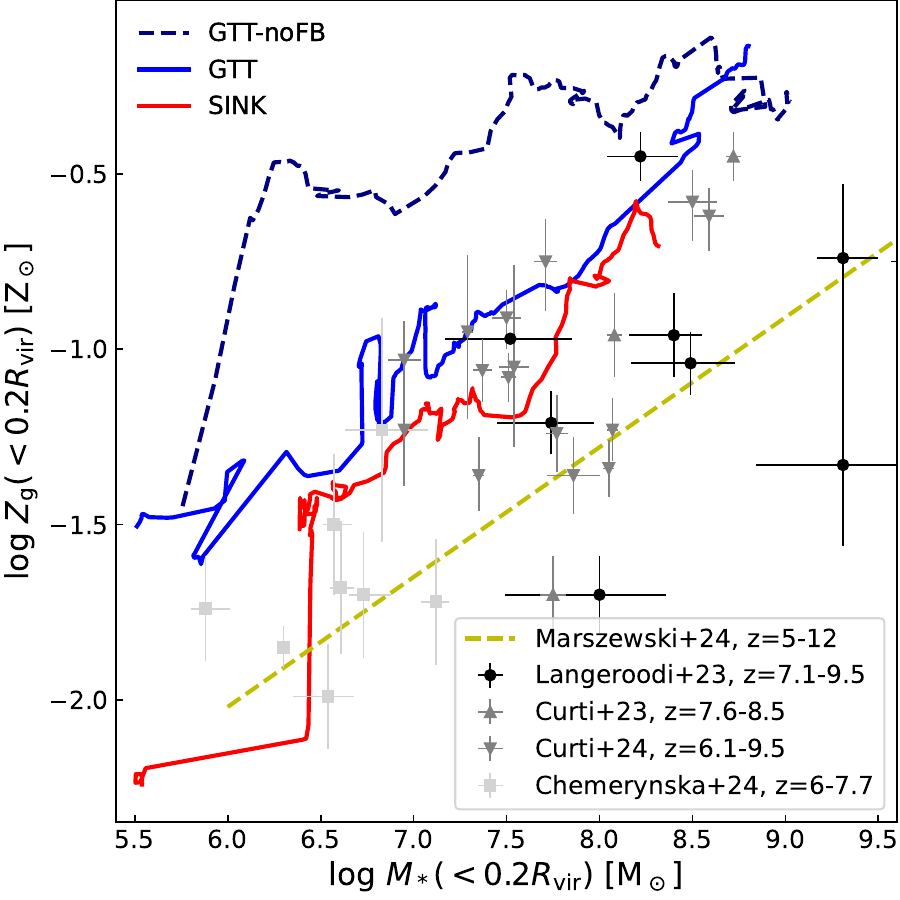}
    \vspace{-5mm}
    \caption{Gas-phase metallicity normalized by the solar value ($\Zsun=0.0134$), as a function of stellar mass measured within $0.2\Rvir$. The yellow dashed line shows the best-fit relation from a suite of FIRE-2 simulations \citep{Marszewski2024}. For comparison, three JWST observation measurements at $z>6$ are overplotted with error bars \citep{Langeroodi2023, Curti2023, Curti2024, Chemerynska2024}, wherein the observed oxygen abundance ($12+\log(\mathrm{O/H})$) is converted to gas metallicity assuming solar abundance \citep{Asplund2009}.}
    \label{fig:MZR}
\end{figure}

Since SN explosions are the main drivers of galactic outflows \citep[e.g.,][]{Dekel1986} and the only mechanism through which metal enrichment is achieved in our simulations, the gas-phase metallicity at a given stellar mass (MZR) is a good indicator of how effectively SN feedback is coupled with the ISM \citep[e.g.,][]{Tremonti2004, Dave2011, Muratov2015}. To this end, we show the mass-weighted gas-phase metallicity versus the stellar mass within $0.2\Rvir$ in Fig.~\ref{fig:MZR}. The \gttco\ run serves as an upper-limit case, where all generated metals remain within the galaxy. Including stellar feedback reduces the MZR normalization in the \gtt\ run by a factor of $\sim10$ at low masses, but it converges to nearly the same level as in the \gttco\ run at higher masses. This implies that SN explosions become weaker in the \gtt\ run at deeper gravitational potentials. With stronger SN feedback, the \sink\ run ejects more metals (Fig.~\ref{fig:outflow}) and produces an MZR with lower normalization.
Both feedback runs are consistent with the observed ranges \citep{Langeroodi2023, Curti2023, Curti2024}, although the \sink\ result shows a notably better agreement at lower masses \citep{Chemerynska2024}. The best-fit relation from the FIRE-2 simulations \citep{Marszewski2024}, which shows little redshift evolution over $z=5$--12, lies below the \sink\ normalization across the probed mass range by a factor of $2$--3. Since the Type~II SN metal yield implemented in FIRE-2 is slightly higher than that in this work \citep[$\approx2\Msun$ per explosion, see Appendix A of][for more detail]{Hopkins2018}, the lower MZR normalization indicates that either the FIRE-2 model incorporates inherently stronger feedback or its star formation prescription promotes burstier star formation that enhances feedback coupling more effectively than our current sink-based approach. 

We note that the offset in the MZR between the two feedback runs is relatively small at higher masses ($10^8$--$10^{8.5}\,\Msun$). In this regime, the metal loading factor in the \sink\ run is significantly higher ($\sim10$) than that in the \gtt\ run, which typically suggests a lower gas-phase metallicity in the former. However, more efficient recycling of metals in the \gtt\ run causes a larger fraction of metals to be locked in star particles rather than remaining in the gas phase. As indicated in the following section, star-forming clumps in the \gtt\ run are longer-lived, with lifetimes exceeding the minimum delay time of Type II SNe. This enables a significant fraction of the metals produced by early SN progenitors to be immediately recycled to subsequent generations of stars within the same clump, thereby substantially increasing stellar metallicities. In contrast, clumps in the \sink\ run are disrupted more rapidly, and newly synthesized metals are generally released to the ISM/CGM only after local star formation is quenched. At $z=6$, the mass fraction of metals locked in stars within $R<0.2\Rvir$ is $\sim0.59$ in the \gtt\ run, compared to 0.34 in the \sink\ run. These results suggest that the gas-phase MZR is regulated by the interplay between self-enrichment and SN-driven outflows, with the relative importance of each process governed by the duration of localized star formation.

\begin{figure}
    \includegraphics[width=\linewidth]{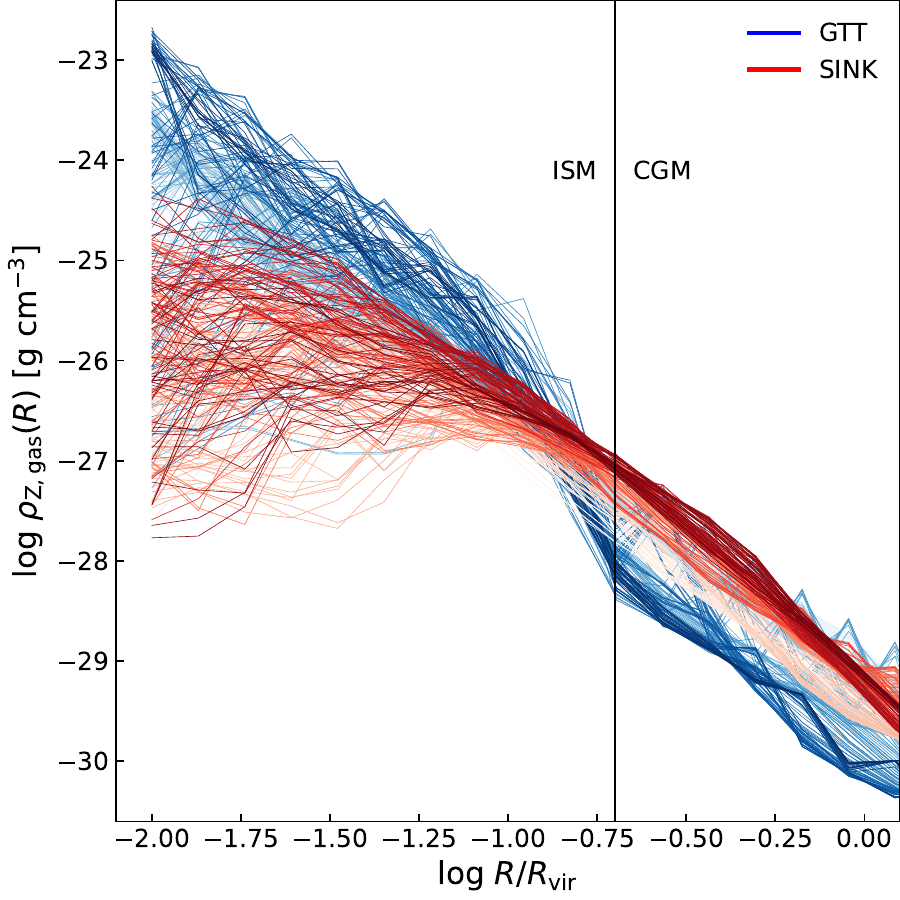}
    \vspace{-5mm}
    \caption{Radial profiles of the gas-phase metal density within shells of thickness $\Delta \log R/\Rvir \sim 0.1$. Results are shown for $6<z<10$, where darker colors represent later epochs. The black vertical line indicates $0.2\Rvir$, which separates the ISM and CGM regions in this study.}
    \label{fig:metal_radial}
\end{figure}

The narrowing gap in the MZR normalization can be further attributed to the distinct spatial distributions of metals in the two simulations. Fig.~\ref{fig:metal_radial} shows the radial profiles of the gas-phase metal density for $6<z<10$, revealing that the metal distribution is highly dependent on the adopted star formation model. In the innermost regions ($<0.1\Rvir$), the \gtt\ run exhibits a significantly steeper density slope with a central metal density $\sim100$ times higher than that of the \sink\ run. The \gtt\ profile also shows a sharp decline between $0.1$ and $0.2\Rvir$, implying that outflows are too weak to enrich the CGM. In contrast, the \sink\ run maintains a more extended metal distribution by effectively transporting metals over larger distances via stronger feedback. Consequently, the density profiles cross between $0.1$ and $0.2\Rvir$, where the higher metal density of the \sink\ run at the interface between the ISM and CGM begins to compensate for its lower central density. Such diverging features in the radial metal gradient are averaged out when integrated within $0.2\Rvir$, which explains the relatively small offset in the volume-averaged MZR shown in Fig.~\ref{fig:MZR}.

\begin{figure}
    \includegraphics[width=\linewidth]{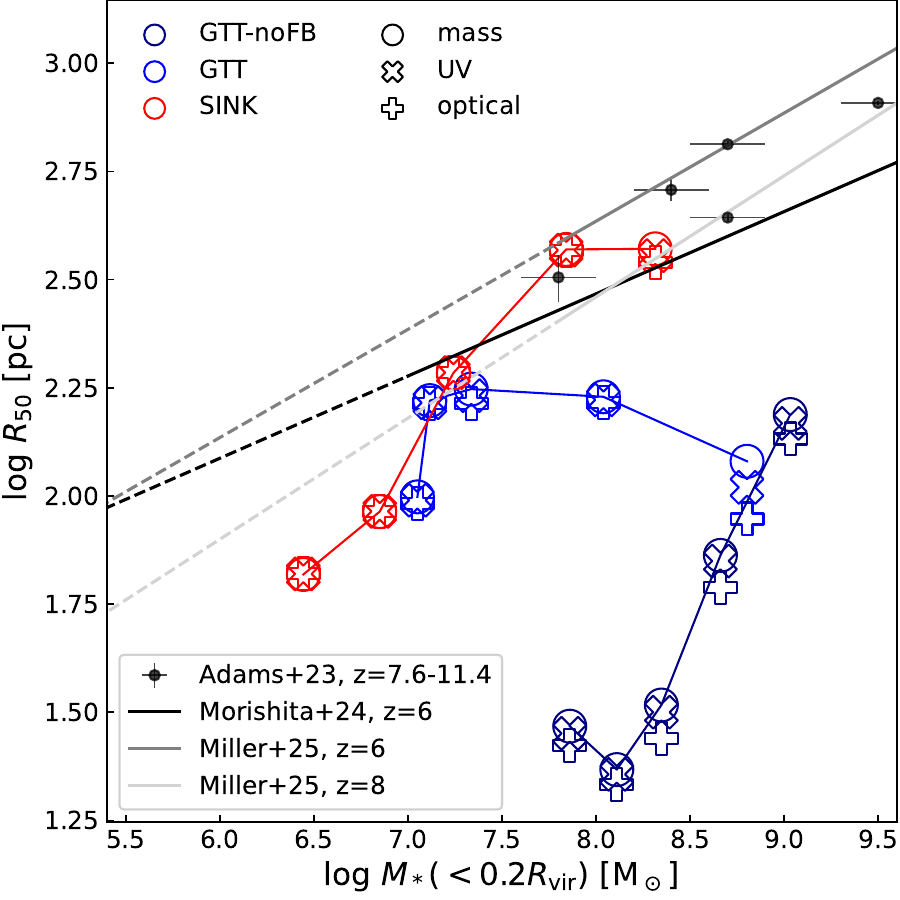}
    \vspace{-5mm}
    \caption{Effective radius of star particles as a function of stellar mass at five different redshifts, $z=10$, 9, 8, 7, and 6. Empty circles and lines represent the half-mass radii, while crosses and plus symbols represent the half-light radii measured in the UV (1450--$1550\,\angstrom$) and optical (4950--$5050\,\angstrom$) bands, respectively. High-redshift observations from \citet{Adams2023} are plotted as filled black circles. The solid black line shows the best-fit relation at $z=6$ from \citet{Morishita2024}, while the gray and light-gray lines are from \citet{Miller2025}. Dashed extensions indicate extrapolations of these observed relations toward lower masses.}
    \label{fig:r50}
\end{figure}

Galaxy size at a given stellar mass provides another critical diagnostic of overcooling, since the central regions, where gas densities and SFRs are highest, are most susceptible to excessive cooling. Fig.~\ref{fig:r50} displays the half-mass radii and the UV and optical half-light radii as a function of stellar mass within $0.2\Rvir$ across five different redshifts, from $z=10$ to 6. The two half-light radii are measured using the integrated luminosity in the UV (1450--$1550\,\angstrom$) and optical (4950--$5050\,\angstrom$) windows. Without feedback (\gttco), the galaxy remains overly compact because of the concentrated star formation at the galaxy center. In this run, the half-light radii are always smaller than the half-mass radius, which indicates that young stars preferentially form within the innermost regions. We also find that the optical half-light radii are consistently smaller than the UV radii in this run. This occurs because the UV emission remains nearly constant during the earliest evolutionary stages for very young stellar populations ($\lesssim8\,\mathrm{Myr}$, with a slight metallicity dependence), whereas the optical luminosity increases as massive stars evolve off the main sequence and transition into the supergiant phase. In the \gtt\ run, the effective radii are larger than those in the \gttco\ run at $z\gtrsim7$; however, they begin to shrink thereafter despite increasing stellar mass. By $z\sim6$, the mass-size relation in the \gtt\ run becomes almost indistinguishable from that of the \gttco\ run, as shown in Fig.~\ref{fig:MZR}. During this contraction phase, both half-light radii decrease below the half-mass radius, which suggests that more massive galaxies are more susceptible to overcooling \citep[e.g.,][]{Kimm2015}.
By performing radiative transfer calculations using RASCAS \citep{Michel-Dansac2020}, we further confirm that the \gtt\ run still exhibits relatively compact sizes when dust attenuation is taken into account (see Appendix~\ref{sec:appendix_rascas} for a more detailed comparison).
In contrast, the effective radii measured using all three methods in the \sink\ run remain mutually consistent until the final redshift, and they are larger than those in the \gtt\ run. The disparity in galaxy size between the two feedback runs widens toward lower redshifts as the \sink\ run maintains a more extended structure. Consequently, the \sink\ run shows better agreement with recent JWST size measurements at similar redshifts \citep{Adams2023, Morishita2024, Miller2025}. 

We calculate the escape fraction of LyC photons (\fesc) at the virial sphere to further quantify the effectiveness of the stellar feedback. Since the majority of LyC photons are emitted at $t\lesssim 5\,\mathrm{Myr}$, \fesc\ serves as an indicator of the rapid dispersal of birth clouds. High escape fractions indicate efficient cloud destruction, whereas overcooled systems exhibit lower values due to higher neutral gas columns. Following the ray-tracing method described in \citet{Kimm2022}, we determine \fesc\ by comparing the intrinsic SED of the star particles to the attenuated spectra sampled along 512 lines of sight. For each direction, we integrate the column densities of \ion{H}{i}, $\mathrm{H_2}$, helium, and metals from the position of each star particle to the virial radius. Following \citet{Laursen2009}, the dust content is determined by calculating the pseudo dust number density in each cell as $n_\mathrm{d,i}=(n_\mathrm{HI}+n_\mathrm{H2}+0.01n_\mathrm{HII})Z_\mathrm{i}/Z_\mathrm{ref}$. We adopt a normalization of $Z_\mathrm{ref}=0.0084$ to ensure that the dust-to-metal ratio is consistent with that of the Small Magellanic Cloud \citep{Roman-Duval2022}.

\begin{figure}
    \includegraphics[width=\linewidth]{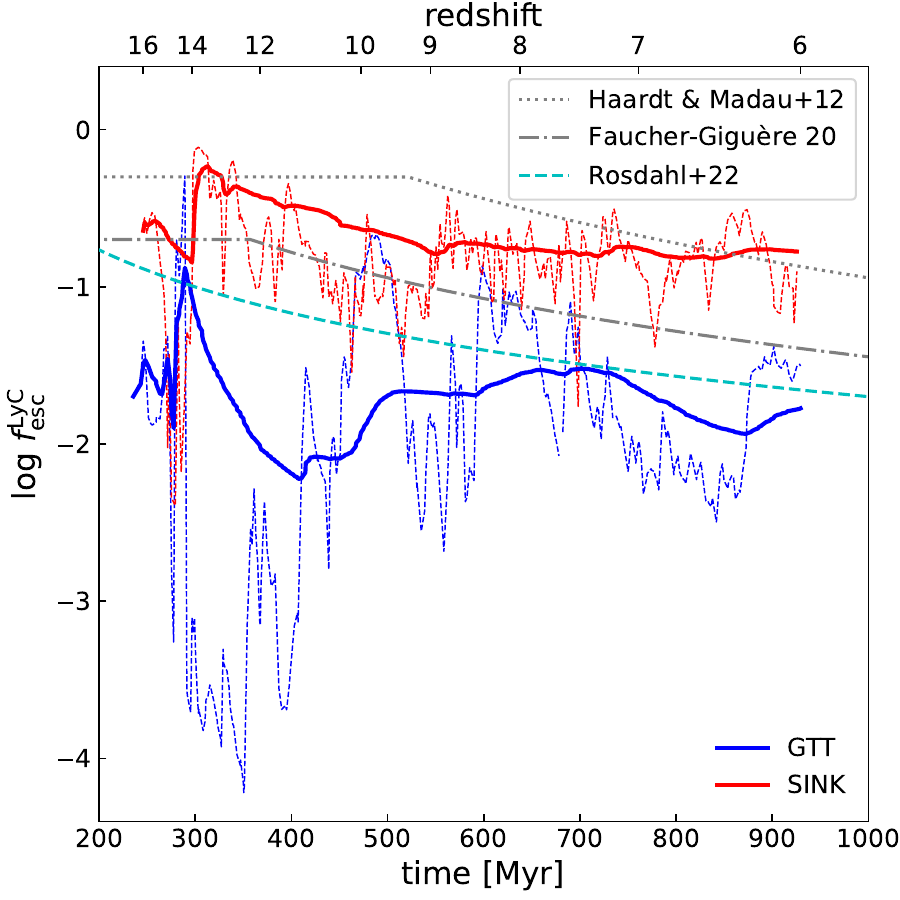}
    \vspace{-5mm}
    \caption{Escape fraction of LyC photons measured at $\Rvir$. The thin dashed and thick solid lines represent instantaneous measurements at each snapshot and luminosity-weighted integrated values, respectively. Two gray lines with different linestyles are models from \citet{Haardt2012} and \citet{Faucher2020} with a maximum value of 0.5 and 0.2, respectively, and the cyan line is the best-fit result from $\mathrm{SPHINX^{20}}$ \citep{Rosdahl2022}.}
    \label{fig:fesc}
\end{figure}

The evolution of the LyC escape fraction over time for the two fiducial simulations is presented in Fig.~\ref{fig:fesc}. Both runs experience a gradual decline in \fesc\ over time; however, it is systematically lower by nearly an order of magnitude in the \gtt\ run. This value is lower than that of $\mathrm{SPHINX^{20}}$, where stellar feedback is artificially enhanced, unlike in our \gtt\ run. Conversely, the \sink\ run exhibits a markedly higher \fesc\ than $\mathrm{SPHINX^{20}}$, even though their stellar-to-halo mass relations are comparable (Fig.~\ref{fig:SMHM}). By $z=6$, the luminosity-weighted escape fraction in the \sink\ run is $\left<\fesc\right>=16.7\%$, which is an order of magnitude larger than that in \gtt\ (1.7\%) even without feedback boosting. This value places the model between the two UV background reionization models of \citet{Haardt2012} and \citet{Faucher2020}. The higher galactic \fesc\ in the \sink\ run, which is primarily determined by photon absorption within star-forming regions, suggests that the sink-based model clears the surrounding dense gas more efficiently or on a shorter timescale than in $\mathrm{SPHINX^{20}}$.

\subsection{Evolution of star clusters and gas clumps}
\label{sec:clump&cluster}

Having established the strong dependence of galaxy properties on the star formation model, we now focus on smaller scales to explore the model effect on clump and cluster properties. To this end, we start by analyzing star clusters in the two fiducial simulations identified using the DBSCAN method \citep{Ester1996}. Each cluster is required to contain at least 10 Pop~II star particles (i.e., $10^4\,\Msun$ before mass loss due to SN explosions), and the separation between any two member particles must be less than 10 pc. We construct cluster trees by linking two star clusters in consecutive snapshots only if more than 50\% of the member stars in the earlier snapshot are retained in the subsequent snapshot.

\begin{figure}
    \includegraphics[width=\linewidth]{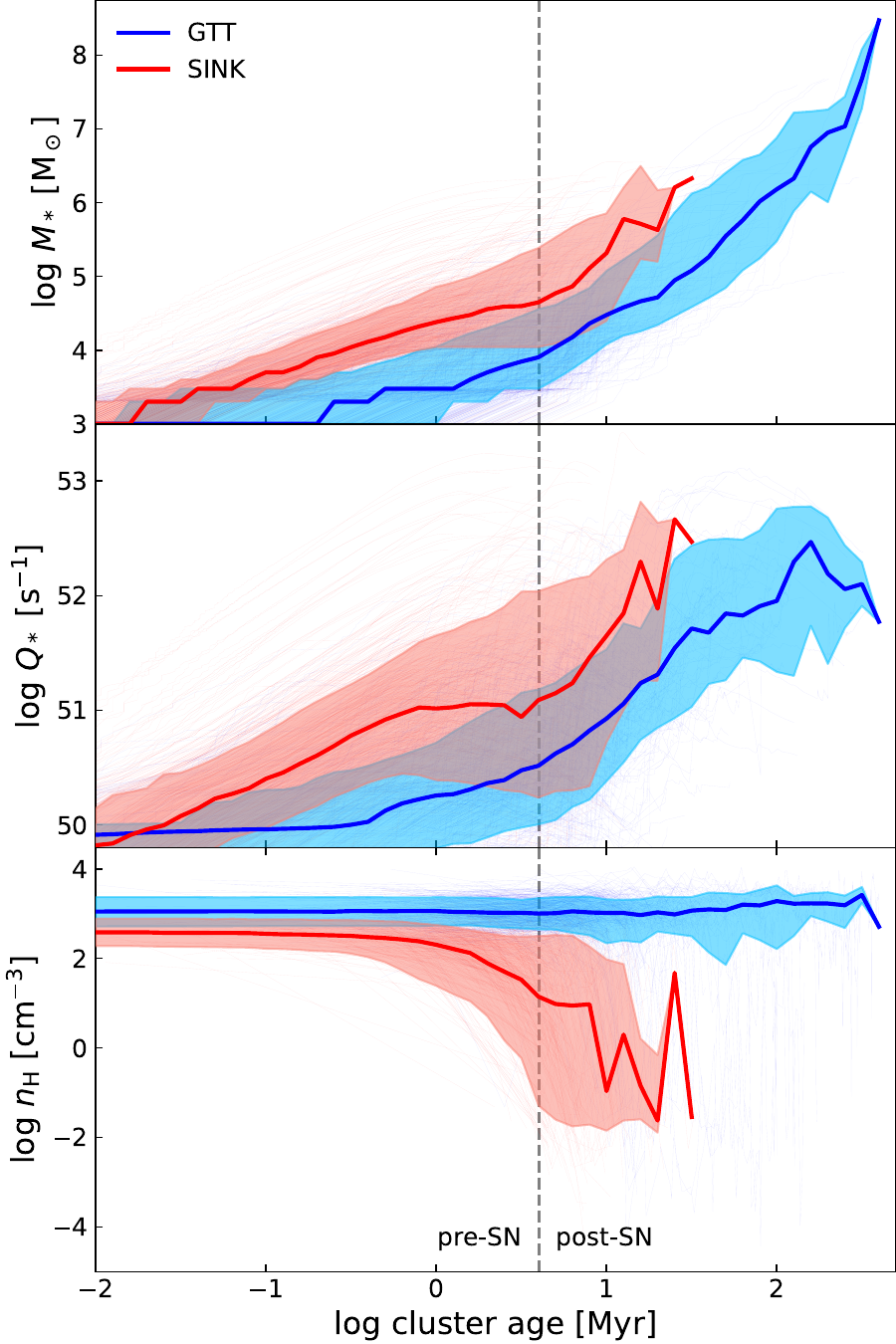}
    \vspace{-5mm}
    \caption{Top: mass of a star cluster as a function of its age. Thin lines show the evolutionary track of individual cluster trees, whereas the thick lines and shaded regions correspond to the median and the central 16--84\% distribution of the entire population. Middle: ionizing photon ($<912\,\angstrom$) production rate from member stars of each star cluster. Bottom: average hydrogen number density around each star cluster. The density is computed within the volume of a $(3\dxmin)^3$ cube, centered at the cluster's center of mass. The gray vertical line in all three panels marks $t=4\,\mathrm{Myr}$, which is approximately the minimum lifetime of SN progenitors.}
    \label{fig:cluster}
\end{figure}

\begin{figure}
    \includegraphics[width=\linewidth]{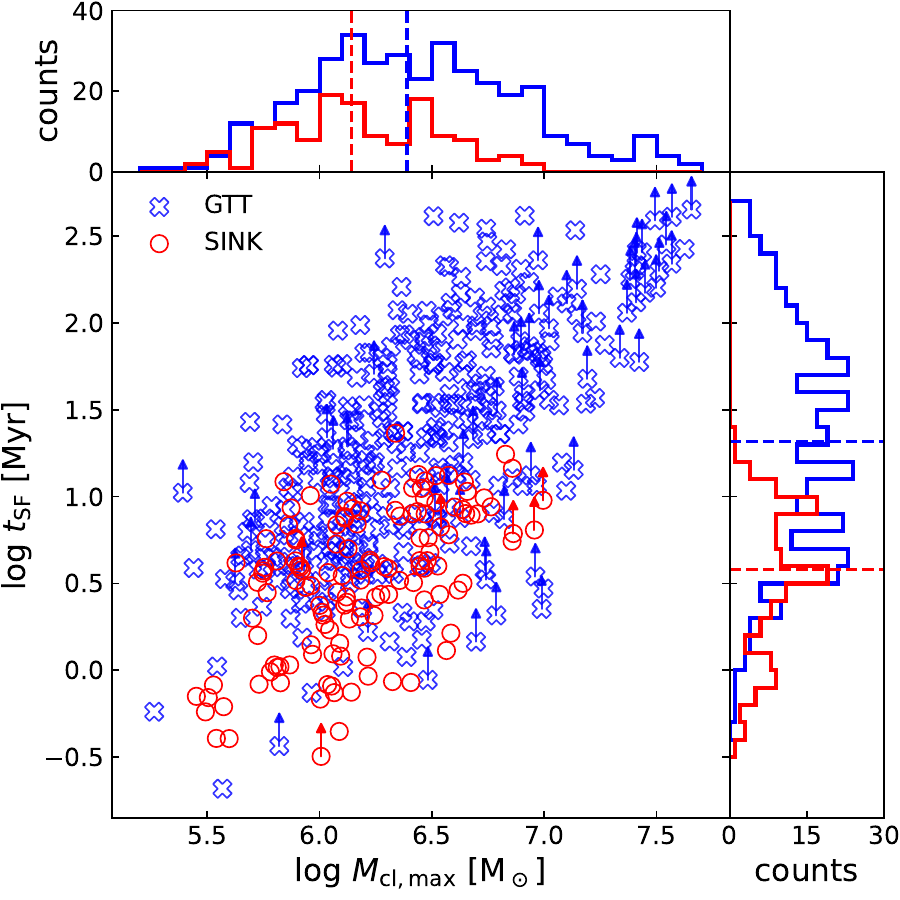}
    \vspace{-5mm}
    \caption{Star-forming timescale ($t_\mathrm{SF}$) as a function of the maximum clump mass reached during its lifetime ($M_\mathrm{cl,max}$) for the \gtt\ (blue crosses) and \sink\ runs (red circles). We use arrows to mark the lower bound of $t_\mathrm{SF}$ for clump trees that are linked to the final snapshot so that $t_\mathrm{SF}$ cannot be determined. For clumps involved in merger events, only the progenitor with the longest $t_\mathrm{SF}$ is shown. The top and the right panels display the number distributions of $M_\mathrm{cl,max}$ and $t_\mathrm{SF}$, respectively. The dashed lines mark the median values of each distribution.}
    \label{fig:clump}
\end{figure}

The top panel of Fig.~\ref{fig:cluster} shows the mass growth histories of individual star clusters (thin lines), median mass at a given age (thick lines), and the central 68\% range (shaded regions). For cluster trees involving mergers, we only track the most massive progenitors to avoid redundancies. Cluster age is defined as the time elapsed since the formation of the first member star such that all clusters begin with $M_*=1000\,\Msun$ (i.e., one Pop~II star particle) at $t=0$. Each line is drawn until the last member star particle is formed. We find a noticeable difference in growth rates: star clusters in the \sink\ run grow faster during the first few Myr and quench earlier as their surrounding clumps are rapidly disrupted by feedback. This indicates that star formation is considerably burstier in the \sink\ run on short timescales. In contrast, the growth of star clusters is much slower in the \gtt\ run during the early evolutionary stages, particularly before the first SN explodes. Clusters start to grow faster thereafter; however, star formation is not effectively quenched as in the \sink\ run. This sustained star formation in the \gtt\ run eventually results in greater cluster masses, with clusters with stellar mass above $10^7\,\Msun$ forming stars for over 100 Myr while remaining embedded in massive gas clumps ($>10^7\,\Msun$, Fig.~\ref{fig:clump}).

The middle panel shows the production rate of ionizing photons ($Q_*$) emitted by cluster member stars, whereas the bottom panel shows the mean hydrogen number density ($n_\mathrm{H}$) within a $(3\dxmin)^3$ cube centered on each cluster. In the \sink\ run, the total $Q_*$ increases steeply and drives the expansion of the \ion{H}{ii} region within the first $\sim1\,\mathrm{Myr}$. Although not shown in the figure, $Q_*$ exceeds the \ion{H}{ii} recombination rate, which implies that an excess of ionizing photons can escape birth clouds without being absorbed by neutral hydrogen (Fig.~\ref{fig:fesc}). Consequently, the median gas density decreases as the \ion{H}{ii} regions expand because of strong radiation feedback. This is primarily driven by photoionization heating rather than gas consumption by accretion. As discussed in Sect.~\ref{sec:bursty}, the clump dispersal occurs on a considerably shorter timescale than the gas depletion time. The density then further decreases by several orders of magnitude once SNe begin to explode at $t\gtrsim4\,\mathrm{Myr}$. Furthermore, this gas dispersal is enhanced by clustered SN feedback. In contrast, because star formation is significantly slower in the \gtt\ run, the median $Q_*$ increases by only $\sim0.3\,\mathrm{dex}$ during the first 4 Myr, before any SNe explode. Moreover, it is more difficult for star-forming environments to be ionized by radiation because of the higher average gas density around young clusters.
We find that the surrounding gas cannot be fully ionized with such low $Q_*$ values, because the Str{\"o}mgren radius $R_\mathrm{S}=(3Q_*/4\pi\alpha_\mathrm{B}n^2_\mathrm{H})^{1/3}$ is smaller than the radius of the selected volume of $(3\dxmin)^3$, where we assume a constant recombination coefficient of $\alpha_\mathrm{B}=2.6\times10^{-13}\,\mathrm{cm^3\,s^{-1}}$. The entire clump can be ionized when $Q_* > 4 \pi R^3_\mathrm{cl}\,\alpha_\mathrm{B}\, n^2_\mathrm{cl} / 3$, where $R_\mathrm{cl}$ and $n_\mathrm{cl}$ are the radius and hydrogen number density of the clump, respectively. Assuming a constant $\alpha_\mathrm{B}$, this inequality can be written as
\begin{equation}
\label{eq:Qstar}
    Q_* \gtrsim 3.2\times10^{43}\,\mathrm{s^{-1}}\,\left(\frac{R_\mathrm{cl}}{1\,\mathrm{pc}}\right)^3\,\left(\frac{n_\mathrm{cl}}{1\,\cmq}\right)^2.
\end{equation}
The radius of a sphere whose volume is equivalent to $(3\dxmin)^3$ ranges between 10.6 and 21.2 pc, whereas the initial median $n_\mathrm{H}$ values for the \gtt\ and \sink\ runs are 1100 and $380\,\cmq$, respectively, as shown in the bottom panel. Substituting these values into Eq.~\eqref{eq:Qstar} suggests that $Q_*$ must be higher than $4.6\times10^{52}$--$3.7\times10^{53}\,\mathrm{s^{-1}}$ and $5.5\times10^{51}$--$4.4\times10^{52}\,\mathrm{s^{-1}}$ in the \gtt\ and \sink\ runs, respectively. Owing to slower star formation, most star clusters in the \gtt\ run cannot reach such values at early ages;, which explains why the impact of radiation feedback remains weak and the global escape fraction is low (Fig.~\ref{fig:fesc}). Consequently, the gas density around clusters in the \gtt\ run barely decreases within their lifetime. Even after the start of SN explosions, more than 50\% of the cluster population remains embedded in high-density environments similar to their birthplaces. These results demonstrate that both radiation and SN feedback are significantly weaker in the \gtt\ run than in the \sink\ run.

To quantify the destruction of gas clumps, we examine their temporal evolution by constructing clump merger trees based on the IDs of member star particles. A descendant is assigned when a clump shares the largest fraction of particles with another clump in a subsequent snapshot, provided this fraction exceeds 10\%. For our analysis, we retain clumps that (i) contain only in-situ stars, (ii) host both in-situ and ex-situ stars but are dominated by the in-situ population, or (iii) are dominated by ex-situ stars provided they are older than 50 Myr (i.e., no remaining Type~II SN progenitors). These selection criteria are designed to exclude clumps whose destruction may be affected by external processes. Gas clumps are assumed to be destroyed at the final snapshot of each tree, which introduces a small uncertainty in the estimated star-forming timescale of $\sim2\,\mathrm{Myr}$ set by the snapshot cadence.

Fig.~\ref{fig:clump} shows the star-forming timescale ($t_\mathrm{SF}$), defined as the time interval between the formation epoch of the first and last in-situ Pop~II stars, as a function of the maximum clump mass, $M_\mathrm{cl,max}$. Since we cannot determine the exact $t_\mathrm{SF}$ for clump trees that are still forming stars at the final snapshot of the simulation, we mark these clumps with upward-pointing arrows to represent the lower limits on $t_\mathrm{SF}$. In both runs, the two quantities exhibit a positive correlation, with more massive clumps forming stars and surviving longer. However, the distributions of both $t_\mathrm{SF}$ and $M_\mathrm{cl,max}$ in the \gtt\ run shift toward larger values relative to those in the \sink\ run. The median value of $t_\mathrm{SF}\approx3.8\,\mathrm{Myr}$ in the \sink\ run is comparable to the minimum lifetime of Type~II SN progenitors, indicating that radiation feedback is sufficiently strong to disrupt star-forming clumps. The clump with the maximum $t_\mathrm{SF}$ survives for $\sim30\,\mathrm{Myr}$, during which gas is supplied over an extended period via clump mergers. However, its mass remains limited because continued accretion is followed by rapid disruption driven by stronger feedback. In comparison, clumps in the \gtt\ run survive for considerably longer times, with a median $t_\mathrm{SF}$ of $\sim15\,\mathrm{Myr}$. A similar conclusion is reached for clumps in isolated disk simulations \citep{HanD2026}. Although $\sim50\%$ of early-born SN progenitors are expected to explode by this time, the prolonged survival of clumps indicates that feedback is not strong enough to achieve self-regulation on clump scales in the \gtt\ run. The most distinctive difference between the \gtt\ and \sink\ run is the presence of numerous clumps with $t_\mathrm{SF}>100\,\mathrm{Myr}$, which is several orders of magnitude longer than the free-fall time of a uniform cloud with a density of $n_\mathrm{H}=10^4\,\cmq$ (i.e., the typical density of star-forming sites in the \gtt\ run). In addition, gas clumps in the \gtt\ run tend to reach considerably higher masses as the rate of star formation lags behind rapid gas inflow, and this makes them even more resistant to disruption.

\section{Discussion}
\label{sec:discussion}
\subsection{Effect of enhanced resolution on sink-based model}
\label{sec:res}

\begin{figure*}
    \includegraphics[width=\textwidth]{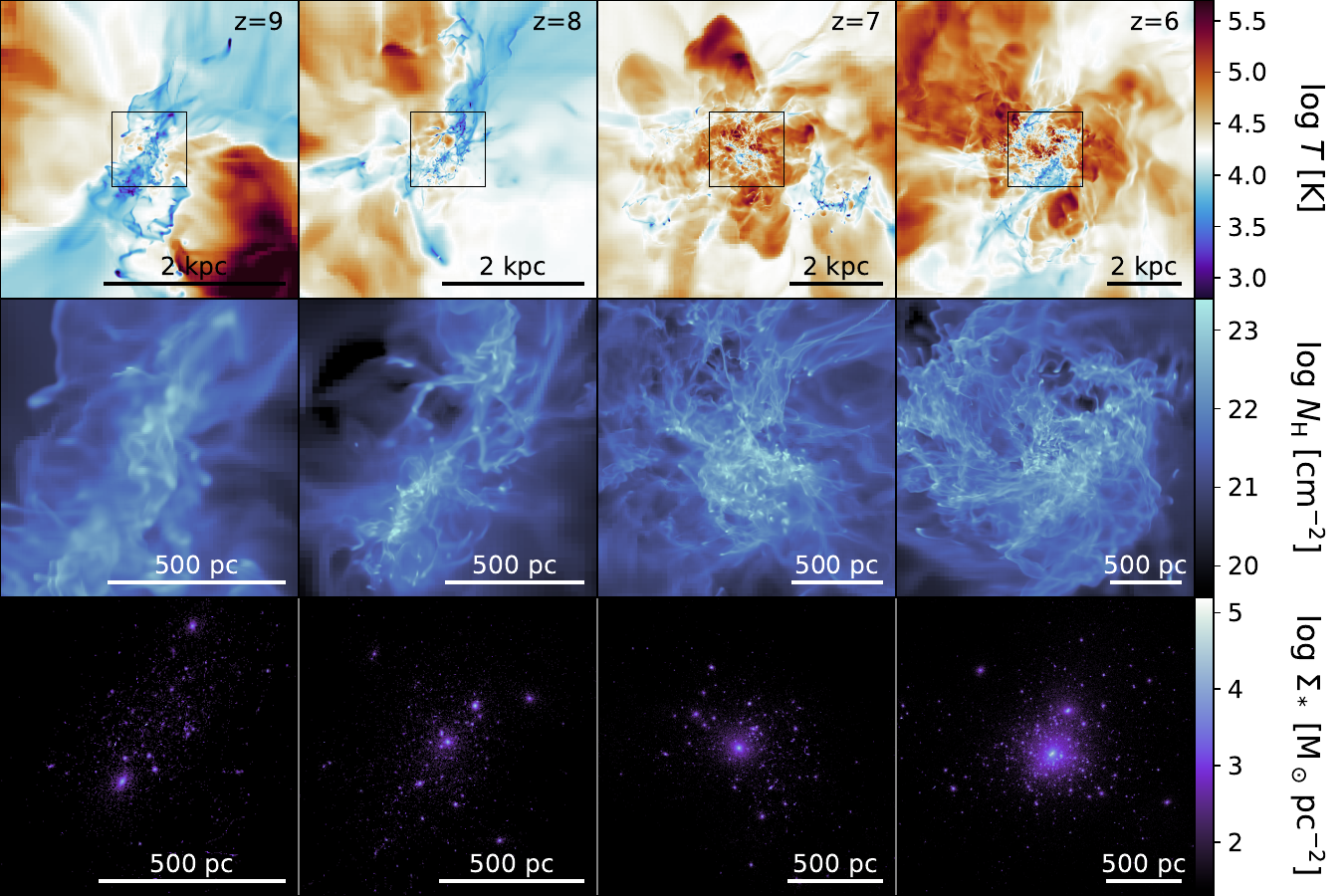}
    \vspace{-5mm}
    \caption{Same as Fig.~\ref{fig:map}, but for the \sinkhr\ run. The gas structure remains turbulent as in the fiducial \sink\ run, with suppressed star formation at the center relative to the \gtt\ run.}
    \label{fig:map_res}
    \vspace{-3mm}
\end{figure*}

\begin{figure}
    \includegraphics[width=\linewidth]{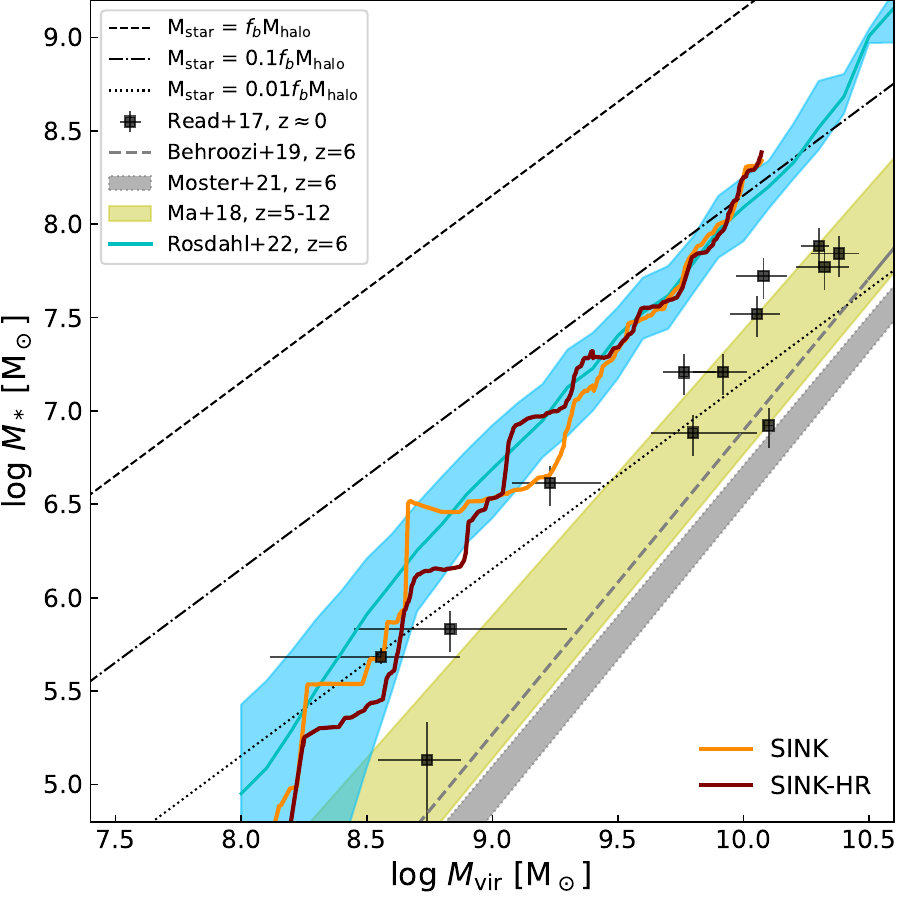}
    \vspace{-5mm}
    \caption{Same as Fig.~\ref{fig:SMHM}, but for the \sinkhr\ run. The two \sink\ runs show better convergence at larger halo masses as the stochasticity in star formation is averaged out.} 
    \label{fig:SMHM_res}
\end{figure}

\begin{figure}
    \includegraphics[width=\linewidth]{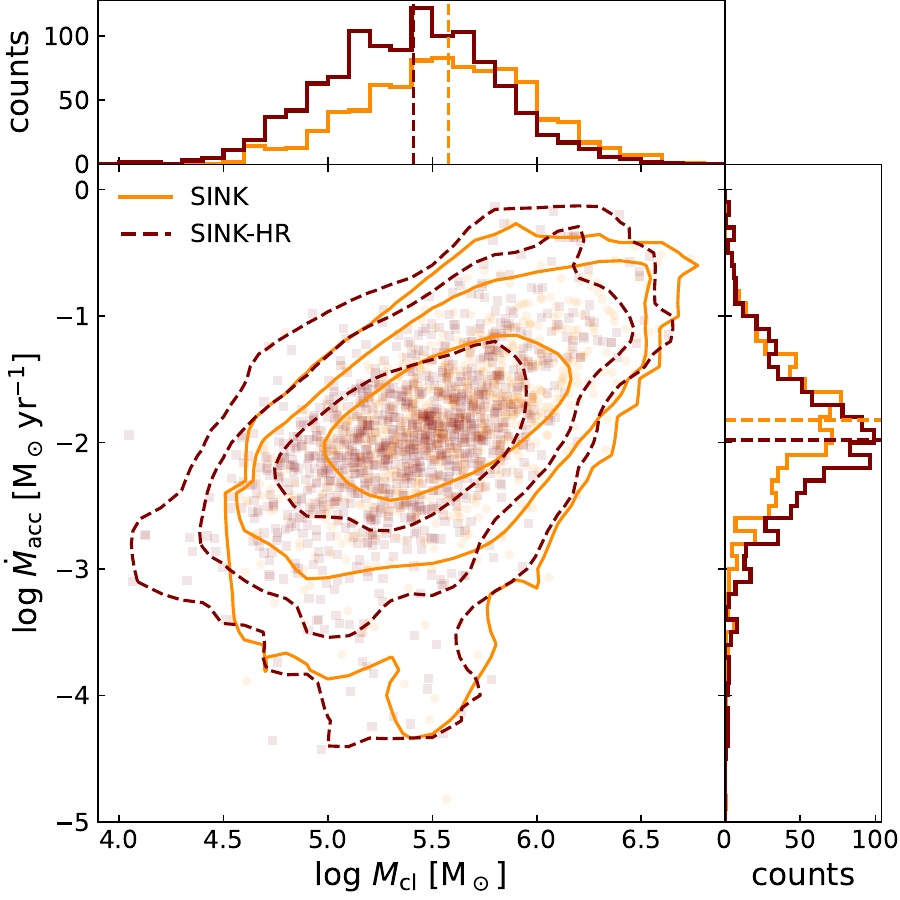}
    \vspace{-5mm}
    \caption{Accretion rate of sink particles as a function of their host clump mass. The three contours enclose the 16th, 50th, and 84th percentiles derived from a Gaussian kernel density estimate of underlying scatter points. The histograms on the top and right panels are one-dimensional distributions of two quantities, with dashed lines indicating their median values.}
    \label{fig:Macc-Mcl_res}
\end{figure}

\begin{figure*}
    \includegraphics[width=\textwidth]{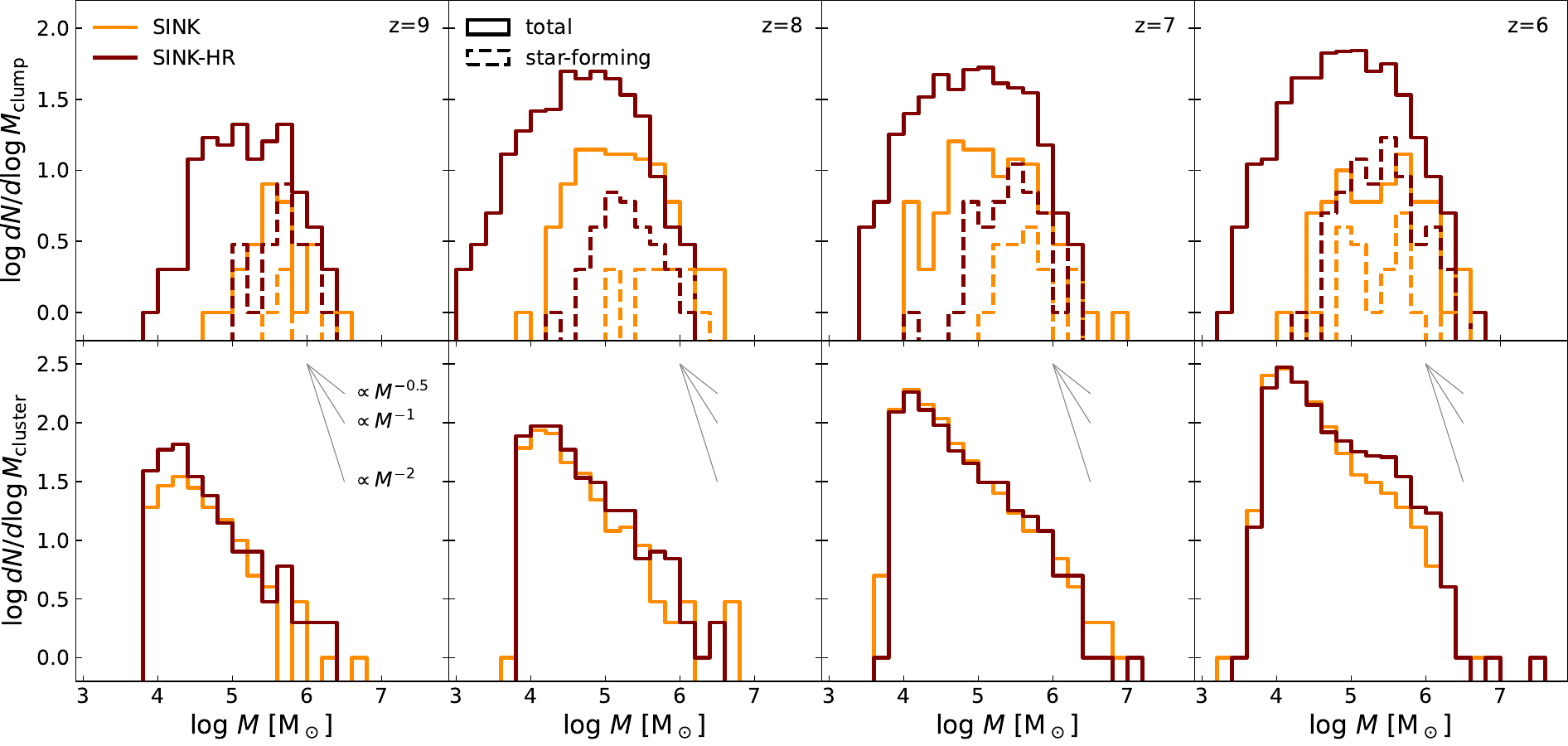}
    \vspace{-5mm}
    \caption{Mass function of gas clumps (top) and star clusters (bottom) in the two \sink\ simulations at four different epochs. Solid histograms represent total populations, whereas dotted lines in the top panels indicate the mass function of clumps currently forming stars.}
    \label{fig:mass_function}
\end{figure*}

We compare the fiducial \sink\ run with \sinkhr, which has a minimum cell size of $\dxmin=4\,\mathrm{pc}$ at $z=6$, to assess the robustness of our results against numerical resolution. Fig.~\ref{fig:map_res} shows that the gas distribution remains clumpy and turbulent, similar to the \sink\ run, rather than developing the rotational structure seen in the \gtt\ run. We also find that several galaxy properties of the \sinkhr\ run, including SFR and gas-phase metallicity, are well converged with those of the \sink\ run throughout the simulation time. The UV luminosity at $z=6$ is higher by a factor of 4 because star formation from the merged satellite occurs slightly later in the \sinkhr\ run (Table~\ref{table2}); however, this enhancement is temporary. Our clump tree analysis of the \sinkhr\ run further confirms that all star-forming clumps are destroyed within 20 Myr, again consistent with the \sink\ run. While the gas properties show little difference, the stellar distribution in the \sinkhr\ run differs slightly from that in the \sink\ run. As shown in the bottom panels, the \sinkhr\ run exhibits a large number of compact, bright peaks with stellar mass surface densities above $10^{3.5}\,\Msun\,\mathrm{pc^{-2}}$. This indicates enhanced small-scale fragmentation at higher resolution, which leads to a larger population of smaller gas clumps.

The global SFH exhibits robust convergence despite small-scale differences in stellar mass surface densities. Fig.~\ref{fig:SMHM_res} presents the stellar-to-halo mass relations of the two \sink\ runs, which nearly overlap at $\Mvir\ge10^{9.5}\,\Msun$. However, at lower masses, the two evolutionary tracks frequently deviate from each other. We attribute this early divergence to stochastic differences in the formation times of gas clumps and sink particles between the two resolutions. Stellar mass growth is dominated by a few clumps when the halo mass is small, and their formation times are not fully synchronized across simulations. As the halo grows, these timing differences are gradually averaged out as the halo hosts a larger number of gas clumps at various evolutionary stages.

To investigate the convergence of the accretion process, we show in Fig.~\ref{fig:Macc-Mcl_res} the instantaneous accretion rate of sink particles as a function of their host clump mass. Most clumps contain only one sink particle; when multiple sink particles are present, we plot the total accretion rate. In both runs, the total accretion rate scales roughly with the clump mass as $\log \dot M_\mathrm{acc}=a\log M_\mathrm{cl}+b$, implying that star formation is more active in more massive clumps. We find that both the slope ($a=0.79$ vs. 0.80) and normalization ($b=-6.26$ vs. -6.30) are well converged, suggesting that the accretion rate at fixed clump mass is nearly identical across the two resolution runs. The main difference appears in the stacked clump mass distribution (top panel), which shifts toward lower masses in the \sinkhr\ run and results in a median accretion rate lower by $\sim30\%$ (right panel).

We examine clump mass functions at four different epochs to understand why global SFR and total stellar mass converge despite differences in the clump mass distributions between the two resolution runs (Fig.~\ref{fig:mass_function}). As expected, the primary difference between the two runs appears at the low-mass end of the clump mass function. However, the additional low-mass clumps in the \sinkhr\ run do not contribute significantly to the global SFR for two reasons. First, although the higher resolution increases fragmentation and the emergence of more small-scale clumps, most of these structures do not reach the densities required to trigger sink formation. Consequently, the mass functions of clumps hosting actively accreting sink particles (dashed lines) are more similar between the two runs than those of the full clump population. Second, even when stars occasionally form within these low-mass clumps, their contribution to the total stellar mass remains negligible. As shown in Fig.~\ref{fig:Macc-Mcl_res}, most star formation is driven by sink particles with high accretion rates in massive clumps, whose distribution is not significantly affected by the increase in resolution.

Cluster mass functions within clumps (bottom panels of Fig.~\ref{fig:mass_function}) exhibit good numerical convergence between the two simulations. In both runs, the mass functions follow a power law with slope $\alpha$ ranging from -1 to -0.5 when expressed as $dN/d\log{M_\mathrm{cluster}}\propto M_\mathrm{cluster}^\alpha$. In the equivalent form $dN/dM_\mathrm{cluster}\propto M_\mathrm{cluster}^{\alpha-1}$, this corresponds to a slope between $-2$ and $-1.5$. This range is broadly consistent with the star cluster mass function inferred from the observations of nearby galaxies \citep[e.g.,][]{Lada2003} and with theoretical expectations \citep[e.g.,][]{Elmegreen1997}. The normalization of the mass functions converges well despite the difference in the number of identified gas clumps at the low-mass end. This consistency suggests that the sink-based star formation model operates robustly across the examined resolution range.

\subsection{Physical origin of bursty star formation in the model with sink particles}
\label{sec:bursty}

Observational studies using CO and $\mathrm{H}\alpha$ emissions suggest that the typical lifetime of GMCs is $\sim10$--$30\,\mathrm{Myr}$ \citep{Kruijssen2019, Chevance2020, Chevance2022}, a timescale comparable to or shorter than the lifetime of Type~II SN progenitors. This implies that the star-forming timescale ($t_\mathrm{SF}$) should be even shorter, requiring stellar feedback to disperse clumps before runaway cooling can sustain long-lived collapse. In contrast, the \gtt\ run produces numerous long-lived and overmassive clumps, as indicated by the upward-pointing arrows in Fig.~\ref{fig:clump}, thereby providing clear evidence of overcooling (Appendix~\ref{sec:appendix_overcooling}).

To understand the efficient disruption of GMCs and the origin of bursty star formation, we first examine why overcooling persists and why star formation remains relatively smooth in the \gtt\ run. To this end, we compute the net cooling rate ($\dot E_\mathrm{cool}$) and the photoionization heating rate ($\dot E_\mathrm{PH}$) within each clump. For $\dot E_\mathrm{cool}$, we account for all cooling processes implemented in the code, including radiative cooling of primordial species and metals. For $\dot E_\mathrm{PH}$, we compute the total excess energy of LyC photons remaining after the ionization of \ion{H}{i}, \ion{He}{i}, or \ion{He}{ii} (i.e., $E_\mathrm{LyC}-13.6,\,24.59$, or 54.42 eV, respectively). The number densities of the primordial species within each clump are used to estimate the interaction probability. We assume that all ionizing photons are absorbed within each clump, making $\dot E_\mathrm{PH}$ an upper limit for the heating rate from stellar radiation.

\begin{figure}
    \includegraphics[width=\linewidth]{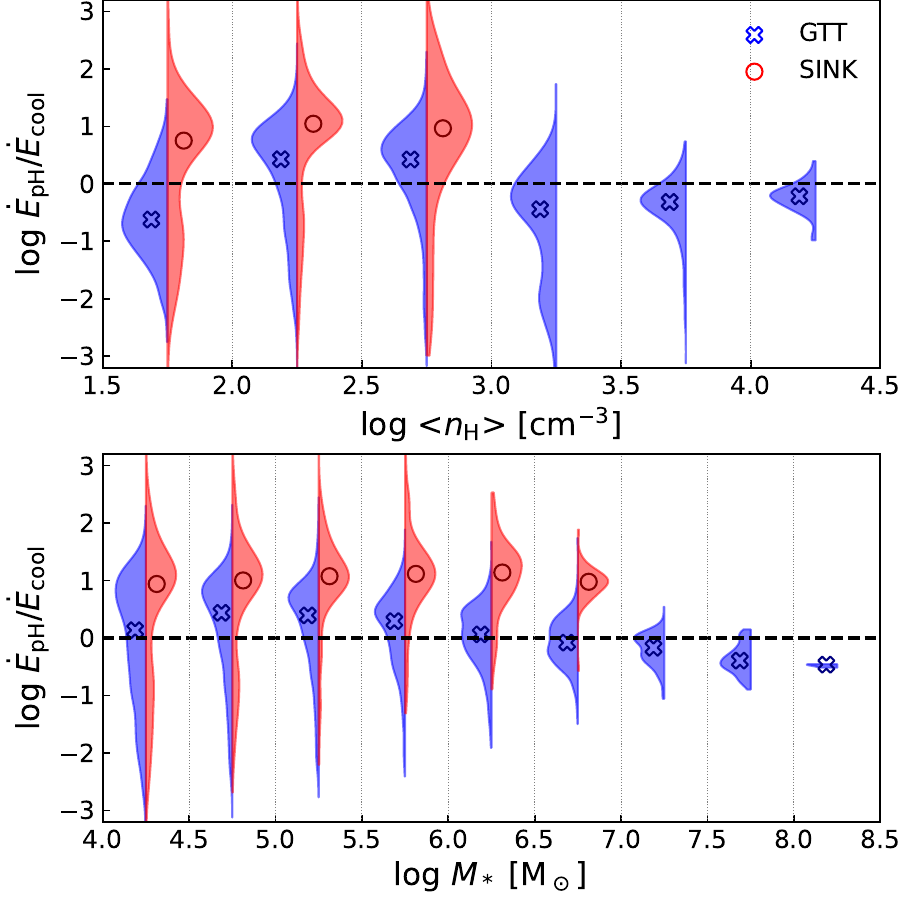}
    \vspace{-5mm}
    \caption{PDF of the ratio between the photoionization heating and cooling rates of individual clumps ($\dot E_\mathrm{PH} / \dot E_\mathrm{cool}$), as a function of the average hydrogen number density (top) and the stellar mass (bottom). The entire clump population is divided into different subsets based on gas density or stellar mass with both quantities uniformly binned in logarithmic space by 0.5 dex. Circles and crosses indicate the median values for the \sink\ and \gtt\ runs, respectively. Empty bins for the \sink\ run are cases where no clumps fall within the corresponding density or stellar mass range.}
    \label{fig:clump_violin}
\end{figure}

Fig.~\ref{fig:clump_violin} shows the ratio of photoionization heating to cooling rates ($\dot E_\mathrm{PH} / \dot E_\mathrm{cool}$) for all clumps that host stellar masses $\ge 10^4\,\Msun$, as a function of average clump density and stellar mass. Because gas cooling becomes increasingly efficient with increasing gas density ($\dot E_\mathrm{cool}\propto n^2_\mathrm{H}$), the cooling rate in the \gtt\ run eventually exceeds the photoionization heating rate in most clumps once $\left\langle n_\mathrm{H}\right\rangle \gtrsim 1000\,\cmq$. Consequently, the energy injected by ionizing photons is rapidly radiated away and cannot effectively heat the clump. This inefficient heating sustains star formation and allows several clumps to host massive star clusters ($\ge10^7\,\Msun$). In contrast, the median ratio $\dot E_\mathrm{PH} / \dot E_\mathrm{cool}$ in the \sink\ run remains around 10 across all density and mass bins. Correspondingly, the \sink\ run exhibits few clumps with $n_\mathrm{H}>1000\,\cmq$ or stellar masses exceeding $10^7\,\Msun$.
We remind the reader that the effect of SN feedback is not considered in this analysis. As shown in Figs.~\ref{fig:cluster} and~\ref{fig:clump}, local quenching and clump disruption typically occur before the onset of most SNe in the \sink\ run. This suggests that the inclusion of SN feedback would not change the results qualitatively. Conversely, SNe may contribute to clump dispersal in the \gtt\ run, given its longer $t_\mathrm{SF}$. Nevertheless, the high gas densities sustained in the \gtt\ run would likely lead to a rapid radiative loss of SN energy, preventing them from effectively suppressing persistent overcooling.

\begin{figure}
    \includegraphics[width=\linewidth]{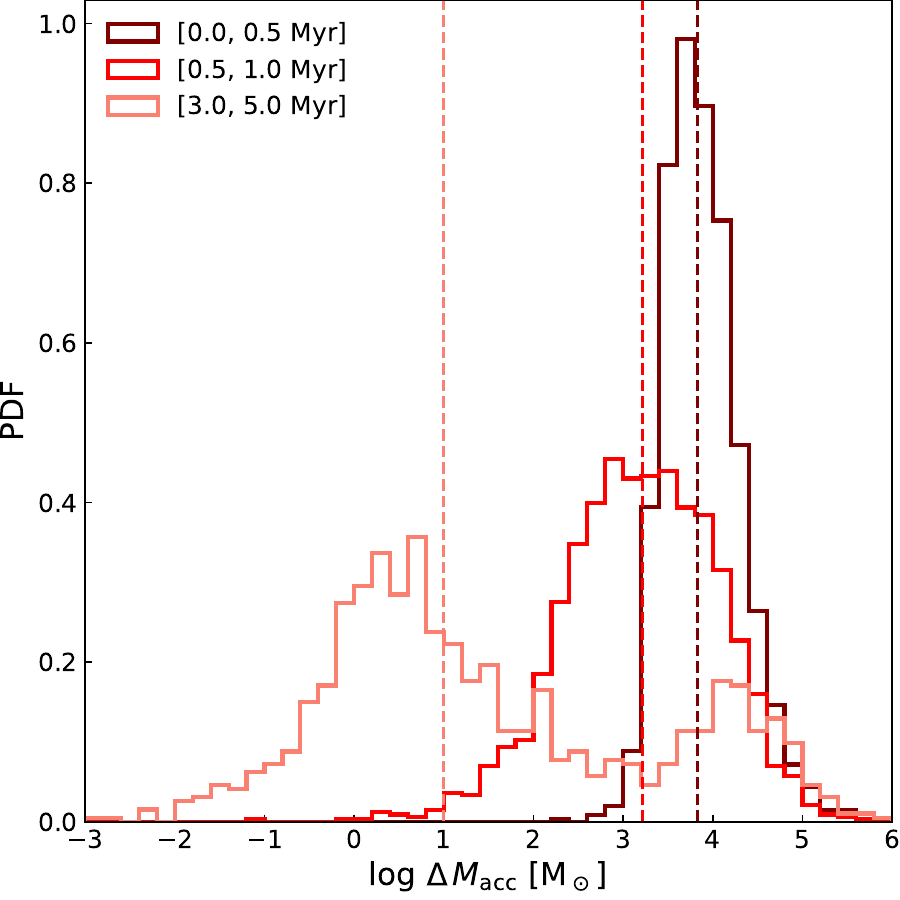}
    \vspace{-5mm}
    \caption{PDF of the total gas mass accreted by individual sink particles during three different sink age intervals: the first 0.5 Myr after sink formation (maroon), 0.5--1 Myr (red), and 3--5 Myr (salmon). The vertical dashed lines indicate the median value of each distribution.}
    \label{fig:Macc}
\end{figure}

Rapid, bursty star formation in the \sink\ run plays a key role in dispersing clumps before they evolve into denser or more massive structures. Fig.~\ref{fig:Macc} shows the PDF of the total gas mass accreted by individual sink particles, $\Delta M_\mathrm{acc}=\int^{t_2}_{t_1} \dot M_\mathrm{acc}\,dt$, over three sink age intervals. The accretion rate peaks immediately after sink formation and declines with sink age. The initial strong accretion occurs because sink formation requires the parent gas clump to be gravitationally bound and undergoing collapsing, which drives efficient gas inflows. During this initial bursty phase ($\sim0.5\,\mathrm{Myr}$), a typical sink accretes $\sim7000\,\Msun$ of gas at a rate of $\sim0.014\,\Msunyr$. This accretion rate corresponds to a median clump scale efficiency of $\varepsilon_\mathrm{ff,cl}=\dot{M}_\mathrm{acc}\,t_\mathrm{ff} / M_\mathrm{cl} \sim 21\%$, assuming a uniform density for parent clumps when estimating their free fall time. Although Eq.~\eqref{eq:epsff} often yields a cell scale $\epsff$ greater than 0.3 in the \gtt\ run, it does not imply a similarly high efficiency at the clump scale because only a small fraction of cells within each clump are actively forming stars.
Given a typical clump mass of $3\times10^{5}\,\Msun$, the accretion could, in principle, continue for $M_\mathrm{cl}/\dot M_\mathrm{acc}\sim 30\,\mathrm{Myr}$ (Fig.~\ref{fig:Macc-Mcl_res}). However, within the next $\sim0.5\,\mathrm{Myr}$, the median accretion rate declines rapidly by a factor of $\sim4$. This decline is primarily driven by radiation feedback from young stars, which suppresses further gas infall \citep[see Fig.~6 of][]{Kang2025}. 
Finally, gas accretion and star formation are nearly quenched once sink particles become $3$--$5\,\mathrm{Myr}$ old. This reflects clump dispersal driven by cumulative radiation feedback over the preceding few Myr, together with SN explosions from the earliest-born progenitors spawned by the same sink. Sink particles that continue to accrete at this age are typically embedded in massive clumps ($\ge10^{6.5}\,\Msun$), whose $t_\mathrm{SF}$ can extend to $\sim10\,\mathrm{Myr}$. At ages $>5\,\mathrm{Myr}$, accretion rates remain close to zero unless the host clumps merge and provide additional gas for star formation. This sequence of rapid accretion followed by efficient quenching illustrates how bursty star formation on short timescales can promote self-regulation via coherent stellar feedback.

In contrast, star formation in the \gtt\ model is more intermittent (Fig.~\ref{fig:cluster}), and the resulting feedback is correspondingly less coherent. This stochastic behavior primarily originates from the stringent star formation criteria of the model, which suppresses continuous star-formation events within the same cell. To investigate how an individual star formation event instantaneously alters the thermodynamic state of its host cell, we restart the \gtt\ run at $z=7$ and evolve it for a short interval ($\sim0.3\,\mathrm{Myr}$), recording outputs at every coarse time step ($\sim 800\,\mathrm{yr}$). We analyze the gas properties of the star-forming cells immediately after 493 distinct star-forming events using this high-cadence dataset. We find that the formation of a new star particle increases the sound speed by $\sim20\%$ through photoionization heating, often shifting the host cell out of the star formation regime defined in Eq.~\eqref{eq:GTT}. Although this feedback is not sufficiently energetic to disrupt the entire clump (Fig.~\ref{fig:clump_violin}), it delays subsequent star formation until the gas cools and turbulence dissipates. This slow star formation prevents a rapid decline in gas density and inhibits coherent feedback, thereby explaining why the \gtt\ model produces numerous long-lived clumps.

\begin{figure}
    \includegraphics[width=\linewidth]{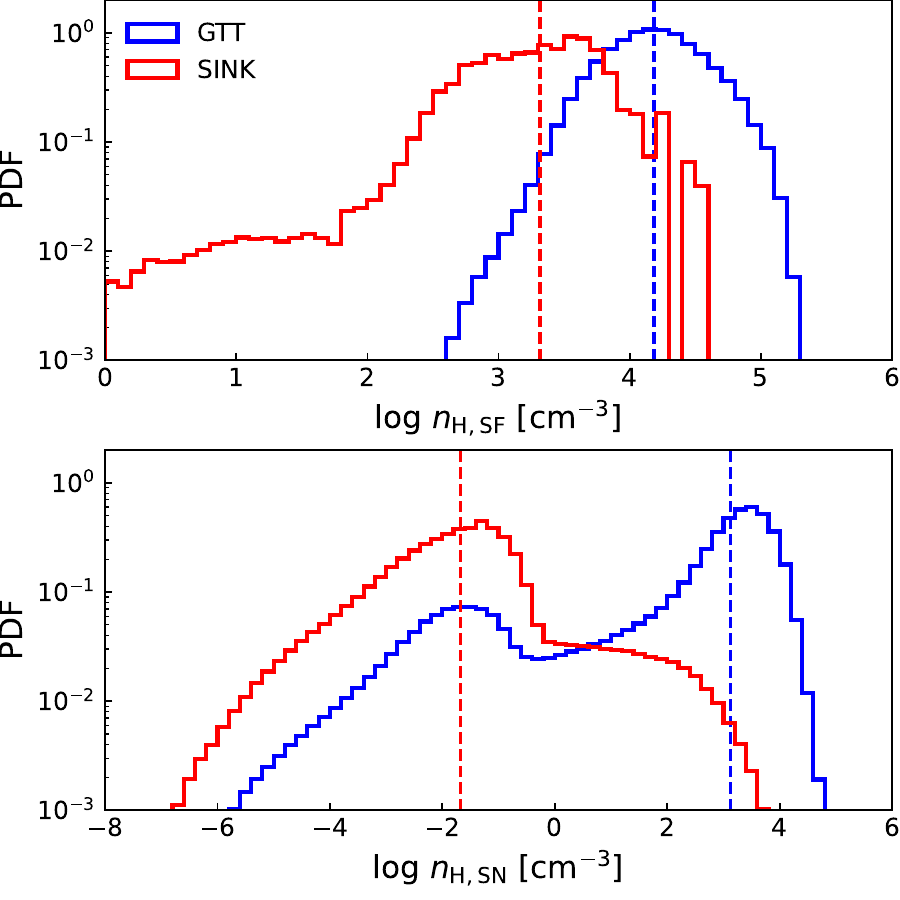}
    \vspace{-5mm}
    \caption{PDF of the hydrogen number densities of star-forming cells (top) and SN explosion sites (bottom). The vertical lines represent the median density of each distribution, which are color coded to match the histograms.}
    \label{fig:SF+SN_hist}
\end{figure}

The contrasting star formation modes and resulting feedback efficiencies are clearly illustrated in Fig.~\ref{fig:SF+SN_hist}. The top panel shows the PDFs of hydrogen number densities at the time of star formation ($n_\mathrm{H,SF}$). For the \sink\ run, the PDF is weighted by the accretion rate of each sink particle to represent the typical density where stars form. In the \sink\ run, stars form at densities that are lower by a factor of $\sim7$ than those in the \gtt\ run \citep[$n_\mathrm{H,SF}=10^{3.4}$ vs. $10^{4.2}\,\cmq$, see also][]{Kang2025}. As discussed above, this difference arises because the \gtt\ run, governed by Eq.~\eqref{eq:GTT}, imposes a more stringent thermo-turbulent requirement for collapse, whereas in the \sink\ run, converging gas flows can develop at lower densities and efficiently convert gas into stars, thereby reducing $n_\mathrm{H,SF}$. This difference extends to environments in which SNe explode, as shown in the bottom panel. The median density at which SNe explode ($n_\mathrm{H,SN}$) differs even more significantly than $n_\mathrm{H,SF}$ ($n_\mathrm{H,SN} \approx 10^{-1.7}$ vs. $10^{3.1}\,\cmq$). Radiation feedback in the \sink\ run pre-processes star-forming regions more strongly, clearing dense gas before the first SN occurs. The five orders of magnitude lower median $n_\mathrm{H,SN}$ in the \sink\ run relative to that in the \gtt\ run increases the momentum injected by SN by a factor of $\sim 3.7$ (Eq.~\eqref{eq:SN}). By integrating over all SN events, we further find that the average momentum per SN is similarly larger in the \sink\ run, resulting in a larger total injected momentum despite its lower overall SFR.

In summary, a strong collapse at the onset of sink formation drives the bursty nature of the \sink\ run. Rapid gas accretion onto sink particles induces clustered star formation within $\sim 1\,\mathrm{Myr}$ and generates intense ionizing radiation, terminating star formation in most clumps before the first SNe explode. In contrast, the \gtt\ run exhibits an intermittent, extended, and inefficient mode of star formation even during active gas collapse. A strict physical criterion in the \gtt\ model limits rapid gas conversion, preventing a burst-and-quench cycle. Although star formation persists at the clump scale, it remains too slow to disperse the parent cloud \citep[see also][]{HanD2026}.

\subsection{Interplay between the local and global efficiencies}

\subsubsection{Reduced accretion efficiency in the sink-based model}
\label{sec:acc0.5}

The sink accretion scheme employed in this work was validated against several analytical benchmarks, such as Bondi accretion \citep{Bleuler2014} and the collapse of an isothermal sphere \citep{Kang2025}. However, in the context of cosmological simulations, the gas structure surrounding the sink is often marginally resolved, potentially rendering the assumption that the total mass flux within the accretion zone is instantly incorporated into the sink inaccurate.
Moreover, our current model neglects small-scale feedback mechanisms, specifically prestellar jets \citep{Cunningham2011, Machida2012, Guszejnov2021}, which could be critical for determining the net accretion efficiency. Indeed, \citet{FK12} suggested that an efficiency of $\varepsilon_\mathrm{acc}=0.5$ is required to reconcile simulated results with the observed IMF, particularly when explicit feedback mechanisms are not included. 
Given these uncertainties, it is important to understand how the efficacy of self-regulation via radiation feedback and resulting stellar masses depend on the burstiness of star formation. To address this issue, we perform a controlled experiment by restarting the \sink\ run at $z=7.05$ ($t=754\,\mathrm{Myr}$), a period corresponding to a local minimum in the SFR. We artificially reduce the accretion rates by a factor of 2, effectively setting $\varepsilon_\mathrm{acc}=0.5$. In this configuration, the remaining 50\% of inflowing gas is retained within the accretion volume rather than being ejected, thereby increasing the local gas density and allowing us to isolate the effects of reduced accretion on subsequent feedback cycles.

\begin{figure}
    \includegraphics[width=\linewidth]{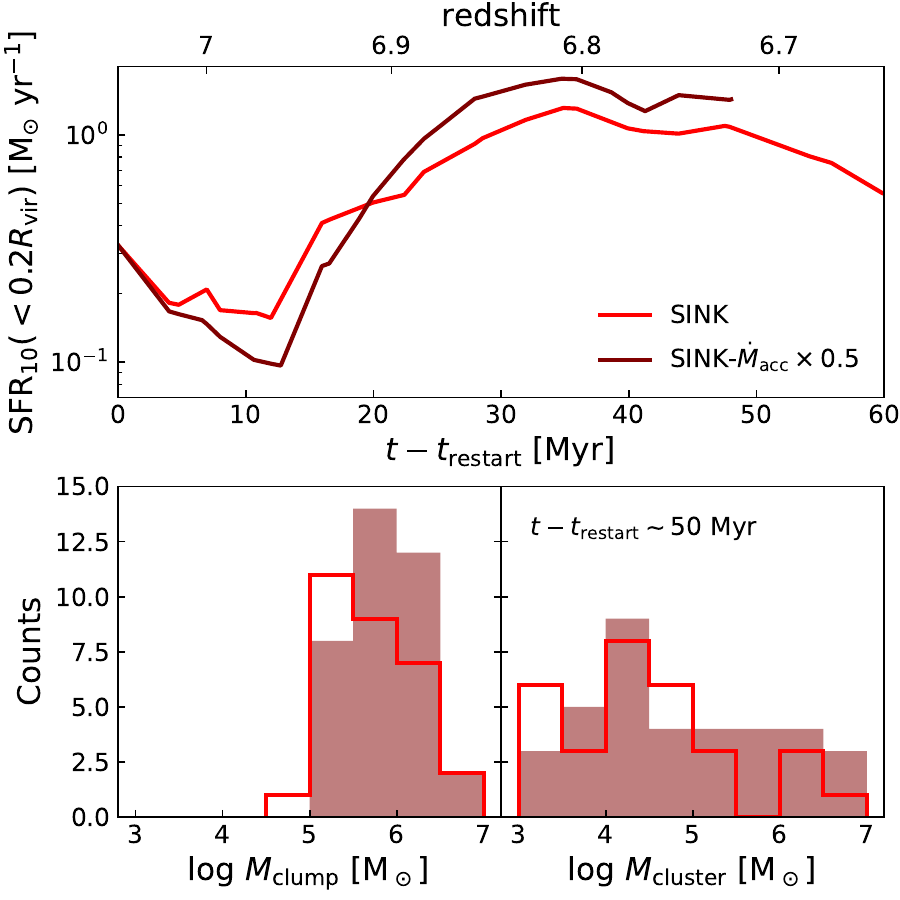}
    \vspace{-5mm}
    \caption{Top: SFR of the two simulations as a function of time since restart. Bottom: mass distribution of gas clumps (left) and their associated star clusters (right) at $\sim50\,\mathrm{Myr}$ after the restart. Only clumps or clusters that host at least one star particle formed after the restart are included.}
    \label{fig:Macc_reduced}
\end{figure}

The top panel of Fig.~\ref{fig:Macc_reduced} shows the SFR measured within the central region of the main halo, which demonstrates that the effect of accretion efficiency on the galactic SFH is non-negligible. At early times ($t-t_\mathrm{restart} \lesssim 15\,\mathrm{Myr}$), the simulation with reduced accretion efficiency exhibits a lower SFR because the reduced accretion rate limits the rapid depletion of the gas reservoir relative to the fiducial \sink\ run. At later times, the SFR in the reduced-efficiency run increases rapidly and exceeds that of the \sink\ run at $t-t_\mathrm{restart} \gtrsim 20\,\mathrm{Myr}$. This behavior arises because the 50\% of gas that was not immediately accreted remains within the accretion zone, increasing the local gas density and eventually enhancing subsequent accretion.

We present the mass distributions of gas clumps and star clusters at 50 Myr after the restart in the bottom panels to qualitatively assess the effect of the reduced efficiency. The most massive cluster in the reduced-efficiency run becomes larger, increasing the maximum clump mass. Despite the enhanced burstiness at later times, clump dispersal proceeds more slowly than that in the fiducial run. The initial suppression of accretion delays early feedback, giving clumps more time to collapse into more strongly bound structures that are more resistant to subsequent feedback. Consequently, the reduced efficiency run produces $\sim20\%$ more stars during the re-simulated period.

These results confirm that slow star formation in regions of converging flow leads to higher gas densities and promotes the growth of massive star-forming clouds. Therefore, simply reducing accretion efficiency is unlikely to suppress star formation to the level required by the observations. However, explicit modeling of prestellar feedback can remove gas from central regions and reduce local gas densities even when accretion rates are limited \citep{Wang2010, Federrath2015, Appel2022}. Incorporating such a physical process may be essential for establishing a more robust self-regulation mechanism and reconciling the simulated stellar-to-halo mass relation with observational constraints.

\subsubsection{Testing a maximally bursty model with Schmidt law}

\begin{figure*}
    \includegraphics[width=\textwidth]{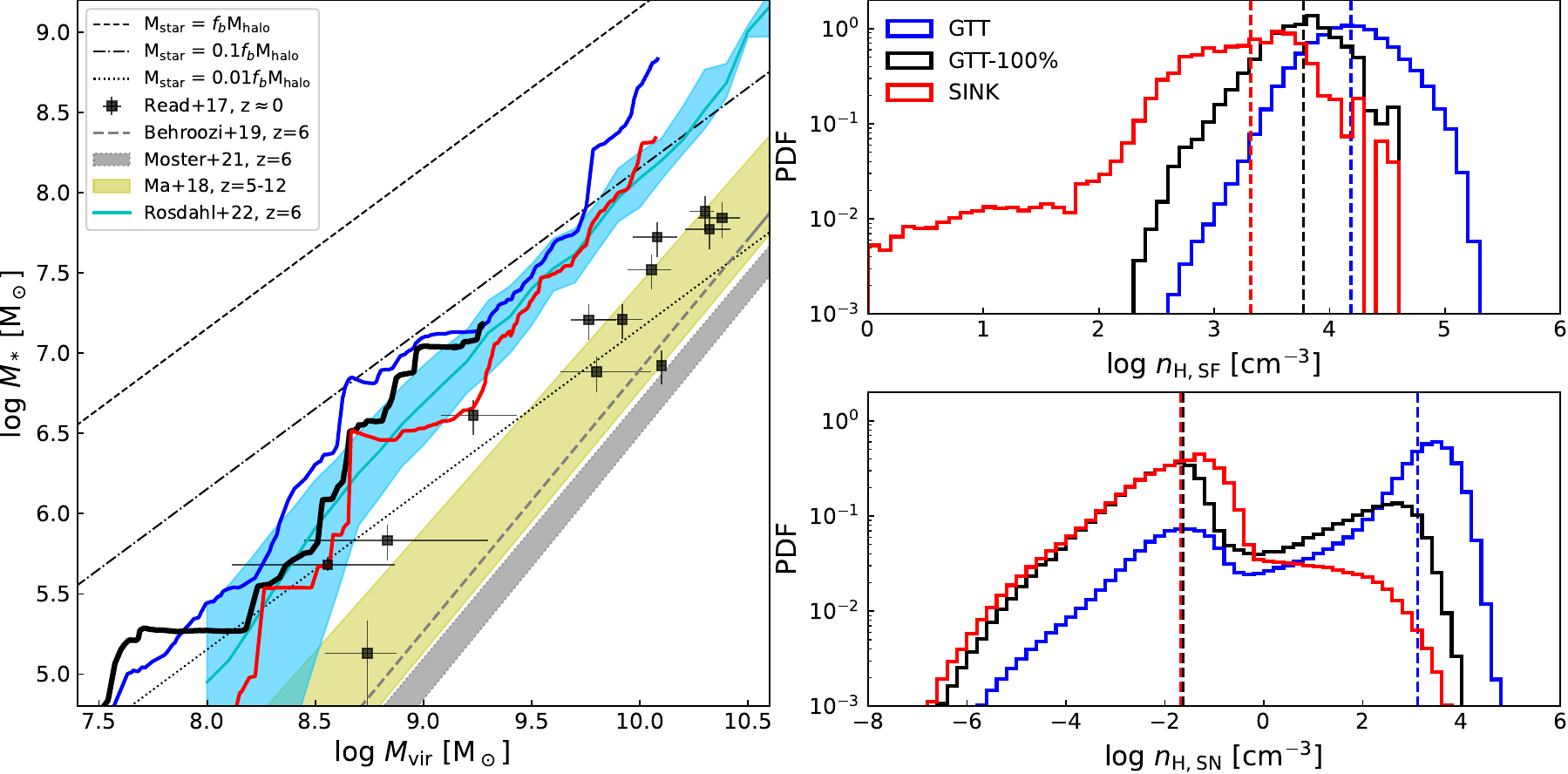}
    \vspace{-5mm}
    \caption{Same diagnostics as Figs.~\ref{fig:SMHM} (left) and \ref{fig:SF+SN_hist} (right), but including the $\epsff=100\%$ run in black lines.}
    \label{fig:eps100}
\end{figure*}

As discussed in the previous subsection, the accretion rate in our fiducial \sink\ run is determined directly from the inward mass flux across the accretion volume. A natural question that arises is whether this numerical treatment makes the \sink\ model conceptually identical to a standard Schmidt law with a local efficiency of $\epsff=100\%$. To address this and investigate how the sink-based model differs from a maximally bursty Schmidt-law prescription, we conduct another experiment by running a \gtt\ simulation with $\epsff=100\%$. To facilitate a more direct comparison, we adopt star formation criteria in this test run that match the physical conditions required for sink formation\footnote{Due to limited computational resources, this run was carried out only down to $z=9$.}. More specifically, we bypass Eq.~\eqref{eq:GTT} and instead apply the density criterion described in Sect.~\ref{sec:SINK} ($\rho_\mathrm{gas} > 8.86 c^2_\mathrm{s}/\pi G \dxminsq$) to candidate cells with $\alpha_\mathrm{vir}<1$ and converging flows.

The results down to $z=9$ indicate that this simple model regulates star formation more effectively than the fiducial \gtt\ model, but still does not reproduce the behavior of the \sink\ run (Fig.~\ref{fig:eps100}). At early times, the stellar-to-halo mass ratio in the $\epsff=100\%$ run grows more slowly than in the \gtt\ run and remains closer to that of the \sink\ run (left panel), suggesting that the rapid conversion of locally collapsing gas into stars indeed enhances feedback. However, as the halo grows, the stellar mass in this run gradually increases above that of the \sink\ run and approaches the trend of the fiducial \gtt\ model. The density distribution of star-forming sites measured at $z>9$ (right panel) supports this interpretation. Because the original turbulent Jeans length condition is replaced by a sink-like density threshold and a virial check, the $\epsff=100\%$ run forms stars at lower densities than the fiducial \gtt\ run. Its SN density distribution has a median value comparable to that of the \sink\ run, indicating that many SNe occur only after the surrounding gas has been partially cleared by radiation feedback. However, the $\epsff=100\%$ run retains a more prominent high-density tail than the \sink\ run, suggesting that radiative pre-processing before SNe is less effective, allowing some dense, long-lived star-forming clumps to survive until the first SNe explode. Thus, although the combination of sink-like star formation criteria and a maximized local efficiency improves feedback regulation during the early stages of galaxy evolution, it does not suppress high-density SN environments as efficiently as the \sink\ model and therefore fails to maintain the coherent burst--quench cycle required to regulate star formation at later times.

The $\epsff=100\%$ experiment therefore demonstrates that simply maximizing the local efficiency in a Schmidt-law model is insufficient to reproduce the behavior of the \sink\ run. Replacing the turbulent Jeans length criterion with the sink density threshold allows star formation to occur at lower densities than in the fiducial \gtt\ run, but this alone does not produce burstier star formation. By construction, our Schmidt-law model treats each star formation event as a stochastic, cell-based process. Once a star particle forms, subsequent events depend on whether the same cell or neighboring regions again satisfy the local criteria. Because radiation feedback immediately heats the surrounding gas at the onset of star formation, the local thermodynamic conditions are rapidly altered, making subsequent star formation less likely. In the \sink\ model, by contrast, once a sink particle forms at the density peak of a collapsing region, subsequent star formation is regulated by continued gas accretion onto the sink particle. Consequently, the collapsing region remains a persistent site of star formation, naturally producing spatially and temporally clustered star formation. The \sink\ model is therefore not equivalent to a Schmidt-law prescription with $\epsff=100\%$. Instead, its stronger regulation arises because it couples star formation directly to the evolution of a collapsing gas reservoir. In this sense, the \sink\ model provides a physically motivated subgrid description of unresolved gravitational collapse that is not captured by conventional Schmidt law prescriptions.

\subsection{Galaxy-scale burstiness and UV variability}
While the bursty star formation discussed throughout this paper refers to rapid and clustered star formation on GMC and clump scales, observational measures of burstiness are typically defined using galaxy-integrated quantities such as temporal variations in SFR or UV luminosity, which are not necessarily equivalent. Nevertheless, clustered star formation on small scales found in the \sink\ run drives wider fluctuations in the global star formation activity of the galaxy. Motivated by recent studies linking UV variability to the abundance of UV-bright galaxies at high redshifts, we examine whether the different star formation behaviors found in the \sink\ and \gtt\ runs lead to differences in galaxy-scale burstiness and UV variability.

\citet{Shen2023} quantified the UV variability, $\sigma_\mathrm{UV}$, by introducing scatter in $M_\mathrm{UV}$ at fixed halo mass, thereby effectively modeling the distribution $P(M_\mathrm{UV}|M_\mathrm{halo})$ when constructing the UV luminosity function. In a similar spirit, but using a slightly different approach, \citet{Kravtsov2024} modeled stochastic fluctuations in the SFR of galaxies and propagated them into variations in UV luminosity. These studies therefore characterized $\sigma_\mathrm{UV}$ in a population-wide sense. Since our simulations follow a single zoom-in halo, we cannot directly measure the full distribution of $M_\mathrm{UV}$ at fixed halo mass. Instead, following the approach of \citet{Semenov2025bursty}, we estimate the temporal fluctuation of $M_\mathrm{UV}$ as a proxy for $\sigma_\mathrm{UV}$. In their work, $\sigma_\mathrm{UV}$ was computed as the standard deviation of $2.5\log\mathrm{SFR}$ over the past 100 Myr, assuming that the UV luminosity is proportional to the SFR. Here, we directly compute $M_\mathrm{UV}$ in the AB magnitude system using our adopted SED model. Instead of adopting a fixed time window, we bin the evolutionary history by halo mass and evaluate the scatter in $M_\mathrm{UV}$ within each bin. This provides a single-trajectory proxy for the UV variability at a given halo mass scale, which can be directly compared with the scatter invoked in empirical UV luminosity function models.

\begin{figure}
    \includegraphics[width=\linewidth]{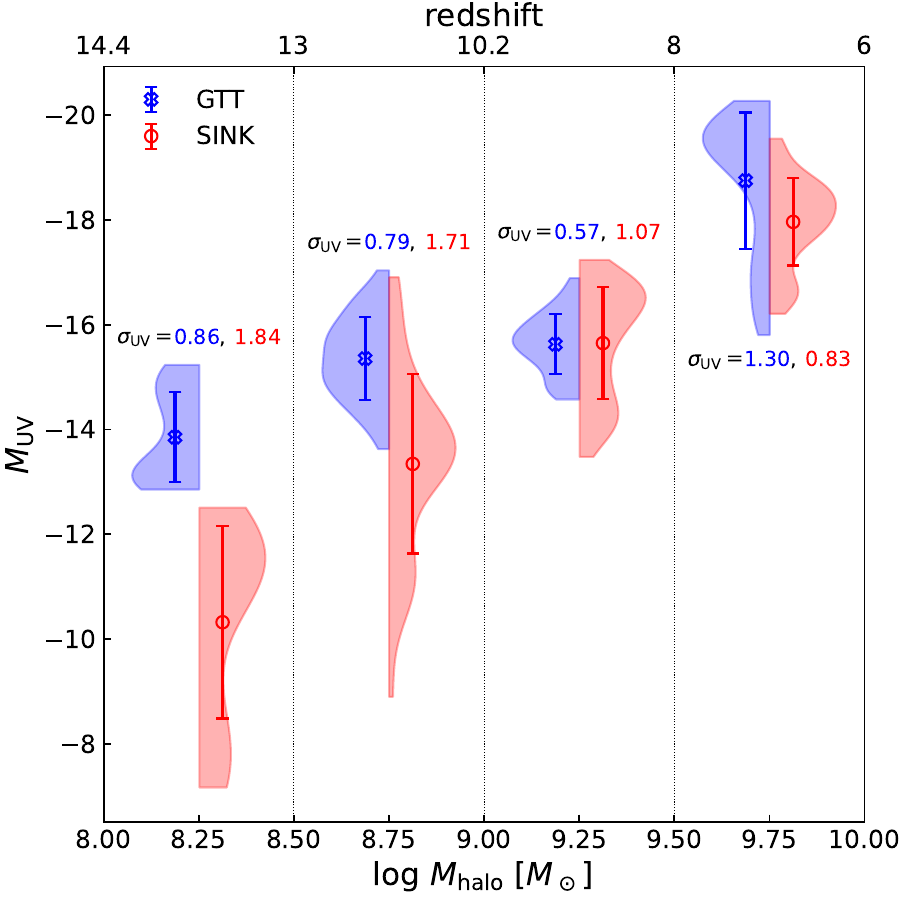}
    \vspace{-5mm}
    \caption{Distribution of UV absolute magnitudes, $M_\mathrm{UV}$, in different halo mass bins for the \gtt\ and \sink\ runs. Blue and red half-violins show the distributions of $M_\mathrm{UV}$ in the \gtt\ and \sink\ runs, respectively. Symbols and vertical error bars indicate the mean and standard deviation of $M_\mathrm{UV}$ in each bin, with the corresponding values of $\sigma_\mathrm{UV}$ annotated in blue and red texts. The top axis shows the approximate redshift corresponding to the cosmic time at each halo mass bin.}
    \label{fig:sigma_UV}
\end{figure}

Fig.~\ref{fig:sigma_UV} shows the distribution of $M_\mathrm{UV}$ in each halo mass bin. Overall, the \sink\ run is systematically fainter than the \gtt\ run at fixed halo mass, reflecting its lower stellar mass and SFR. The value of $\sigma_\mathrm{UV}$ generally decreases with increasing halo mass in both runs. This trend suggests that UV fluctuations are more prominent at earlier evolutionary stages, when the galaxy is less massive and individual star-forming events contribute a larger fraction of total UV luminosity. As the halo grows, the galaxy contains a larger number of star-forming regions, and the galaxy-integrated UV luminosity becomes averaged over increasing number of stellar populations, reducing relative variability. An exception is found in the most massive bin, where the \gtt\ run shows the largest scatter among the four bins, exceeding that of the \sink\ run within the same mass range. This increase in the \gtt\ run appears to be associated with a particularly UV-bright phase near the end of the simulation, during which the SFR becomes even higher than that in the \gttco\ run (Fig.~\ref{fig:SFR}) because of a merging satellite.

Despite being less UV luminous on average, the \sink\ run exhibits a larger $\sigma_\mathrm{UV}$ at $M_\mathrm{halo}\le10^{9.5}\,\Msun$ ($z\ge8$). This implies that, for galaxies with comparable mean UV luminosities or stellar masses, the larger temporal fluctuation in the \sink\ run could enhance the probability of scattering galaxies into the bright end of the UV luminosity function \citep[e.g.,][]{Ren2019, Sun2023}. The values measured in the \sink\ run are broadly comparable to the level of scatter invoked in empirical and semi-empirical models of high-redshift UV luminosity functions, where $\sigma_\mathrm{UV}\sim1$--2 is often required to boost the abundance of UV-bright galaxies at $z\ge10$ \citep{Shen2023, Kravtsov2024}. However, because our measurement follows a single evolutionary track rather than an ensemble of galaxies at fixed halo mass, it should be interpreted as a temporal proxy for the contribution of bursty star formation to the population-wide scatter. A larger sample of simulated galaxies would be required for a more robust statistical comparison and for directly constructing the UV luminosity function.

\subsection{Comparison with previous works}
Many theoretical studies based on GMC-scale simulations with sub-parsec resolution have found that star formation is short-lived, with GMC disruption occurring immediately after its cessation. For example, \citet{KimJG2018} showed that clouds spanning a wide range of initial gas surface densities are disrupted within $\lesssim4\tff$, primarily due to photoionization-driven expansion of ionized gas. Similarly, simulations of high surface density clouds \citep{Kimm2022, Menon2025} found that star formation terminates even earlier ($\sim3\,\mathrm{Myr}$), as clouds collapse and are dispersed more rapidly, rendering Type~II SN explosions largely ineffective in regulating star formation within GMCs. Consequently, LyC photons can efficiently escape from their birth clouds, playing a key role in setting the thermal properties of the ISM and driving cosmic reionization at high redshifts.

These findings indicate that bursty star formation is essential for disrupting GMCs through early feedback and suggest that larger-scale galaxy formation simulations affected by the overcooling problem may fail to capture this behavior.
Our experiments support the idea that reproducing bursty clump-scale star formation is critical for enabling radiation feedback to disrupt star-forming clouds as effectively as in GMC simulations. 
A similar conclusion was reached by \citet{HanD2026}, who found in isolated disk galaxy simulations that the same sink-based model produces a stellar mass a factor of 2 lower than that of the \gtt\ model. Their comparison of different feedback channels further showed that a radiation-only run yields a stellar mass 2.4 times lower than that of an SN-only run, demonstrating that radiation feedback can become the dominant regulatory mechanism when star formation is sufficiently bursty.

Turning to the Schmidt-type model, the SFR-weighted average $\epsff$ at the cell scale in our \gtt\ simulation is $\approx22\%$, with a sharp peak around 30\% due to frequent star formation in transonic to subsonic turbulence \citep[see Fig.~1 of][]{Kang2025}. This local efficiency appears sufficiently high to sustain global self-regulation, given that \citet{Agertz2015, Agertz2016} showed that increasing $\epsff$ from 1\% to 10\% can be sufficient for feedback-driven regulation. Similarly, \citet{Semenov2018} found that the transition from a dynamics-regulated to a self-regulated regime occurs at $\epsff\gtrsim1\%$, above which the global depletion timescale becomes insensitive to the precise value of $\epsff$ and is instead governed by stellar feedback. Re-simulating a FIRE galaxy with a wide range of $\epsff$, \citet{Orr2018} reported that the global SFR remains largely unchanged over $1\%<\epsff<10^4\,\%$, further supporting that star formation is regulated by feedback.
However, our results with the variable-$\epsff$ model differ from previous studies that achieved regulation with lower, constant efficiencies. Although isolating the origin of this discrepancy is difficult due to differences in numerical methods, resolution, and subgrid prescriptions, \citet{Nunez-Castineyra2021} demonstrated that increasing $\varepsilon_\mathrm{acc}$ in Eq.~\eqref{eq:epsff} can reduce the stellar mass. In their disk galaxy simulations, increasing $\varepsilon_\mathrm{acc}$ from 0.09 to a maximum value of 1 led to more efficient disruption of long-lived star-forming clumps and alleviated the formation of an excessively bright bulge. Nevertheless, even with $\varepsilon_\mathrm{acc}=1$, the resulting stellar mass remained higher than that in a model with constant $\epsff=0.09$. This suggests that simply increasing $\varepsilon_\mathrm{acc}$ within the multi-freefall framework is unlikely to fully resolve the overcooling problem, and that self-regulation arises from the interplay of multiple factors rather than from $\epsff$ alone.

Within the context of star formation modeling, the criteria used to trigger star formation may be as important as $\epsff$ itself \citep{Hopkins2013SF, Nobels2024}. Previous simulations suggest that the weak sensitivity of global SFRs to $\epsff$ is achieved when star formation is restricted to dense, locally self-gravitating gas, so that clustered star formation and coherent feedback arise naturally \citep{Hopkins2014, Hopkins2018}. This highlights the importance of identifying collapsing density peaks, rather than simply increasing the local efficiency. However, the formation and identification of such peaks can themselves depend on the hydrodynamic method. In their code comparison study, \citet{Hu2023} found that Lagrangian codes produced burstier SFHs, larger SN-driven bubbles, and stronger galactic outflows than the Eulerian \textsc{ramses} run even when the same physical model was adopted. They suggested that unresolved collapse in AMR calculations can be stabilized or smoothed at the cell scale, limiting the formation of sharply peaked, self-gravitating structures.
Beyond this numerical effect, the outcome of a gravitational boundedness check also depends on how local gas properties are estimated. Accurate estimates of the local sound speed and turbulent velocity are therefore critical, because these quantities determine whether gas is identified as locally self-gravitating and therefore directly affect the ability of the model to select collapsing density peaks. In this regard, future studies incorporating more detailed chemical networks for self-consistent cooling \citep[e.g.,][]{Katz2022}, along with explicit subgrid turbulence models \citep{Semenov2025turb}, will be necessary to assess whether improved estimates of local conditions can alleviate the overcooling problem seen in our \gtt\ run.

\section{Summary}
\label{sec:summary}
In this study, we simulate the evolution of a dwarf DMH with a mass of $10^{10}\,\Msun$ at $z=6$ using two different star formation models. We demonstrate that bursty star formation can partially mitigate the overcooling problem by enhancing the spatial and temporal clustering of star formation and the associated stellar feedback. The main results are summarized as follows:

\begin{itemize}
    \item In the \sink\ run, young sink particles are embedded in strongly convergent flows and exhibit high accretion rates. This leads to highly clustered star formation on timescales of $\sim\,1 \, \mathrm{Myr}$, creating star clusters with masses of $M_\mathrm{cluster}\sim10^4$--$10^6\,\Msun$ (Figs.~\ref{fig:mass_function},~\ref{fig:Macc}). In contrast, star formation in the \gtt\ model occurs at a considerably slower rate.
    
    \item The differing modes of star formation lead to substantial differences in local SFRs, which in turn systematically alter the impact of radiative feedback. In the \sink\ run, the collective ionizing radiation provides a sufficient photon budget to fully ionize and disperse star-forming clumps within $\sim\,3 \mathrm{Myr}$ (Figs.~\ref{fig:clump},~\ref{fig:clump_violin}). In contrast, in the \gtt\ run, many star-forming clumps survive for significantly longer periods, sometimes exceeding 100 Myr, and account for more than $\sim70\%$ of the star formation.
    
    \item A significant fraction (17\%) of LyC radiation escapes from the host DMH in the \sink\ run because star-forming regions are rapidly disrupted (Fig.~\ref{fig:fesc}). In the \gtt\ run, however, long-lived clumps efficiently absorb ionizing radiation, resulting in a lower global LyC escape fraction (1.7\%).
    
    \item SNe in the \sink\ run explode in low-density cavities that are pre-processed by efficient radiation feedback. Consequently, the momentum injected into the ISM per explosion is increased by a factor of $\sim3.7$ compared to the \gtt\ run, where SNe explode in denser environments, driving stronger galactic outflows (Figs.~\ref{fig:outflow},~\ref{fig:SF+SN_hist}).
    
    \item Owing to the enhanced feedback efficiency associated with bursty star formation, the \sink\ run produces a total stellar mass that is $\sim3.2$ times lower than that in the \gtt\ run (Fig.~\ref{fig:SMHM}). The resulting rotation curve rises smoothly instead of exhibiting the centrally peaked profile often associated with overcooling (Fig.~\ref{fig:Vcirc}). The discrepancy between the two runs becomes more pronounced at higher halo masses, as feedback in the \gtt\ model becomes increasingly ineffective at deeper gravitational potentials (Figs.~\ref{fig:SMHM},~\ref{fig:SFR}).
    
    \item The impact of different feedback strengths is also reflected in observable galaxy properties. Stronger feedback in the \sink\ run redistributes metals more efficiently and produces a more extended stellar distribution. This results in a lower normalization of the mass-metallicity relation and a larger effective radius at a fixed stellar mass, in close agreement with recent high-redshift JWST observations (Figs.~\ref{fig:MZR},~\ref{fig:r50}).
    
    \item The sink-based model exhibits robust numerical convergence across two different resolutions. Although the \sinkhr\ run resolves a greater number of small-scale clumps, the cluster mass function and total stellar mass remain consistent with those of the fiducial \sink\ run (Figs.~\ref{fig:SMHM_res},~\ref{fig:mass_function}).
    
    \item Artificially reducing accretion rates does not reduce the stellar mass (Fig.~\ref{fig:Macc_reduced}). Instead, slower star formation during the early stages of clump evolution allows gas clumps to become denser, eventually boosting the SFR in the later stages. Conversely, even fixing $\epsff=100\%$ in the \gtt\ run fails to reproduce the regulation achieved by the \sink\ run, demonstrating that local efficiency alone is insufficient without clustered star formation and feedback (Fig.~\ref{fig:eps100}).
\end{itemize}

Overall, our results suggest that the sink-particle star formation model provides a promising avenue for mitigating the overcooling problem in high-redshift galaxy formation. However, despite its relative success compared with the Schmidt-type model, the present study adopts several simplifications that warrant further investigation. As discussed above, prestellar feedback may reduce the gas accretion efficiency onto sink particles and thereby alter the early evolution of star-forming clouds. In addition, non-thermal components such as magnetic fields and cosmic rays may provide important pressure support in the ISM. Because magnetic field strength increases with gas density, magnetic fields may suppress gas accretion onto sink particles. Cosmic rays are expected to drive gentler and more extended outflows and regulate the long-term gas supply from the CGM to the ISM \citep[e.g.,][]{Girichidis2016,Farcy2022,Armillotta2024}. Future studies incorporating these processes will be essential for developing a more complete and physically robust picture of galaxy evolution.

\begin{acknowledgements}
CK and TK were supported by the National Research Foundation of Korea (RS-2022-NR070872 and RS-2025-00516961). TK was also supported by the Yonsei Fellowship, funded by Lee Youn Jae, and acted as the corresponding author. DH is also partially supported by the National Research
Foundation (NRF) of Korea grant funded by the Korea government (MSIT, RS2022-NR068800). We thank Yohan Dubois, Romain Teyssier, and Beomchan Koh for helpful discussions that improved this work, and Leo Michel-Dansac for developing and making RASCAS open source.
The supercomputing time for numerical simulations was kindly provided by KISTI (KSC-2024-CRE-0200), and large data transfer was supported by KREONET, which is managed and operated by KISTI. This work was also performed using the DiRAC Data Intensive service at Leicester, operated by the University of Leicester IT Services, which forms part of the STFC DiRAC HPC Facility (www.dirac.ac.uk). The equipment was funded by BEIS capital funding via STFC capital grants ST/K000373/1 and ST/R002363/1 and STFC DiRAC Operations grant ST/R001014/1. DiRAC is part of the National e-Infrastructure.
\end{acknowledgements}

\bibliographystyle{bibtex/aa}
\bibliography{reference}

@ARTICLE{Girichidis2016,
       author = {{Girichidis}, Philipp and {Naab}, Thorsten and {Walch}, Stefanie and {Hanasz}, Micha{\l} and {Mac Low}, Mordecai-Mark and {Ostriker}, Jeremiah P. and {Gatto}, Andrea and {Peters}, Thomas and {W{\"u}nsch}, Richard and {Glover}, Simon C.~O. and {Klessen}, Ralf S. and {Clark}, Paul C. and {Baczynski}, Christian},
        title = "{Launching Cosmic-Ray-driven Outflows from the Magnetized Interstellar Medium}",
      journal = {\apjl},
         year = 2016,
        month = jan,
       volume = {816},
       number = {2},
          eid = {L19},
        pages = {L19},
          doi = {10.3847/2041-8205/816/2/L19},
archivePrefix = {arXiv},
       eprint = {1509.07247},
 primaryClass = {astro-ph.GA},
       adsurl = {https://ui.adsabs.harvard.edu/abs/2016ApJ...816L..19G}
}

@ARTICLE{Farcy2022,
       author = {{Farcy}, Marion and {Rosdahl}, Joakim and {Dubois}, Yohan and {Blaizot}, J{\'e}r{\'e}my and {Martin-Alvarez}, Sergio},
        title = "{Radiation-magnetohydrodynamics simulations of cosmic ray feedback in disc galaxies}",
      journal = {\mnras},
         year = 2022,
        month = jul,
       volume = {513},
       number = {4},
        pages = {5000-5019},
          doi = {10.1093/mnras/stac1196},
archivePrefix = {arXiv},
       eprint = {2202.01245},
 primaryClass = {astro-ph.GA},
       adsurl = {https://ui.adsabs.harvard.edu/abs/2022MNRAS.513.5000F}
}

@ARTICLE{Armillotta2024,
       author = {{Armillotta}, Lucia and {Ostriker}, Eve C. and {Kim}, Chang-Goo and {Jiang}, Yan-Fei},
        title = "{Cosmic-Ray Acceleration of Galactic Outflows in Multiphase Gas}",
      journal = {\apj},
         year = 2024,
        month = mar,
       volume = {964},
       number = {1},
          eid = {99},
        pages = {99},
          doi = {10.3847/1538-4357/ad1e5c},
archivePrefix = {arXiv},
       eprint = {2401.04169},
 primaryClass = {astro-ph.GA},
       adsurl = {https://ui.adsabs.harvard.edu/abs/2024ApJ...964...99A}
}

@ARTICLE{Bleuler2015,
       author = {{Bleuler}, Andreas and {Teyssier}, Romain and {Carassou}, S{\'e}bastien and {Martizzi}, Davide},
        title = "{PHEW: a parallel segmentation algorithm for three-dimensional AMR datasets. Application to structure detection in self-gravitating flows}",
      journal = {Computational Astrophysics and Cosmology},
         year = 2015,
        month = jun,
       volume = {2},
          eid = {5},
        pages = {5},
          doi = {10.1186/s40668-015-0009-7},
archivePrefix = {arXiv},
       eprint = {1412.0510},
 primaryClass = {astro-ph.IM},
       adsurl = {https://ui.adsabs.harvard.edu/abs/2015ComAC...2....5B}
}

@inproceedings{Ester1996,
  author    = {{Ester}, Martin and {Kriegel}, Hans-Peter and {Sander}, J{\"o}rg and {Xu}, Xiaowei},
  title     = {A Density-Based Algorithm for Discovering Clusters in Large Spatial Databases with Noise},
  booktitle = {Proceedings of the Second International Conference on Knowledge Discovery and Data Mining (KDD'96)},
  year      = {1996},
  pages     = {226--231},
  publisher = {AAAI Press},
}

@ARTICLE{Stopyra2021,
       author = {{Stopyra}, Stephen and {Pontzen}, Andrew and {Peiris}, Hiranya and {Roth}, Nina and {Rey}, Martin P.},
        title = "{GenetIC{\textemdash}A New Initial Conditions Generator to Support Genetically Modified Zoom Simulations}",
      journal = {\apjs},
         year = 2021,
        month = feb,
       volume = {252},
       number = {2},
          eid = {28},
        pages = {28},
          doi = {10.3847/1538-4365/abcd94},
archivePrefix = {arXiv},
       eprint = {2006.01841},
 primaryClass = {astro-ph.IM},
       adsurl = {https://ui.adsabs.harvard.edu/abs/2021ApJS..252...28S}
}

@ARTICLE{Rosdahl2015,
       author = {{Rosdahl}, J. and {Teyssier}, R.},
        title = "{A scheme for radiation pressure and photon diffusion with the M1 closure in RAMSES-RT}",
      journal = {\mnras},
         year = 2015,
        month = jun,
       volume = {449},
       number = {4},
        pages = {4380-4403},
          doi = {10.1093/mnras/stv567},
archivePrefix = {arXiv},
       eprint = {1411.6440},
 primaryClass = {astro-ph.IM},
       adsurl = {https://ui.adsabs.harvard.edu/abs/2015MNRAS.449.4380R}
}

@ARTICLE{Teyssier2002,
       author = {{Teyssier}, R.},
        title = "{Cosmological hydrodynamics with adaptive mesh refinement. A new high resolution code called RAMSES}",
      journal = {\aap},
         year = 2002,
        month = apr,
       volume = {385},
        pages = {337-364},
          doi = {10.1051/0004-6361:20011817},
archivePrefix = {arXiv},
       eprint = {astro-ph/0111367},
 primaryClass = {astro-ph},
       adsurl = {https://ui.adsabs.harvard.edu/abs/2002A&A...385..337T}
}

@ARTICLE{Pontzen2021,
       author = {{Pontzen}, Andrew and {Rey}, Martin P. and {Cadiou}, Corentin and {Agertz}, Oscar and {Teyssier}, Romain and {Read}, Justin and {Orkney}, Matthew D.~A.},
        title = "{EDGE: a new approach to suppressing numerical diffusion in adaptive mesh simulations of galaxy formation}",
      journal = {\mnras},
         year = 2021,
        month = feb,
       volume = {501},
       number = {2},
        pages = {1755-1765},
          doi = {10.1093/mnras/staa3645},
archivePrefix = {arXiv},
       eprint = {2009.03313},
 primaryClass = {astro-ph.GA},
       adsurl = {https://ui.adsabs.harvard.edu/abs/2021MNRAS.501.1755P}
}

@ARTICLE{Planck2020,
       author = {{Planck Collaboration} and {Aghanim}, N. and {Akrami}, Y. and {Ashdown}, M. and {Aumont}, J. and {Baccigalupi}, C. and {Ballardini}, M. and {Banday}, A.~J. and {Barreiro}, R.~B. and {Bartolo}, N. and {Basak}, S. and {Battye}, R. and {Benabed}, K. and {Bernard}, J. -P. and {Bersanelli}, M. and {Bielewicz}, P. and {Bock}, J.~J. and {Bond}, J.~R. and {Borrill}, J. and {Bouchet}, F.~R. and {Boulanger}, F. and {Bucher}, M. and {Burigana}, C. and {Butler}, R.~C. and {Calabrese}, E. and {Cardoso}, J. -F. and {Carron}, J. and {Challinor}, A. and {Chiang}, H.~C. and {Chluba}, J. and {Colombo}, L.~P.~L. and {Combet}, C. and {Contreras}, D. and {Crill}, B.~P. and {Cuttaia}, F. and {de Bernardis}, P. and {de Zotti}, G. and {Delabrouille}, J. and {Delouis}, J. -M. and {Di Valentino}, E. and {Diego}, J.~M. and {Dor{\'e}}, O. and {Douspis}, M. and {Ducout}, A. and {Dupac}, X. and {Dusini}, S. and {Efstathiou}, G. and {Elsner}, F. and {En{\ss}lin}, T.~A. and {Eriksen}, H.~K. and {Fantaye}, Y. and {Farhang}, M. and {Fergusson}, J. and {Fernandez-Cobos}, R. and {Finelli}, F. and {Forastieri}, F. and {Frailis}, M. and {Fraisse}, A.~A. and {Franceschi}, E. and {Frolov}, A. and {Galeotta}, S. and {Galli}, S. and {Ganga}, K. and {G{\'e}nova-Santos}, R.~T. and {Gerbino}, M. and {Ghosh}, T. and {Gonz{\'a}lez-Nuevo}, J. and {G{\'o}rski}, K.~M. and {Gratton}, S. and {Gruppuso}, A. and {Gudmundsson}, J.~E. and {Hamann}, J. and {Handley}, W. and {Hansen}, F.~K. and {Herranz}, D. and {Hildebrandt}, S.~R. and {Hivon}, E. and {Huang}, Z. and {Jaffe}, A.~H. and {Jones}, W.~C. and {Karakci}, A. and {Keih{\"a}nen}, E. and {Keskitalo}, R. and {Kiiveri}, K. and {Kim}, J. and {Kisner}, T.~S. and {Knox}, L. and {Krachmalnicoff}, N. and {Kunz}, M. and {Kurki-Suonio}, H. and {Lagache}, G. and {Lamarre}, J. -M. and {Lasenby}, A. and {Lattanzi}, M. and {Lawrence}, C.~R. and {Le Jeune}, M. and {Lemos}, P. and {Lesgourgues}, J. and {Levrier}, F. and {Lewis}, A. and {Liguori}, M. and {Lilje}, P.~B. and {Lilley}, M. and {Lindholm}, V. and {L{\'o}pez-Caniego}, M. and {Lubin}, P.~M. and {Ma}, Y. -Z. and {Mac{\'\i}as-P{\'e}rez}, J.~F. and {Maggio}, G. and {Maino}, D. and {Mandolesi}, N. and {Mangilli}, A. and {Marcos-Caballero}, A. and {Maris}, M. and {Martin}, P.~G. and {Martinelli}, M. and {Mart{\'\i}nez-Gonz{\'a}lez}, E. and {Matarrese}, S. and {Mauri}, N. and {McEwen}, J.~D. and {Meinhold}, P.~R. and {Melchiorri}, A. and {Mennella}, A. and {Migliaccio}, M. and {Millea}, M. and {Mitra}, S. and {Miville-Desch{\^e}nes}, M. -A. and {Molinari}, D. and {Montier}, L. and {Morgante}, G. and {Moss}, A. and {Natoli}, P. and {N{\o}rgaard-Nielsen}, H.~U. and {Pagano}, L. and {Paoletti}, D. and {Partridge}, B. and {Patanchon}, G. and {Peiris}, H.~V. and {Perrotta}, F. and {Pettorino}, V. and {Piacentini}, F. and {Polastri}, L. and {Polenta}, G. and {Puget}, J. -L. and {Rachen}, J.~P. and {Reinecke}, M. and {Remazeilles}, M. and {Renzi}, A. and {Rocha}, G. and {Rosset}, C. and {Roudier}, G. and {Rubi{\~n}o-Mart{\'\i}n}, J.~A. and {Ruiz-Granados}, B. and {Salvati}, L. and {Sandri}, M. and {Savelainen}, M. and {Scott}, D. and {Shellard}, E.~P.~S. and {Sirignano}, C. and {Sirri}, G. and {Spencer}, L.~D. and {Sunyaev}, R. and {Suur-Uski}, A. -S. and {Tauber}, J.~A. and {Tavagnacco}, D. and {Tenti}, M. and {Toffolatti}, L. and {Tomasi}, M. and {Trombetti}, T. and {Valenziano}, L. and {Valiviita}, J. and {Van Tent}, B. and {Vibert}, L. and {Vielva}, P. and {Villa}, F. and {Vittorio}, N. and {Wandelt}, B.~D. and {Wehus}, I.~K. and {White}, M. and {White}, S.~D.~M. and {Zacchei}, A. and {Zonca}, A.},
        title = "{Planck 2018 results. VI. Cosmological parameters}",
      journal = {\aap},
         year = 2020,
        month = sep,
       volume = {641},
          eid = {A6},
        pages = {A6},
          doi = {10.1051/0004-6361/201833910},
archivePrefix = {arXiv},
       eprint = {1807.06209},
 primaryClass = {astro-ph.CO},
       adsurl = {https://ui.adsabs.harvard.edu/abs/2020A&A...641A...6P}
}

@ARTICLE{Tweed2009,
       author = {{Tweed}, D. and {Devriendt}, J. and {Blaizot}, J. and {Colombi}, S. and {Slyz}, A.},
        title = "{Building merger trees from cosmological N-body simulations. Towards improving galaxy formation models using subhaloes}",
      journal = {\aap},
         year = 2009,
        month = nov,
       volume = {506},
       number = {2},
        pages = {647-660},
          doi = {10.1051/0004-6361/200911787},
archivePrefix = {arXiv},
       eprint = {0902.0679},
 primaryClass = {astro-ph.CO},
       adsurl = {https://ui.adsabs.harvard.edu/abs/2009A&A...506..647T}
}

@ARTICLE{Bryan1998,
       author = {{Bryan}, Greg L. and {Norman}, Michael L.},
        title = "{Statistical Properties of X-Ray Clusters: Analytic and Numerical Comparisons}",
      journal = {\apj},
         year = 1998,
        month = mar,
       volume = {495},
       number = {1},
        pages = {80-99},
          doi = {10.1086/305262},
archivePrefix = {arXiv},
       eprint = {astro-ph/9710107},
 primaryClass = {astro-ph},
       adsurl = {https://ui.adsabs.harvard.edu/abs/1998ApJ...495...80B}
}

@ARTICLE{Snaith2018,
       author = {{Snaith}, Owain N. and {Park}, Changbom and {Kim}, Juhan and {Rosdahl}, Joakim},
        title = "{Resolution convergence in cosmological hydrodynamical simulations using adaptive mesh refinement}",
      journal = {\mnras},
         year = 2018,
        month = jun,
       volume = {477},
       number = {1},
        pages = {983-1003},
          doi = {10.1093/mnras/sty673},
archivePrefix = {arXiv},
       eprint = {1803.08061},
 primaryClass = {astro-ph.CO},
       adsurl = {https://ui.adsabs.harvard.edu/abs/2018MNRAS.477..983S}
}

@ARTICLE{Kang2025,
       author = {{Kang}, Cheonsu and {Kimm}, Taysun and {Han}, Daniel and {Katz}, Harley and {Devriendt}, Julien and {Slyz}, Adrianne and {Teyssier}, Romain},
        title = "{Impact of star formation models on the growth of simulated galaxies at high redshifts}",
      journal = {\aap},
         year = 2025,
        month = jan,
       volume = {693},
          eid = {A149},
        pages = {A149},
          doi = {10.1051/0004-6361/202451502},
archivePrefix = {arXiv},
       eprint = {2407.12090},
 primaryClass = {astro-ph.GA},
       adsurl = {https://ui.adsabs.harvard.edu/abs/2025A&A...693A.149K}
}

@ARTICLE{Bleuler2014,
       author = {{Bleuler}, Andreas and {Teyssier}, Romain},
        title = "{Towards a more realistic sink particle algorithm for the RAMSES CODE}",
      journal = {\mnras},
         year = 2014,
        month = dec,
       volume = {445},
       number = {4},
        pages = {4015-4036},
          doi = {10.1093/mnras/stu2005},
archivePrefix = {arXiv},
       eprint = {1409.6528},
 primaryClass = {astro-ph.SR},
       adsurl = {https://ui.adsabs.harvard.edu/abs/2014MNRAS.445.4015B}
}

@ARTICLE{Kimm2017,
       author = {{Kimm}, Taysun and {Katz}, Harley and {Haehnelt}, Martin and {Rosdahl}, Joakim and {Devriendt}, Julien and {Slyz}, Adrianne},
        title = "{Feedback-regulated star formation and escape of LyC photons from mini-haloes during reionization}",
      journal = {\mnras},
         year = 2017,
        month = apr,
       volume = {466},
       number = {4},
        pages = {4826-4846},
          doi = {10.1093/mnras/stx052},
archivePrefix = {arXiv},
       eprint = {1608.04762},
 primaryClass = {astro-ph.GA},
       adsurl = {https://ui.adsabs.harvard.edu/abs/2017MNRAS.466.4826K}
}

@ARTICLE{Asplund2009,
       author = {{Asplund}, Martin and {Grevesse}, Nicolas and {Sauval}, A. Jacques and {Scott}, Pat},
        title = "{The Chemical Composition of the Sun}",
      journal = {\araa},
         year = 2009,
        month = sep,
       volume = {47},
       number = {1},
        pages = {481-522},
          doi = {10.1146/annurev.astro.46.060407.145222},
archivePrefix = {arXiv},
       eprint = {0909.0948},
 primaryClass = {astro-ph.SR},
       adsurl = {https://ui.adsabs.harvard.edu/abs/2009ARA&A..47..481A}
}

@ARTICLE{KM05,
       author = {{Krumholz}, Mark R. and {McKee}, Christopher F.},
        title = "{A General Theory of Turbulence-regulated Star Formation, from Spirals to Ultraluminous Infrared Galaxies}",
      journal = {\apj},
         year = 2005,
        month = sep,
       volume = {630},
       number = {1},
        pages = {250-268},
          doi = {10.1086/431734},
archivePrefix = {arXiv},
       eprint = {astro-ph/0505177},
 primaryClass = {astro-ph},
       adsurl = {https://ui.adsabs.harvard.edu/abs/2005ApJ...630..250K}
}

@ARTICLE{PN11,
       author = {{Padoan}, Paolo and {Nordlund}, {\r{A}}ke},
        title = "{The Star Formation Rate of Supersonic Magnetohydrodynamic Turbulence}",
      journal = {\apj},
         year = 2011,
        month = mar,
       volume = {730},
       number = {1},
          eid = {40},
        pages = {40},
          doi = {10.1088/0004-637X/730/1/40},
archivePrefix = {arXiv},
       eprint = {0907.0248},
 primaryClass = {astro-ph.GA},
       adsurl = {https://ui.adsabs.harvard.edu/abs/2011ApJ...730...40P}
}

@ARTICLE{HC11,
       author = {{Hennebelle}, Patrick and {Chabrier}, Gilles},
        title = "{Analytical Star Formation Rate from Gravoturbulent Fragmentation}",
      journal = {\apjl},
         year = 2011,
        month = dec,
       volume = {743},
       number = {2},
          eid = {L29},
        pages = {L29},
          doi = {10.1088/2041-8205/743/2/L29},
archivePrefix = {arXiv},
       eprint = {1110.0033},
 primaryClass = {astro-ph.GA},
       adsurl = {https://ui.adsabs.harvard.edu/abs/2011ApJ...743L..29H}
}

@ARTICLE{FK12,
       author = {{Federrath}, Christoph and {Klessen}, Ralf S.},
        title = "{The Star Formation Rate of Turbulent Magnetized Clouds: Comparing Theory, Simulations, and Observations}",
      journal = {\apj},
         year = 2012,
        month = dec,
       volume = {761},
       number = {2},
          eid = {156},
        pages = {156},
          doi = {10.1088/0004-637X/761/2/156},
archivePrefix = {arXiv},
       eprint = {1209.2856},
 primaryClass = {astro-ph.SR},
       adsurl = {https://ui.adsabs.harvard.edu/abs/2012ApJ...761..156F}
}

@ARTICLE{Semenov2016,
       author = {{Semenov}, Vadim A. and {Kravtsov}, Andrey V. and {Gnedin}, Nickolay Y.},
        title = "{Nonuniversal Star Formation Efficiency in Turbulent ISM}",
      journal = {\apj},
         year = 2016,
        month = aug,
       volume = {826},
       number = {2},
          eid = {200},
        pages = {200},
          doi = {10.3847/0004-637X/826/2/200},
archivePrefix = {arXiv},
       eprint = {1512.03101},
 primaryClass = {astro-ph.GA},
       adsurl = {https://ui.adsabs.harvard.edu/abs/2016ApJ...826..200S}
}

@INPROCEEDINGS{Bonazzola1986,
       author = {{Bonazzola}, S. and {Falgarone}, E. and {Heyvaerts}, J. and {Perault}, M. and {Puget}, J.~L.},
        title = "{Jeans criterion in a turbulent medium}",
    booktitle = {Interstellar Processes: Abstracts of Contributed Papers},
         year = 1986,
       editor = {{Hollenbach}, D.~J. and {Thronson}, H.~A., Jr.},
        month = oct,
        pages = {41},
       adsurl = {https://ui.adsabs.harvard.edu/abs/1986inpr.conf...41B}
}

@ARTICLE{Gong2013,
       author = {{Gong}, Hao and {Ostriker}, Eve C.},
        title = "{Implementation of Sink Particles in the Athena Code}",
      journal = {\apjs},
         year = 2013,
        month = jan,
       volume = {204},
       number = {1},
          eid = {8},
        pages = {8},
          doi = {10.1088/0067-0049/204/1/8},
archivePrefix = {arXiv},
       eprint = {1211.6425},
 primaryClass = {astro-ph.IM},
       adsurl = {https://ui.adsabs.harvard.edu/abs/2013ApJS..204....8G}
}

@ARTICLE{Hennebelle2024,
       author = {{Hennebelle}, Patrick and {Brucy}, No{\'e} and {Colman}, Tine},
        title = "{Inefficient star formation in high Mach number environments: I. The turbulent support analytical model}",
      journal = {\aap},
         year = 2024,
        month = oct,
       volume = {690},
          eid = {A43},
        pages = {A43},
          doi = {10.1051/0004-6361/202450524},
archivePrefix = {arXiv},
       eprint = {2404.17368},
 primaryClass = {astro-ph.GA},
       adsurl = {https://ui.adsabs.harvard.edu/abs/2024A&A...690A..43H}
}

@ARTICLE{Brucy2024,
       author = {{Brucy}, No{\'e} and {Hennebelle}, Patrick and {Colman}, Tine and {Klessen}, Ralf S. and {Le Yhuelic}, Corentin},
        title = "{Inefficient star formation in high Mach number environments: II. Numerical simulations and comparison with analytical models}",
      journal = {\aap},
         year = 2024,
        month = oct,
       volume = {690},
          eid = {A44},
        pages = {A44},
          doi = {10.1051/0004-6361/202450525},
archivePrefix = {arXiv},
       eprint = {2404.17374},
 primaryClass = {astro-ph.GA},
       adsurl = {https://ui.adsabs.harvard.edu/abs/2024A&A...690A..44B}
}

@ARTICLE{Stanway2018,
       author = {{Stanway}, E.~R. and {Eldridge}, J.~J.},
        title = "{Re-evaluating old stellar populations}",
      journal = {\mnras},
         year = 2018,
        month = sep,
       volume = {479},
       number = {1},
        pages = {75-93},
          doi = {10.1093/mnras/sty1353},
archivePrefix = {arXiv},
       eprint = {1805.08784},
 primaryClass = {astro-ph.GA},
       adsurl = {https://ui.adsabs.harvard.edu/abs/2018MNRAS.479...75S}
}

@ARTICLE{Schaerer2002,
       author = {{Schaerer}, D.},
        title = "{On the properties of massive Population III stars and metal-free stellar populations}",
      journal = {\aap},
         year = 2002,
        month = jan,
       volume = {382},
        pages = {28-42},
          doi = {10.1051/0004-6361:20011619},
archivePrefix = {arXiv},
       eprint = {astro-ph/0110697},
 primaryClass = {astro-ph},
       adsurl = {https://ui.adsabs.harvard.edu/abs/2002A&A...382...28S}
}

@ARTICLE{Kroupa2002,
       author = {{Kroupa}, Pavel},
        title = "{The Initial Mass Function of Stars: Evidence for Uniformity in Variable Systems}",
      journal = {Science},
         year = 2002,
        month = jan,
       volume = {295},
       number = {5552},
        pages = {82-91},
          doi = {10.1126/science.1067524},
archivePrefix = {arXiv},
       eprint = {astro-ph/0201098},
 primaryClass = {astro-ph},
       adsurl = {https://ui.adsabs.harvard.edu/abs/2002Sci...295...82K}
}

@ARTICLE{Leitherer1999,
       author = {{Leitherer}, Claus and {Schaerer}, Daniel and {Goldader}, Jeffrey D. and {Delgado}, Rosa M. Gonz{\'a}lez and {Robert}, Carmelle and {Kune}, Denis Foo and {de Mello}, Du{\'\i}lia F. and {Devost}, Daniel and {Heckman}, Timothy M.},
        title = "{Starburst99: Synthesis Models for Galaxies with Active Star Formation}",
      journal = {\apjs},
         year = 1999,
        month = jul,
       volume = {123},
       number = {1},
        pages = {3-40},
          doi = {10.1086/313233},
archivePrefix = {arXiv},
       eprint = {astro-ph/9902334},
 primaryClass = {astro-ph},
       adsurl = {https://ui.adsabs.harvard.edu/abs/1999ApJS..123....3L}
}

@ARTICLE{Kimm2014,
       author = {{Kimm}, Taysun and {Cen}, Renyue},
        title = "{Escape Fraction of Ionizing Photons during Reionization: Effects due to Supernova Feedback and Runaway OB Stars}",
      journal = {\apj},
         year = 2014,
        month = jun,
       volume = {788},
       number = {2},
          eid = {121},
        pages = {121},
          doi = {10.1088/0004-637X/788/2/121},
archivePrefix = {arXiv},
       eprint = {1405.0552},
 primaryClass = {astro-ph.GA},
       adsurl = {https://ui.adsabs.harvard.edu/abs/2014ApJ...788..121K}
}

@ARTICLE{Kimm2015,
       author = {{Kimm}, Taysun and {Cen}, Renyue and {Devriendt}, Julien and {Dubois}, Yohan and {Slyz}, Adrianne},
        title = "{Towards simulating star formation in turbulent high-z galaxies with mechanical supernova feedback}",
      journal = {\mnras},
         year = 2015,
        month = aug,
       volume = {451},
       number = {3},
        pages = {2900-2921},
          doi = {10.1093/mnras/stv1211},
archivePrefix = {arXiv},
       eprint = {1501.05655},
 primaryClass = {astro-ph.GA},
       adsurl = {https://ui.adsabs.harvard.edu/abs/2015MNRAS.451.2900K}
}

@ARTICLE{Nomoto2006,
       author = {{Nomoto}, Ken'ichi and {Tominaga}, Nozomu and {Umeda}, Hideyuki and {Kobayashi}, Chiaki and {Maeda}, Keiichi},
        title = "{Nucleosynthesis yields of core-collapse supernovae and hypernovae, and galactic chemical evolution}",
      journal = {\nphysa},
         year = 2006,
        month = oct,
       volume = {777},
        pages = {424-458},
          doi = {10.1016/j.nuclphysa.2006.05.008},
archivePrefix = {arXiv},
       eprint = {astro-ph/0605725},
 primaryClass = {astro-ph},
       adsurl = {https://ui.adsabs.harvard.edu/abs/2006NuPhA.777..424N}
}

@ARTICLE{Heger2002,
       author = {{Heger}, A. and {Woosley}, S.~E.},
        title = "{The Nucleosynthetic Signature of Population III}",
      journal = {\apj},
         year = 2002,
        month = mar,
       volume = {567},
       number = {1},
        pages = {532-543},
          doi = {10.1086/338487},
archivePrefix = {arXiv},
       eprint = {astro-ph/0107037},
 primaryClass = {astro-ph},
       adsurl = {https://ui.adsabs.harvard.edu/abs/2002ApJ...567..532H}
}

@ARTICLE{Schmidt1959,
       author = {{Schmidt}, Maarten},
        title = "{The Rate of Star Formation.}",
      journal = {\apj},
         year = 1959,
        month = mar,
       volume = {129},
        pages = {243},
          doi = {10.1086/146614},
       adsurl = {https://ui.adsabs.harvard.edu/abs/1959ApJ...129..243S}
}

@ARTICLE{Read2017,
       author = {{Read}, J.~I. and {Iorio}, G. and {Agertz}, O. and {Fraternali}, F.},
        title = "{The stellar mass-halo mass relation of isolated field dwarfs: a critical test of {\ensuremath{\Lambda}}CDM at the edge of galaxy formation}",
      journal = {\mnras},
         year = 2017,
        month = may,
       volume = {467},
       number = {2},
        pages = {2019-2038},
          doi = {10.1093/mnras/stx147},
archivePrefix = {arXiv},
       eprint = {1607.03127},
 primaryClass = {astro-ph.GA},
       adsurl = {https://ui.adsabs.harvard.edu/abs/2017MNRAS.467.2019R}
}

@ARTICLE{Rosdahl2022,
       author = {{Rosdahl}, Joakim and {Blaizot}, J{\'e}r{\'e}my and {Katz}, Harley and {Kimm}, Taysun and {Garel}, Thibault and {Haehnelt}, Martin and {Keating}, Laura C. and {Martin-Alvarez}, Sergio and {Michel-Dansac}, L{\'e}o and {Ocvirk}, Pierre},
        title = "{LyC escape from SPHINX galaxies in the Epoch of Reionization}",
      journal = {\mnras},
         year = 2022,
        month = sep,
       volume = {515},
       number = {2},
        pages = {2386-2414},
          doi = {10.1093/mnras/stac1942},
archivePrefix = {arXiv},
       eprint = {2207.03232},
 primaryClass = {astro-ph.GA},
       adsurl = {https://ui.adsabs.harvard.edu/abs/2022MNRAS.515.2386R}
}

@ARTICLE{Behroozi2013,
       author = {{Behroozi}, Peter S. and {Wechsler}, Risa H. and {Conroy}, Charlie},
        title = "{The Average Star Formation Histories of Galaxies in Dark Matter Halos from z = 0-8}",
      journal = {\apj},
         year = 2013,
        month = jun,
       volume = {770},
       number = {1},
          eid = {57},
        pages = {57},
          doi = {10.1088/0004-637X/770/1/57},
archivePrefix = {arXiv},
       eprint = {1207.6105},
 primaryClass = {astro-ph.CO},
       adsurl = {https://ui.adsabs.harvard.edu/abs/2013ApJ...770...57B}
}

@ARTICLE{Behroozi2019,
       author = {{Behroozi}, Peter and {Wechsler}, Risa H. and {Hearin}, Andrew P. and {Conroy}, Charlie},
        title = "{UNIVERSEMACHINE: The correlation between galaxy growth and dark matter halo assembly from z = 0-10}",
      journal = {\mnras},
         year = 2019,
        month = sep,
       volume = {488},
       number = {3},
        pages = {3143-3194},
          doi = {10.1093/mnras/stz1182},
archivePrefix = {arXiv},
       eprint = {1806.07893},
 primaryClass = {astro-ph.GA},
       adsurl = {https://ui.adsabs.harvard.edu/abs/2019MNRAS.488.3143B}
}

@ARTICLE{Moster2013,
       author = {{Moster}, Benjamin P. and {Naab}, Thorsten and {White}, Simon D.~M.},
        title = "{Galactic star formation and accretion histories from matching galaxies to dark matter haloes}",
      journal = {\mnras},
         year = 2013,
        month = feb,
       volume = {428},
       number = {4},
        pages = {3121-3138},
          doi = {10.1093/mnras/sts261},
archivePrefix = {arXiv},
       eprint = {1205.5807},
 primaryClass = {astro-ph.CO},
       adsurl = {https://ui.adsabs.harvard.edu/abs/2013MNRAS.428.3121M}
}

@ARTICLE{Moster2021,
       author = {{Moster}, Benjamin P. and {Naab}, Thorsten and {Lindstr{\"o}m}, Magnus and {O'Leary}, Joseph A.},
        title = "{GalaxyNet: connecting galaxies and dark matter haloes with deep neural networks and reinforcement learning in large volumes}",
      journal = {\mnras},
         year = 2021,
        month = oct,
       volume = {507},
       number = {2},
        pages = {2115-2136},
          doi = {10.1093/mnras/stab1449},
archivePrefix = {arXiv},
       eprint = {2005.12276},
 primaryClass = {astro-ph.GA},
       adsurl = {https://ui.adsabs.harvard.edu/abs/2021MNRAS.507.2115M}
}

@ARTICLE{Ma2018,
       author = {{Ma}, Xiangcheng and {Hopkins}, Philip F. and {Garrison-Kimmel}, Shea and {Faucher-Gigu{\`e}re}, Claude-Andr{\'e} and {Quataert}, Eliot and {Boylan-Kolchin}, Michael and {Hayward}, Christopher C. and {Feldmann}, Robert and {Kere{\v{s}}}, Du{\v{s}}an},
        title = "{Simulating galaxies in the reionization era with FIRE-2: galaxy scaling relations, stellar mass functions, and luminosity functions}",
      journal = {\mnras},
         year = 2018,
        month = aug,
       volume = {478},
       number = {2},
        pages = {1694-1715},
          doi = {10.1093/mnras/sty1024},
archivePrefix = {arXiv},
       eprint = {1706.06605},
 primaryClass = {astro-ph.GA},
       adsurl = {https://ui.adsabs.harvard.edu/abs/2018MNRAS.478.1694M}
}

@ARTICLE{Marszewski2024,
       author = {{Marszewski}, Andrew and {Sun}, Guochao and {Faucher-Gigu{\`e}re}, Claude-Andr{\'e} and {Hayward}, Christopher C. and {Feldmann}, Robert},
        title = "{The High-Redshift Gas-Phase Mass{\textendash}Metallicity Relation in FIRE-2}",
      journal = {\apjl},
         year = 2024,
        month = jun,
       volume = {967},
       number = {2},
          eid = {L41},
        pages = {L41},
          doi = {10.3847/2041-8213/ad4cee},
archivePrefix = {arXiv},
       eprint = {2403.08853},
 primaryClass = {astro-ph.GA},
       adsurl = {https://ui.adsabs.harvard.edu/abs/2024ApJ...967L..41M}
}

@ARTICLE{Chemerynska2024,
       author = {{Chemerynska}, Iryna and {Atek}, Hakim and {Dayal}, Pratika and {Furtak}, Lukas J. and {Feldmann}, Robert and {Greene}, Jenny E. and {Maseda}, Michael V. and {Nanayakkara}, Themiya and {Oesch}, Pascal A. and {Fujimoto}, Seiji and {Labb{\'e}}, Ivo and {Bezanson}, Rachel and {Brammer}, Gabriel and {Cutler}, Sam E. and {Leja}, Joel and {Pan}, Richard and {Price}, Sedona H. and {Wang}, Bingjie and {Weaver}, John R. and {Whitaker}, Katherine E.},
        title = "{The Extreme Low-mass End of the Mass{\textendash}Metallicity Relation at z {\ensuremath{\sim}} 7}",
      journal = {\apjl},
         year = 2024,
        month = nov,
       volume = {976},
       number = {1},
          eid = {L15},
        pages = {L15},
          doi = {10.3847/2041-8213/ad8dc9},
archivePrefix = {arXiv},
       eprint = {2407.17110},
 primaryClass = {astro-ph.GA},
       adsurl = {https://ui.adsabs.harvard.edu/abs/2024ApJ...976L..15C}
}

@ARTICLE{Curti2023,
       author = {{Curti}, Mirko and {D'Eugenio}, Francesco and {Carniani}, Stefano and {Maiolino}, Roberto and {Sandles}, Lester and {Witstok}, Joris and {Baker}, William M. and {Bennett}, Jake S. and {Piotrowska}, Joanna M. and {Tacchella}, Sandro and {Charlot}, Stephane and {Nakajima}, Kimihiko and {Maheson}, Gabriel and {Mannucci}, Filippo and {Amiri}, Amirnezam and {Arribas}, Santiago and {Belfiore}, Francesco and {Bonaventura}, Nina R. and {Bunker}, Andrew J. and {Chevallard}, Jacopo and {Cresci}, Giovanni and {Curtis-Lake}, Emma and {Hayden-Pawson}, Connor and {Jones}, Gareth C. and {Kumari}, Nimisha and {Laseter}, Isaac and {Looser}, Tobias J. and {Marconi}, Alessandro and {Maseda}, Michael V. and {Scholtz}, Jan and {Smit}, Renske and {{\"U}bler}, Hannah and {Wallace}, Imaan E.~B.},
        title = "{The chemical enrichment in the early Universe as probed by JWST via direct metallicity measurements at z {\ensuremath{\sim}} 8}",
      journal = {\mnras},
         year = 2023,
        month = jan,
       volume = {518},
       number = {1},
        pages = {425-438},
          doi = {10.1093/mnras/stac2737},
archivePrefix = {arXiv},
       eprint = {2207.12375},
 primaryClass = {astro-ph.GA},
       adsurl = {https://ui.adsabs.harvard.edu/abs/2023MNRAS.518..425C}
}

@ARTICLE{Curti2024,
       author = {{Curti}, Mirko and {Maiolino}, Roberto and {Curtis-Lake}, Emma and {Chevallard}, Jacopo and {Carniani}, Stefano and {D'Eugenio}, Francesco and {Looser}, Tobias J. and {Scholtz}, Jan and {Charlot}, Stephane and {Cameron}, Alex and {{\"U}bler}, Hannah and {Witstok}, Joris and {Boyett}, Kristian and {Laseter}, Isaac and {Sandles}, Lester and {Arribas}, Santiago and {Bunker}, Andrew and {Giardino}, Giovanna and {Maseda}, Michael V. and {Rawle}, Tim and {Rodr{\'\i}guez Del Pino}, Bruno and {Smit}, Renske and {Willott}, Chris J. and {Eisenstein}, Daniel J. and {Hausen}, Ryan and {Johnson}, Benjamin and {Rieke}, Marcia and {Robertson}, Brant and {Tacchella}, Sandro and {Williams}, Christina C. and {Willmer}, Christopher and {Baker}, William M. and {Bhatawdekar}, Rachana and {Egami}, Eiichi and {Helton}, Jakob M. and {Ji}, Zhiyuan and {Kumari}, Nimisha and {Perna}, Michele and {Shivaei}, Irene and {Sun}, Fengwu},
        title = "{JADES: Insights into the low-mass end of the mass-metallicity-SFR relation at 3 < z < 10 from deep JWST/NIRSpec spectroscopy}",
      journal = {\aap},
         year = 2024,
        month = apr,
       volume = {684},
          eid = {A75},
        pages = {A75},
          doi = {10.1051/0004-6361/202346698},
archivePrefix = {arXiv},
       eprint = {2304.08516},
 primaryClass = {astro-ph.GA},
       adsurl = {https://ui.adsabs.harvard.edu/abs/2024A&A...684A..75C}
}

@ARTICLE{Langeroodi2023,
       author = {{Langeroodi}, Danial and {Hjorth}, Jens and {Chen}, Wenlei and {Kelly}, Patrick L. and {Williams}, Hayley and {Lin}, Yu-Heng and {Scarlata}, Claudia and {Zitrin}, Adi and {Broadhurst}, Tom and {Diego}, Jose M. and {Huang}, Xiaosheng and {Filippenko}, Alexei V. and {Foley}, Ryan J. and {Jha}, Saurabh and {Koekemoer}, Anton M. and {Oguri}, Masamune and {Perez-Fournon}, Ismael and {Pierel}, Justin and {Poidevin}, Frederick and {Strolger}, Lou},
        title = "{Evolution of the Mass-Metallicity Relation from Redshift z {\ensuremath{\approx}} 8 to the Local Universe}",
      journal = {\apj},
         year = 2023,
        month = nov,
       volume = {957},
       number = {1},
          eid = {39},
        pages = {39},
          doi = {10.3847/1538-4357/acdbc1},
archivePrefix = {arXiv},
       eprint = {2212.02491},
 primaryClass = {astro-ph.GA},
       adsurl = {https://ui.adsabs.harvard.edu/abs/2023ApJ...957...39L}
}

@ARTICLE{Miller2025,
       author = {{Miller}, Tim B. and {Suess}, Katherine A. and {Setton}, David J. and {Price}, Sedona H. and {Labbe}, Ivo and {Bezanson}, Rachel and {Brammer}, Gabriel and {Cutler}, Sam E. and {Furtak}, Lukas J. and {Leja}, Joel and {Pan}, Richard and {Wang}, Bingjie and {Weaver}, John R. and {Whitaker}, Katherine E. and {Dayal}, Pratika and {de Graaff}, Anna and {Feldmann}, Robert and {Greene}, Jenny E. and {Fujimoto}, S. and {Maseda}, Michael V. and {Nanayakkara}, Themiya and {Nelson}, Erica J. and {van Dokkum}, Pieter and {Zitrin}, Adi},
        title = "{JWST UNCOVERs the Optical Size{\textendash}Stellar Mass Relation at 4 < z < 8: Rapid Growth in the Sizes of Low-mass Galaxies in the First Billion Years of the Universe}",
      journal = {\apj},
         year = 2025,
        month = aug,
       volume = {988},
       number = {2},
          eid = {196},
        pages = {196},
          doi = {10.3847/1538-4357/ade438},
archivePrefix = {arXiv},
       eprint = {2412.06957},
 primaryClass = {astro-ph.GA},
       adsurl = {https://ui.adsabs.harvard.edu/abs/2025ApJ...988..196M}
}

@ARTICLE{Morishita2024,
       author = {{Morishita}, Takahiro and {Stiavelli}, Massimo and {Chary}, Ranga-Ram and {Trenti}, Michele and {Bergamini}, Pietro and {Chiaberge}, Marco and {Leethochawalit}, Nicha and {Roberts-Borsani}, Guido and {Shen}, Xuejian and {Treu}, Tommaso},
        title = "{Enhanced Subkiloparsec-scale Star Formation: Results from a JWST Size Analysis of 341 Galaxies at 5 < z < 14}",
      journal = {\apj},
         year = 2024,
        month = mar,
       volume = {963},
       number = {1},
          eid = {9},
        pages = {9},
          doi = {10.3847/1538-4357/ad1404},
archivePrefix = {arXiv},
       eprint = {2308.05018},
 primaryClass = {astro-ph.GA},
       adsurl = {https://ui.adsabs.harvard.edu/abs/2024ApJ...963....9M}
}

@ARTICLE{Adams2023,
       author = {{Adams}, N.~J. and {Conselice}, C.~J. and {Ferreira}, L. and {Austin}, D. and {Trussler}, J.~A.~A. and {Juod{\v{z}}balis}, I. and {Wilkins}, S.~M. and {Caruana}, J. and {Dayal}, P. and {Verma}, A. and {Vijayan}, A.~P.},
        title = "{Discovery and properties of ultra-high redshift galaxies (9 < z < 12) in the JWST ERO SMACS 0723 Field}",
      journal = {\mnras},
         year = 2023,
        month = jan,
       volume = {518},
       number = {3},
        pages = {4755-4766},
          doi = {10.1093/mnras/stac3347},
archivePrefix = {arXiv},
       eprint = {2207.11217},
 primaryClass = {astro-ph.GA},
       adsurl = {https://ui.adsabs.harvard.edu/abs/2023MNRAS.518.4755A}
}

@ARTICLE{Rosdahl2013,
       author = {{Rosdahl}, J. and {Blaizot}, J. and {Aubert}, D. and {Stranex}, T. and {Teyssier}, R.},
        title = "{RAMSES-RT: radiation hydrodynamics in the cosmological context}",
      journal = {\mnras},
         year = 2013,
        month = dec,
       volume = {436},
       number = {3},
        pages = {2188-2231},
          doi = {10.1093/mnras/stt1722},
archivePrefix = {arXiv},
       eprint = {1304.7126},
 primaryClass = {astro-ph.CO},
       adsurl = {https://ui.adsabs.harvard.edu/abs/2013MNRAS.436.2188R}
}

@ARTICLE{Toro1994,
       author = {{Toro}, E.~F. and {Spruce}, M. and {Speares}, W.},
        title = "{Restoration of the contact surface in the HLL-Riemann solver}",
      journal = {Shock Waves},
         year = 1994,
        month = jul,
       volume = {4},
       number = {1},
        pages = {25-34},
          doi = {10.1007/BF01414629},
       adsurl = {https://ui.adsabs.harvard.edu/abs/1994ShWav...4...25T}
}

@ARTICLE{Guillet2011,
       author = {{Guillet}, Thomas and {Teyssier}, Romain},
        title = "{A simple multigrid scheme for solving the Poisson equation with arbitrary domain boundaries}",
      journal = {Journal of Computational Physics},
         year = 2011,
        month = jun,
       volume = {230},
       number = {12},
        pages = {4756-4771},
          doi = {10.1016/j.jcp.2011.02.044},
archivePrefix = {arXiv},
       eprint = {1104.1703},
 primaryClass = {physics.comp-ph},
       adsurl = {https://ui.adsabs.harvard.edu/abs/2011JCoPh.230.4756G}
}

@ARTICLE{Levermore1984,
       author = {{Levermore}, C.~D.},
        title = "{Relating Eddington factors to flux limiters.}",
      journal = {\jqsrt},
         year = 1984,
        month = feb,
       volume = {31},
       number = {2},
        pages = {149-160},
          doi = {10.1016/0022-4073(84)90112-2},
       adsurl = {https://ui.adsabs.harvard.edu/abs/1984JQSRT..31..149L}
}

@ARTICLE{Katz2017,
       author = {{Katz}, Harley and {Kimm}, Taysun and {Sijacki}, Debora and {Haehnelt}, Martin G.},
        title = "{Interpreting ALMA observations of the ISM during the epoch of reionization}",
      journal = {\mnras},
         year = 2017,
        month = jul,
       volume = {468},
       number = {4},
        pages = {4831-4861},
          doi = {10.1093/mnras/stx608},
archivePrefix = {arXiv},
       eprint = {1612.01786},
 primaryClass = {astro-ph.GA},
       adsurl = {https://ui.adsabs.harvard.edu/abs/2017MNRAS.468.4831K}
}

@ARTICLE{Ferland1998,
       author = {{Ferland}, G.~J. and {Korista}, K.~T. and {Verner}, D.~A. and {Ferguson}, J.~W. and {Kingdon}, J.~B. and {Verner}, E.~M.},
        title = "{CLOUDY 90: Numerical Simulation of Plasmas and Their Spectra}",
      journal = {\pasp},
         year = 1998,
        month = jul,
       volume = {110},
       number = {749},
        pages = {761-778},
          doi = {10.1086/316190},
       adsurl = {https://ui.adsabs.harvard.edu/abs/1998PASP..110..761F}
}

@ARTICLE{Rosen1995,
       author = {{Rosen}, Alexander and {Bregman}, Joel N.},
        title = "{Global Models of the Interstellar Medium in Disk Galaxies}",
      journal = {\apj},
         year = 1995,
        month = feb,
       volume = {440},
        pages = {634},
          doi = {10.1086/175303},
       adsurl = {https://ui.adsabs.harvard.edu/abs/1995ApJ...440..634R}
}

@ARTICLE{Haardt2012,
       author = {{Haardt}, Francesco and {Madau}, Piero},
        title = "{Radiative Transfer in a Clumpy Universe. IV. New Synthesis Models of the Cosmic UV/X-Ray Background}",
      journal = {\apj},
         year = 2012,
        month = feb,
       volume = {746},
       number = {2},
          eid = {125},
        pages = {125},
          doi = {10.1088/0004-637X/746/2/125},
archivePrefix = {arXiv},
       eprint = {1105.2039},
 primaryClass = {astro-ph.CO},
       adsurl = {https://ui.adsabs.harvard.edu/abs/2012ApJ...746..125H}
}

@ARTICLE{Faucher2020,
       author = {{Faucher-Gigu{\`e}re}, Claude-Andr{\'e}},
        title = "{A cosmic UV/X-ray background model update}",
      journal = {\mnras},
         year = 2020,
        month = apr,
       volume = {493},
       number = {2},
        pages = {1614-1632},
          doi = {10.1093/mnras/staa302},
archivePrefix = {arXiv},
       eprint = {1903.08657},
 primaryClass = {astro-ph.CO},
       adsurl = {https://ui.adsabs.harvard.edu/abs/2020MNRAS.493.1614F}
}

@ARTICLE{Laursen2009,
       author = {{Laursen}, Peter and {Sommer-Larsen}, Jesper and {Andersen}, Anja C.},
        title = "{Ly{\ensuremath{\alpha}} Radiative Transfer with Dust: Escape Fractions from Simulated High-Redshift Galaxies}",
      journal = {\apj},
         year = 2009,
        month = oct,
       volume = {704},
       number = {2},
        pages = {1640-1656},
          doi = {10.1088/0004-637X/704/2/1640},
archivePrefix = {arXiv},
       eprint = {0907.2698},
 primaryClass = {astro-ph.CO},
       adsurl = {https://ui.adsabs.harvard.edu/abs/2009ApJ...704.1640L}
}

@ARTICLE{Roman-Duval2022,
       author = {{Roman-Duval}, Julia and {Jenkins}, Edward B. and {Tchernyshyov}, Kirill and {Clark}, Christopher J.~R. and {De Cia}, Annalisa and {Gordon}, Karl D. and {Hamanowicz}, Aleksandra and {Lebouteiller}, Vianney and {Rafelski}, Marc and {Sandstrom}, Karin and {Werk}, Jessica and {Yanchulova Merica-Jones}, Petia},
        title = "{METAL: The Metal Evolution, Transport, and Abundance in the Large Magellanic Cloud Hubble Program. III. Interstellar Depletions, Dust-to-Metal, and Dust-to-Gas Ratios versus Metallicity}",
      journal = {\apj},
         year = 2022,
        month = mar,
       volume = {928},
       number = {1},
          eid = {90},
        pages = {90},
          doi = {10.3847/1538-4357/ac5248},
archivePrefix = {arXiv},
       eprint = {2202.04765},
 primaryClass = {astro-ph.GA},
       adsurl = {https://ui.adsabs.harvard.edu/abs/2022ApJ...928...90R}
}

@ARTICLE{Rosdahl2018,
       author = {{Rosdahl}, Joakim and {Katz}, Harley and {Blaizot}, J{\'e}r{\'e}my and {Kimm}, Taysun and {Michel-Dansac}, L{\'e}o and {Garel}, Thibault and {Haehnelt}, Martin and {Ocvirk}, Pierre and {Teyssier}, Romain},
        title = "{The SPHINX cosmological simulations of the first billion years: the impact of binary stars on reionization}",
      journal = {\mnras},
         year = 2018,
        month = sep,
       volume = {479},
       number = {1},
        pages = {994-1016},
          doi = {10.1093/mnras/sty1655},
archivePrefix = {arXiv},
       eprint = {1801.07259},
 primaryClass = {astro-ph.GA},
       adsurl = {https://ui.adsabs.harvard.edu/abs/2018MNRAS.479..994R}
}

@ARTICLE{Kannan2025,
       author = {{Kannan}, Rahul and {Puchwein}, Ewald and {Smith}, Aaron and {Borrow}, Josh and {Garaldi}, Enrico and {Keating}, Laura and {Vogelsberger}, Mark and {Zier}, Oliver and {McClymont}, William and {Shen}, Xuejian and {Popovic}, Filip and {Tacchella}, Sandro and {Hernquist}, Lars and {Springel}, Volker},
        title = "{Introducing the THESAN-ZOOM project: radiation-hydrodynamic simulations of high-redshift galaxies with a multi-phase interstellar medium}",
      journal = {arXiv e-prints},
         year = 2025,
        month = feb,
          eid = {arXiv:2502.20437},
        pages = {arXiv:2502.20437},
          doi = {10.48550/arXiv.2502.20437},
archivePrefix = {arXiv},
       eprint = {2502.20437},
 primaryClass = {astro-ph.GA},
       adsurl = {https://ui.adsabs.harvard.edu/abs/2025arXiv250220437K}
}

@ARTICLE{Schaye2025,
       author = {{Schaye}, Joop and {Chaikin}, Evgenii and {Schaller}, Matthieu and {Ploeckinger}, Sylvia and {Hu{\v{s}}ko}, Filip and {McGibbon}, Rob and {Trayford}, James W. and {Ben{\'\i}tez-Llambay}, Alejandro and {Correa}, Camila and {Frenk}, Carlos S. and {Richings}, Alexander J. and {Forouhar Moreno}, Victor J. and {Bah{\'e}}, Yannick M. and {Borrow}, Josh and {Durrant}, Anna and {Gebek}, Andrea and {Helly}, John C. and {Jenkins}, Adrian and {Lacey}, Cedric G. and {Ludlow}, Aaron and {Nobels}, Folkert S.~J.},
        title = "{The COLIBRE project: cosmological hydrodynamical simulations of galaxy formation and evolution}",
      journal = {arXiv e-prints},
         year = 2025,
        month = aug,
          eid = {arXiv:2508.21126},
        pages = {arXiv:2508.21126},
          doi = {10.48550/arXiv.2508.21126},
archivePrefix = {arXiv},
       eprint = {2508.21126},
 primaryClass = {astro-ph.GA},
       adsurl = {https://ui.adsabs.harvard.edu/abs/2025arXiv250821126S}
}

@ARTICLE{Agertz2015,
       author = {{Agertz}, Oscar and {Kravtsov}, Andrey V.},
        title = "{On the Interplay between Star Formation and Feedback in Galaxy Formation Simulations}",
      journal = {\apj},
         year = 2015,
        month = may,
       volume = {804},
       number = {1},
          eid = {18},
        pages = {18},
          doi = {10.1088/0004-637X/804/1/18},
archivePrefix = {arXiv},
       eprint = {1404.2613},
 primaryClass = {astro-ph.GA},
       adsurl = {https://ui.adsabs.harvard.edu/abs/2015ApJ...804...18A}
}

@ARTICLE{Agertz2016,
       author = {{Agertz}, Oscar and {Kravtsov}, Andrey V.},
        title = "{The Impact of Stellar Feedback on the Structure, Size, and Morphology of Galaxies in Milky-Way-sized Dark Matter Halos}",
      journal = {\apj},
         year = 2016,
        month = jun,
       volume = {824},
       number = {2},
          eid = {79},
        pages = {79},
          doi = {10.3847/0004-637X/824/2/79},
archivePrefix = {arXiv},
       eprint = {1509.00853},
 primaryClass = {astro-ph.GA},
       adsurl = {https://ui.adsabs.harvard.edu/abs/2016ApJ...824...79A}
}

@ARTICLE{Hopkins2014,
       author = {{Hopkins}, Philip F. and {Kere{\v{s}}}, Du{\v{s}}an and {O{\~n}orbe}, Jos{\'e} and {Faucher-Gigu{\`e}re}, Claude-Andr{\'e} and {Quataert}, Eliot and {Murray}, Norman and {Bullock}, James S.},
        title = "{Galaxies on FIRE (Feedback In Realistic Environments): stellar feedback explains cosmologically inefficient star formation}",
      journal = {\mnras},
         year = 2014,
        month = nov,
       volume = {445},
       number = {1},
        pages = {581-603},
          doi = {10.1093/mnras/stu1738},
archivePrefix = {arXiv},
       eprint = {1311.2073},
 primaryClass = {astro-ph.CO},
       adsurl = {https://ui.adsabs.harvard.edu/abs/2014MNRAS.445..581H}
}

@ARTICLE{Hopkins2018,
       author = {{Hopkins}, Philip F. and {Wetzel}, Andrew and {Kere{\v{s}}}, Du{\v{s}}an and {Faucher-Gigu{\`e}re}, Claude-Andr{\'e} and {Quataert}, Eliot and {Boylan-Kolchin}, Michael and {Murray}, Norman and {Hayward}, Christopher C. and {Garrison-Kimmel}, Shea and {Hummels}, Cameron and {Feldmann}, Robert and {Torrey}, Paul and {Ma}, Xiangcheng and {Angl{\'e}s-Alc{\'a}zar}, Daniel and {Su}, Kung-Yi and {Orr}, Matthew and {Schmitz}, Denise and {Escala}, Ivanna and {Sanderson}, Robyn and {Grudi{\'c}}, Michael Y. and {Hafen}, Zachary and {Kim}, Ji-Hoon and {Fitts}, Alex and {Bullock}, James S. and {Wheeler}, Coral and {Chan}, T.~K. and {Elbert}, Oliver D. and {Narayanan}, Desika},
        title = "{FIRE-2 simulations: physics versus numerics in galaxy formation}",
      journal = {\mnras},
         year = 2018,
        month = oct,
       volume = {480},
       number = {1},
        pages = {800-863},
          doi = {10.1093/mnras/sty1690},
archivePrefix = {arXiv},
       eprint = {1702.06148},
 primaryClass = {astro-ph.GA},
       adsurl = {https://ui.adsabs.harvard.edu/abs/2018MNRAS.480..800H}
}

@ARTICLE{Katz1992,
       author = {{Katz}, Neal},
        title = "{Dissipational Galaxy Formation. II. Effects of Star Formation}",
      journal = {\apj},
         year = 1992,
        month = jun,
       volume = {391},
        pages = {502},
          doi = {10.1086/171366},
       adsurl = {https://ui.adsabs.harvard.edu/abs/1992ApJ...391..502K}
}

@ARTICLE{Girelli2020,
       author = {{Girelli}, G. and {Pozzetti}, L. and {Bolzonella}, M. and {Giocoli}, C. and {Marulli}, F. and {Baldi}, M.},
        title = "{The stellar-to-halo mass relation over the past 12 Gyr. I. Standard {\ensuremath{\Lambda}}CDM model}",
      journal = {\aap},
         year = 2020,
        month = feb,
       volume = {634},
          eid = {A135},
        pages = {A135},
          doi = {10.1051/0004-6361/201936329},
archivePrefix = {arXiv},
       eprint = {2001.02230},
 primaryClass = {astro-ph.CO},
       adsurl = {https://ui.adsabs.harvard.edu/abs/2020A&A...634A.135G}
}

@ARTICLE{Mandelbaum2006,
       author = {{Mandelbaum}, Rachel and {Seljak}, Uro{\v{s}} and {Kauffmann}, Guinevere and {Hirata}, Christopher M. and {Brinkmann}, Jonathan},
        title = "{Galaxy halo masses and satellite fractions from galaxy-galaxy lensing in the Sloan Digital Sky Survey: stellar mass, luminosity, morphology and environment dependencies}",
      journal = {\mnras},
         year = 2006,
        month = may,
       volume = {368},
       number = {2},
        pages = {715-731},
          doi = {10.1111/j.1365-2966.2006.10156.x},
archivePrefix = {arXiv},
       eprint = {astro-ph/0511164},
 primaryClass = {astro-ph},
       adsurl = {https://ui.adsabs.harvard.edu/abs/2006MNRAS.368..715M}
}

@ARTICLE{Leauthaud2012,
       author = {{Leauthaud}, Alexie and {Tinker}, Jeremy and {Bundy}, Kevin and {Behroozi}, Peter S. and {Massey}, Richard and {Rhodes}, Jason and {George}, Matthew R. and {Kneib}, Jean-Paul and {Benson}, Andrew and {Wechsler}, Risa H. and {Busha}, Michael T. and {Capak}, Peter and {Cort{\^e}s}, Marina and {Ilbert}, Olivier and {Koekemoer}, Anton M. and {Le F{\`e}vre}, Oliver and {Lilly}, Simon and {McCracken}, Henry J. and {Salvato}, Mara and {Schrabback}, Tim and {Scoville}, Nick and {Smith}, Tristan and {Taylor}, James E.},
        title = "{New Constraints on the Evolution of the Stellar-to-dark Matter Connection: A Combined Analysis of Galaxy-Galaxy Lensing, Clustering, and Stellar Mass Functions from z = 0.2 to z =1}",
      journal = {\apj},
         year = 2012,
        month = jan,
       volume = {744},
       number = {2},
          eid = {159},
        pages = {159},
          doi = {10.1088/0004-637X/744/2/159},
archivePrefix = {arXiv},
       eprint = {1104.0928},
 primaryClass = {astro-ph.CO},
       adsurl = {https://ui.adsabs.harvard.edu/abs/2012ApJ...744..159L}
}

@ARTICLE{Kennicutt1998,
       author = {{Kennicutt}, Jr., Robert C.},
        title = "{The Global Schmidt Law in Star-forming Galaxies}",
      journal = {\apj},
         year = 1998,
        month = may,
       volume = {498},
       number = {2},
        pages = {541-552},
          doi = {10.1086/305588},
archivePrefix = {arXiv},
       eprint = {astro-ph/9712213},
 primaryClass = {astro-ph},
       adsurl = {https://ui.adsabs.harvard.edu/abs/1998ApJ...498..541K}
}

@ARTICLE{Krumholz2014,
       author = {{Krumholz}, Mark R.},
        title = "{The big problems in star formation: The star formation rate, stellar clustering, and the initial mass function}",
      journal = {\physrep},
         year = 2014,
        month = jun,
       volume = {539},
        pages = {49-134},
          doi = {10.1016/j.physrep.2014.02.001},
archivePrefix = {arXiv},
       eprint = {1402.0867},
 primaryClass = {astro-ph.GA},
       adsurl = {https://ui.adsabs.harvard.edu/abs/2014PhR...539...49K}
}

@ARTICLE{Somerville2015,
       author = {{Somerville}, Rachel S. and {Dav{\'e}}, Romeel},
        title = "{Physical Models of Galaxy Formation in a Cosmological Framework}",
      journal = {\araa},
         year = 2015,
        month = aug,
       volume = {53},
        pages = {51-113},
          doi = {10.1146/annurev-astro-082812-140951},
archivePrefix = {arXiv},
       eprint = {1412.2712},
 primaryClass = {astro-ph.GA},
       adsurl = {https://ui.adsabs.harvard.edu/abs/2015ARA&A..53...51S}
}

@ARTICLE{Naab2017,
       author = {{Naab}, Thorsten and {Ostriker}, Jeremiah P.},
        title = "{Theoretical Challenges in Galaxy Formation}",
      journal = {\araa},
         year = 2017,
        month = aug,
       volume = {55},
       number = {1},
        pages = {59-109},
          doi = {10.1146/annurev-astro-081913-040019},
archivePrefix = {arXiv},
       eprint = {1612.06891},
 primaryClass = {astro-ph.GA},
       adsurl = {https://ui.adsabs.harvard.edu/abs/2017ARA&A..55...59N}
}

@ARTICLE{More2011,
       author = {{More}, Surhud and {van den Bosch}, Frank C. and {Cacciato}, Marcello and {Skibba}, Ramin and {Mo}, H.~J. and {Yang}, Xiaohu},
        title = "{Satellite kinematics - III. Halo masses of central galaxies in SDSS}",
      journal = {\mnras},
         year = 2011,
        month = jan,
       volume = {410},
       number = {1},
        pages = {210-226},
          doi = {10.1111/j.1365-2966.2010.17436.x},
archivePrefix = {arXiv},
       eprint = {1003.3203},
 primaryClass = {astro-ph.CO},
       adsurl = {https://ui.adsabs.harvard.edu/abs/2011MNRAS.410..210M}
}

@ARTICLE{vanUitert2016,
       author = {{van Uitert}, Edo and {Cacciato}, Marcello and {Hoekstra}, Henk and {Brouwer}, Margot and {Sif{\'o}n}, Crist{\'o}bal and {Viola}, Massimo and {Baldry}, Ivan and {Bland-Hawthorn}, Joss and {Brough}, Sarah and {Brown}, M.~J.~I. and {Choi}, Ami and {Driver}, Simon P. and {Erben}, Thomas and {Heymans}, Catherine and {Hildebrandt}, Hendrik and {Joachimi}, Benjamin and {Kuijken}, Konrad and {Liske}, Jochen and {Loveday}, Jon and {McFarland}, John and {Miller}, Lance and {Nakajima}, Reiko and {Peacock}, John and {Radovich}, Mario and {Robotham}, A.~S.~G. and {Schneider}, Peter and {Sikkema}, Gert and {Taylor}, Edward N. and {Verdoes Kleijn}, Gijs},
        title = "{The stellar-to-halo mass relation of GAMA galaxies from 100 deg$^{2}$ of KiDS weak lensing data}",
      journal = {\mnras},
         year = 2016,
        month = jul,
       volume = {459},
       number = {3},
        pages = {3251-3270},
          doi = {10.1093/mnras/stw747},
archivePrefix = {arXiv},
       eprint = {1601.06791},
 primaryClass = {astro-ph.GA},
       adsurl = {https://ui.adsabs.harvard.edu/abs/2016MNRAS.459.3251V}
}

@ARTICLE{Karachentsev2021,
       author = {{Karachentsev}, Igor and {Kashibadze}, Olga},
        title = "{Tracing the local volume galaxy halo-to-stellar mass ratio with satellite kinematics}",
      journal = {Astronomische Nachrichten},
         year = 2021,
        month = aug,
       volume = {342},
       number = {999},
        pages = {999-1023},
          doi = {10.1002/asna.20210018},
archivePrefix = {arXiv},
       eprint = {2109.00336},
 primaryClass = {astro-ph.GA},
       adsurl = {https://ui.adsabs.harvard.edu/abs/2021AN....342..999K}
}

@ARTICLE{Creasey2011,
       author = {{Creasey}, Peter and {Theuns}, Tom and {Bower}, Richard G. and {Lacey}, Cedric G.},
        title = "{Numerical overcooling in shocks}",
      journal = {\mnras},
         year = 2011,
        month = aug,
       volume = {415},
       number = {4},
        pages = {3706-3720},
          doi = {10.1111/j.1365-2966.2011.19001.x},
archivePrefix = {arXiv},
       eprint = {1106.0306},
 primaryClass = {astro-ph.CO},
       adsurl = {https://ui.adsabs.harvard.edu/abs/2011MNRAS.415.3706C}
}

@ARTICLE{Rey2024,
       author = {{Rey}, Martin P. and {Katz}, Harley B. and {Cameron}, Alex J. and {Devriendt}, Julien and {Slyz}, Adrianne},
        title = "{Boosting galactic outflows with enhanced resolution}",
      journal = {\mnras},
         year = 2024,
        month = mar,
       volume = {528},
       number = {3},
        pages = {5412-5431},
          doi = {10.1093/mnras/stae388},
archivePrefix = {arXiv},
       eprint = {2302.08521},
 primaryClass = {astro-ph.GA},
       adsurl = {https://ui.adsabs.harvard.edu/abs/2024MNRAS.528.5412R}
}

@ARTICLE{Ceverino2009,
       author = {{Ceverino}, Daniel and {Klypin}, Anatoly},
        title = "{The Role of Stellar Feedback in the Formation of Galaxies}",
      journal = {\apj},
         year = 2009,
        month = apr,
       volume = {695},
       number = {1},
        pages = {292-309},
          doi = {10.1088/0004-637X/695/1/292},
archivePrefix = {arXiv},
       eprint = {0712.3285},
 primaryClass = {astro-ph},
       adsurl = {https://ui.adsabs.harvard.edu/abs/2009ApJ...695..292C}
}

@ARTICLE{Schaye2015,
       author = {{Schaye}, Joop and {Crain}, Robert A. and {Bower}, Richard G. and {Furlong}, Michelle and {Schaller}, Matthieu and {Theuns}, Tom and {Dalla Vecchia}, Claudio and {Frenk}, Carlos S. and {McCarthy}, I.~G. and {Helly}, John C. and {Jenkins}, Adrian and {Rosas-Guevara}, Y.~M. and {White}, Simon D.~M. and {Baes}, Maarten and {Booth}, C.~M. and {Camps}, Peter and {Navarro}, Julio F. and {Qu}, Yan and {Rahmati}, Alireza and {Sawala}, Till and {Thomas}, Peter A. and {Trayford}, James},
        title = "{The EAGLE project: simulating the evolution and assembly of galaxies and their environments}",
      journal = {\mnras},
         year = 2015,
        month = jan,
       volume = {446},
       number = {1},
        pages = {521-554},
          doi = {10.1093/mnras/stu2058},
archivePrefix = {arXiv},
       eprint = {1407.7040},
 primaryClass = {astro-ph.GA},
       adsurl = {https://ui.adsabs.harvard.edu/abs/2015MNRAS.446..521S}
}

@ARTICLE{Pillepich2018,
       author = {{Pillepich}, Annalisa and {Springel}, Volker and {Nelson}, Dylan and {Genel}, Shy and {Naiman}, Jill and {Pakmor}, R{\"u}diger and {Hernquist}, Lars and {Torrey}, Paul and {Vogelsberger}, Mark and {Weinberger}, Rainer and {Marinacci}, Federico},
        title = "{Simulating galaxy formation with the IllustrisTNG model}",
      journal = {\mnras},
         year = 2018,
        month = jan,
       volume = {473},
       number = {3},
        pages = {4077-4106},
          doi = {10.1093/mnras/stx2656},
archivePrefix = {arXiv},
       eprint = {1703.02970},
 primaryClass = {astro-ph.GA},
       adsurl = {https://ui.adsabs.harvard.edu/abs/2018MNRAS.473.4077P}
}

@ARTICLE{Faucher2018,
       author = {{Faucher-Gigu{\`e}re}, Claude-Andr{\'e}},
        title = "{A model for the origin of bursty star formation in galaxies}",
      journal = {\mnras},
         year = 2018,
        month = jan,
       volume = {473},
       number = {3},
        pages = {3717-3731},
          doi = {10.1093/mnras/stx2595},
archivePrefix = {arXiv},
       eprint = {1701.04824},
 primaryClass = {astro-ph.GA},
       adsurl = {https://ui.adsabs.harvard.edu/abs/2018MNRAS.473.3717F}
}

@ARTICLE{Muratov2015,
       author = {{Muratov}, Alexander L. and {Kere{\v{s}}}, Du{\v{s}}an and {Faucher-Gigu{\`e}re}, Claude-Andr{\'e} and {Hopkins}, Philip F. and {Quataert}, Eliot and {Murray}, Norman},
        title = "{Gusty, gaseous flows of FIRE: galactic winds in cosmological simulations with explicit stellar feedback}",
      journal = {\mnras},
         year = 2015,
        month = dec,
       volume = {454},
       number = {3},
        pages = {2691-2713},
          doi = {10.1093/mnras/stv2126},
archivePrefix = {arXiv},
       eprint = {1501.03155},
 primaryClass = {astro-ph.GA},
       adsurl = {https://ui.adsabs.harvard.edu/abs/2015MNRAS.454.2691M}
}

@ARTICLE{Pandya2021,
       author = {{Pandya}, Viraj and {Fielding}, Drummond B. and {Angl{\'e}s-Alc{\'a}zar}, Daniel and {Somerville}, Rachel S. and {Bryan}, Greg L. and {Hayward}, Christopher C. and {Stern}, Jonathan and {Kim}, Chang-Goo and {Quataert}, Eliot and {Forbes}, John C. and {Faucher-Gigu{\`e}re}, Claude-Andr{\'e} and {Feldmann}, Robert and {Hafen}, Zachary and {Hopkins}, Philip F. and {Kere{\v{s}}}, Du{\v{s}}an and {Murray}, Norman and {Wetzel}, Andrew},
        title = "{Characterizing mass, momentum, energy, and metal outflow rates of multiphase galactic winds in the FIRE-2 cosmological simulations}",
      journal = {\mnras},
         year = 2021,
        month = dec,
       volume = {508},
       number = {2},
        pages = {2979-3008},
          doi = {10.1093/mnras/stab2714},
archivePrefix = {arXiv},
       eprint = {2103.06891},
 primaryClass = {astro-ph.GA},
       adsurl = {https://ui.adsabs.harvard.edu/abs/2021MNRAS.508.2979P}
}

@ARTICLE{Kretschmer2020,
       author = {{Kretschmer}, Michael and {Teyssier}, Romain},
        title = "{Forming early-type galaxies without AGN feedback: a combination of merger-driven outflows and inefficient star formation}",
      journal = {\mnras},
         year = 2020,
        month = feb,
       volume = {492},
       number = {1},
        pages = {1385-1398},
          doi = {10.1093/mnras/stz3495},
archivePrefix = {arXiv},
       eprint = {1906.11836},
 primaryClass = {astro-ph.GA},
       adsurl = {https://ui.adsabs.harvard.edu/abs/2020MNRAS.492.1385K}
}

@ARTICLE{Girma2024,
       author = {{Girma}, Eden and {Teyssier}, Romain},
        title = "{A new star formation recipe for magnetohydrodynamics simulations of galaxy formation}",
      journal = {\mnras},
         year = 2024,
        month = jan,
       volume = {527},
       number = {3},
        pages = {6779-6794},
          doi = {10.1093/mnras/stad3640},
archivePrefix = {arXiv},
       eprint = {2311.10826},
 primaryClass = {astro-ph.GA},
       adsurl = {https://ui.adsabs.harvard.edu/abs/2024MNRAS.527.6779G}
}

@ARTICLE{Molina2012,
       author = {{Molina}, F.~Z. and {Glover}, S.~C.~O. and {Federrath}, C. and {Klessen}, R.~S.},
        title = "{The density variance-Mach number relation in supersonic turbulence - I. Isothermal, magnetized gas}",
      journal = {\mnras},
         year = 2012,
        month = jul,
       volume = {423},
       number = {3},
        pages = {2680-2689},
          doi = {10.1111/j.1365-2966.2012.21075.x},
       adsurl = {https://ui.adsabs.harvard.edu/abs/2012MNRAS.423.2680M}
}

@ARTICLE{Bate1995,
       author = {{Bate}, Matthew R. and {Bonnell}, Ian A. and {Price}, Nigel M.},
        title = "{Modelling accretion in protobinary systems}",
      journal = {\mnras},
         year = 1995,
        month = nov,
       volume = {277},
       number = {2},
        pages = {362-376},
          doi = {10.1093/mnras/277.2.362},
archivePrefix = {arXiv},
       eprint = {astro-ph/9510149},
 primaryClass = {astro-ph},
       adsurl = {https://ui.adsabs.harvard.edu/abs/1995MNRAS.277..362B}
}

@ARTICLE{Krumholz2004,
       author = {{Krumholz}, Mark R. and {McKee}, Christopher F. and {Klein}, Richard I.},
        title = "{Embedding Lagrangian Sink Particles in Eulerian Grids}",
      journal = {\apj},
         year = 2004,
        month = aug,
       volume = {611},
       number = {1},
        pages = {399-412},
          doi = {10.1086/421935},
archivePrefix = {arXiv},
       eprint = {astro-ph/0312612},
 primaryClass = {astro-ph},
       adsurl = {https://ui.adsabs.harvard.edu/abs/2004ApJ...611..399K}
}

@ARTICLE{Federrath2010,
       author = {{Federrath}, Christoph and {Banerjee}, Robi and {Clark}, Paul C. and {Klessen}, Ralf S.},
        title = "{Modeling Collapse and Accretion in Turbulent Gas Clouds: Implementation and Comparison of Sink Particles in AMR and SPH}",
      journal = {\apj},
         year = 2010,
        month = apr,
       volume = {713},
       number = {1},
        pages = {269-290},
          doi = {10.1088/0004-637X/713/1/269},
archivePrefix = {arXiv},
       eprint = {1001.4456},
 primaryClass = {astro-ph.SR},
       adsurl = {https://ui.adsabs.harvard.edu/abs/2010ApJ...713..269F}
}

@INPROCEEDINGS{Nordlund1999,
       author = {{Nordlund}, {\r{A}}. Ke and {Padoan}, Paolo},
        title = "{The Density PDFs of Supersonic Random Flows}",
    booktitle = {Interstellar Turbulence},
         year = 1999,
       editor = {{Franco}, Jose and {Carraminana}, Alberto},
        month = jan,
        pages = {218},
          doi = {10.48550/arXiv.astro-ph/9810074},
archivePrefix = {arXiv},
       eprint = {astro-ph/9810074},
 primaryClass = {astro-ph},
       adsurl = {https://ui.adsabs.harvard.edu/abs/1999intu.conf..218N}
}

@ARTICLE{Vazquez1994,
       author = {{Vazquez-Semadeni}, Enrique},
        title = "{Hierarchical Structure in Nearly Pressureless Flows as a Consequence of Self-similar Statistics}",
      journal = {\apj},
         year = 1994,
        month = mar,
       volume = {423},
        pages = {681},
          doi = {10.1086/173847},
       adsurl = {https://ui.adsabs.harvard.edu/abs/1994ApJ...423..681V}
}

@ARTICLE{Padoan2002,
       author = {{Padoan}, Paolo and {Nordlund}, {\r{A}}ke},
        title = "{The Stellar Initial Mass Function from Turbulent Fragmentation}",
      journal = {\apj},
         year = 2002,
        month = sep,
       volume = {576},
       number = {2},
        pages = {870-879},
          doi = {10.1086/341790},
archivePrefix = {arXiv},
       eprint = {astro-ph/0011465},
 primaryClass = {astro-ph},
       adsurl = {https://ui.adsabs.harvard.edu/abs/2002ApJ...576..870P}
}

@ARTICLE{Federrath2008,
       author = {{Federrath}, Christoph and {Klessen}, Ralf S. and {Schmidt}, Wolfram},
        title = "{The Density Probability Distribution in Compressible Isothermal Turbulence: Solenoidal versus Compressive Forcing}",
      journal = {\apjl},
         year = 2008,
        month = dec,
       volume = {688},
       number = {2},
        pages = {L79},
          doi = {10.1086/595280},
archivePrefix = {arXiv},
       eprint = {0808.0605},
 primaryClass = {astro-ph},
       adsurl = {https://ui.adsabs.harvard.edu/abs/2008ApJ...688L..79F}
}

@ARTICLE{Passot1998,
       author = {{Passot}, Thierry and {V{\'a}zquez-Semadeni}, Enrique},
        title = "{Density probability distribution in one-dimensional polytropic gas dynamics}",
      journal = {\pre},
         year = 1998,
        month = oct,
       volume = {58},
       number = {4},
        pages = {4501-4510},
          doi = {10.1103/PhysRevE.58.4501},
archivePrefix = {arXiv},
       eprint = {physics/9802019},
 primaryClass = {physics.flu-dyn},
       adsurl = {https://ui.adsabs.harvard.edu/abs/1998PhRvE..58.4501P}
}

@ARTICLE{Padoan1997,
       author = {{Padoan}, Paolo and {Nordlund}, Ake and {Jones}, Bernard J.~T.},
        title = "{The universality of the stellar initial mass function}",
      journal = {\mnras},
         year = 1997,
        month = jun,
       volume = {288},
       number = {1},
        pages = {145-152},
          doi = {10.1093/mnras/288.1.145},
archivePrefix = {arXiv},
       eprint = {astro-ph/9703110},
 primaryClass = {astro-ph},
       adsurl = {https://ui.adsabs.harvard.edu/abs/1997MNRAS.288..145P}
}

@ARTICLE{Bromm2002,
       author = {{Bromm}, Volker and {Coppi}, Paolo S. and {Larson}, Richard B.},
        title = "{The Formation of the First Stars. I. The Primordial Star-forming Cloud}",
      journal = {\apj},
         year = 2002,
        month = jan,
       volume = {564},
       number = {1},
        pages = {23-51},
          doi = {10.1086/323947},
archivePrefix = {arXiv},
       eprint = {astro-ph/0102503},
 primaryClass = {astro-ph},
       adsurl = {https://ui.adsabs.harvard.edu/abs/2002ApJ...564...23B}
}

@ARTICLE{Grudic2021,
       author = {{Grudi{\'c}}, Michael Y. and {Guszejnov}, D{\'a}vid and {Hopkins}, Philip F. and {Offner}, Stella S.~R. and {Faucher-Gigu{\`e}re}, Claude-Andr{\'e}},
        title = "{STARFORGE: Towards a comprehensive numerical model of star cluster formation and feedback}",
      journal = {\mnras},
         year = 2021,
        month = sep,
       volume = {506},
       number = {2},
        pages = {2199-2231},
          doi = {10.1093/mnras/stab1347},
archivePrefix = {arXiv},
       eprint = {2010.11254},
 primaryClass = {astro-ph.IM},
       adsurl = {https://ui.adsabs.harvard.edu/abs/2021MNRAS.506.2199G}
}

@ARTICLE{Guszejnov2021,
       author = {{Guszejnov}, D{\'a}vid and {Grudi{\'c}}, Michael Y. and {Hopkins}, Philip F. and {Offner}, Stella S.~R. and {Faucher-Gigu{\`e}re}, Claude-Andr{\'e}},
        title = "{STARFORGE: the effects of protostellar outflows on the IMF}",
      journal = {\mnras},
         year = 2021,
        month = apr,
       volume = {502},
       number = {3},
        pages = {3646-3663},
          doi = {10.1093/mnras/stab278},
archivePrefix = {arXiv},
       eprint = {2010.11249},
 primaryClass = {astro-ph.GA},
       adsurl = {https://ui.adsabs.harvard.edu/abs/2021MNRAS.502.3646G}
}

@ARTICLE{Guszejnov2022IMF,
       author = {{Guszejnov}, D{\'a}vid and {Grudi{\'c}}, Michael Y. and {Offner}, Stella S.~R. and {Faucher-Gigu{\`e}re}, Claude-Andr{\'e} and {Hopkins}, Philip F. and {Rosen}, Anna L.},
        title = "{Effects of the environment and feedback physics on the initial mass function of stars in the STARFORGE simulations}",
      journal = {\mnras},
         year = 2022,
        month = oct,
       volume = {515},
       number = {4},
        pages = {4929-4952},
          doi = {10.1093/mnras/stac2060},
archivePrefix = {arXiv},
       eprint = {2205.10413},
 primaryClass = {astro-ph.GA},
       adsurl = {https://ui.adsabs.harvard.edu/abs/2022MNRAS.515.4929G}
}

@ARTICLE{Guszejnov2022cluster,
       author = {{Guszejnov}, D{\'a}vid and {Markey}, Carleen and {Offner}, Stella S.~R. and {Grudi{\'c}}, Michael Y. and {Faucher-Gigu{\`e}re}, Claude-Andr{\'e} and {Rosen}, Anna L. and {Hopkins}, Philip F.},
        title = "{Cluster assembly and the origin of mass segregation in the STARFORGE simulations}",
      journal = {\mnras},
         year = 2022,
        month = sep,
       volume = {515},
       number = {1},
        pages = {167-184},
          doi = {10.1093/mnras/stac1737},
archivePrefix = {arXiv},
       eprint = {2201.01781},
 primaryClass = {astro-ph.GA},
       adsurl = {https://ui.adsabs.harvard.edu/abs/2022MNRAS.515..167G}
}

@ARTICLE{Federrath2014,
       author = {{Federrath}, Christoph and {Schr{\"o}n}, Martin and {Banerjee}, Robi and {Klessen}, Ralf S.},
        title = "{Modeling Jet and Outflow Feedback during Star Cluster Formation}",
      journal = {\apj},
         year = 2014,
        month = aug,
       volume = {790},
       number = {2},
          eid = {128},
        pages = {128},
          doi = {10.1088/0004-637X/790/2/128},
archivePrefix = {arXiv},
       eprint = {1406.3625},
 primaryClass = {astro-ph.SR},
       adsurl = {https://ui.adsabs.harvard.edu/abs/2014ApJ...790..128F}
}

@ARTICLE{Myers2014,
       author = {{Myers}, Andrew T. and {Klein}, Richard I. and {Krumholz}, Mark R. and {McKee}, Christopher F.},
        title = "{Star cluster formation in turbulent, magnetized dense clumps with radiative and outflow feedback}",
      journal = {\mnras},
         year = 2014,
        month = apr,
       volume = {439},
       number = {4},
        pages = {3420-3438},
          doi = {10.1093/mnras/stu190},
archivePrefix = {arXiv},
       eprint = {1401.6096},
 primaryClass = {astro-ph.GA},
       adsurl = {https://ui.adsabs.harvard.edu/abs/2014MNRAS.439.3420M}
}

@ARTICLE{MacLow2004,
       author = {{Mac Low}, Mordecai-Mark and {Klessen}, Ralf S.},
        title = "{Control of star formation by supersonic turbulence}",
      journal = {Reviews of Modern Physics},
         year = 2004,
        month = jan,
       volume = {76},
       number = {1},
        pages = {125-194},
          doi = {10.1103/RevModPhys.76.125},
archivePrefix = {arXiv},
       eprint = {astro-ph/0301093},
 primaryClass = {astro-ph},
       adsurl = {https://ui.adsabs.harvard.edu/abs/2004RvMP...76..125M}
}

@ARTICLE{Bonnell2011,
       author = {{Bonnell}, I.~A. and {Clarke}, C.~J. and {Bate}, M.~R. and {Pringle}, J.~E.},
        title = "{Accretion in stellar clusters and the initial mass function}",
      journal = {\mnras},
         year = 2001,
        month = jun,
       volume = {324},
       number = {3},
        pages = {573-579},
          doi = {10.1046/j.1365-8711.2001.04311.x},
archivePrefix = {arXiv},
       eprint = {astro-ph/0102121},
 primaryClass = {astro-ph},
       adsurl = {https://ui.adsabs.harvard.edu/abs/2001MNRAS.324..573B}
}

@ARTICLE{Hui1997,
       author = {{Hui}, Lam and {Gnedin}, Nickolay Y.},
        title = "{Equation of state of the photoionized intergalactic medium}",
      journal = {\mnras},
         year = 1997,
        month = nov,
       volume = {292},
       number = {1},
        pages = {27-42},
          doi = {10.1093/mnras/292.1.27},
archivePrefix = {arXiv},
       eprint = {astro-ph/9612232},
 primaryClass = {astro-ph},
       adsurl = {https://ui.adsabs.harvard.edu/abs/1997MNRAS.292...27H}
}

@ARTICLE{Dekel1986,
       author = {{Dekel}, A. and {Silk}, J.},
        title = "{The Origin of Dwarf Galaxies, Cold Dark Matter, and Biased Galaxy Formation}",
      journal = {\apj},
         year = 1986,
        month = apr,
       volume = {303},
        pages = {39},
          doi = {10.1086/164050},
       adsurl = {https://ui.adsabs.harvard.edu/abs/1986ApJ...303...39D}
}

@ARTICLE{Kruijssen2019,
       author = {{Kruijssen}, J.~M. Diederik and {Schruba}, Andreas and {Chevance}, M{\'e}lanie and {Longmore}, Steven N. and {Hygate}, Alexander P.~S. and {Haydon}, Daniel T. and {McLeod}, Anna F. and {Dalcanton}, Julianne J. and {Tacconi}, Linda J. and {van Dishoeck}, Ewine F.},
        title = "{Fast and inefficient star formation due to short-lived molecular clouds and rapid feedback}",
      journal = {\nat},
         year = 2019,
        month = may,
       volume = {569},
       number = {7757},
        pages = {519-522},
          doi = {10.1038/s41586-019-1194-3},
archivePrefix = {arXiv},
       eprint = {1905.08801},
 primaryClass = {astro-ph.GA},
       adsurl = {https://ui.adsabs.harvard.edu/abs/2019Natur.569..519K}
}

@ARTICLE{Chevance2020,
       author = {{Chevance}, M{\'e}lanie and {Kruijssen}, J.~M. Diederik and {Hygate}, Alexander P.~S. and {Schruba}, Andreas and {Longmore}, Steven N. and {Groves}, Brent and {Henshaw}, Jonathan D. and {Herrera}, Cinthya N. and {Hughes}, Annie and {Jeffreson}, Sarah M.~R. and {Lang}, Philipp and {Leroy}, Adam K. and {Meidt}, Sharon E. and {Pety}, J{\'e}r{\^o}me and {Razza}, Alessandro and {Rosolowsky}, Erik and {Schinnerer}, Eva and {Bigiel}, Frank and {Blanc}, Guillermo A. and {Emsellem}, Eric and {Faesi}, Christopher M. and {Glover}, Simon C.~O. and {Haydon}, Daniel T. and {Ho}, I.-Ting and {Kreckel}, Kathryn and {Lee}, Janice C. and {Liu}, Daizhong and {Querejeta}, Miguel and {Saito}, Toshiki and {Sun}, Jiayi and {Usero}, Antonio and {Utomo}, Dyas},
        title = "{The lifecycle of molecular clouds in nearby star-forming disc galaxies}",
      journal = {\mnras},
         year = 2020,
        month = apr,
       volume = {493},
       number = {2},
        pages = {2872-2909},
          doi = {10.1093/mnras/stz3525},
archivePrefix = {arXiv},
       eprint = {1911.03479},
 primaryClass = {astro-ph.GA},
       adsurl = {https://ui.adsabs.harvard.edu/abs/2020MNRAS.493.2872C}
}

@ARTICLE{Chevance2022,
       author = {{Chevance}, M{\'e}lanie and {Kruijssen}, J.~M. Diederik and {Krumholz}, Mark R. and {Groves}, Brent and {Keller}, Benjamin W. and {Hughes}, Annie and {Glover}, Simon C.~O. and {Henshaw}, Jonathan D. and {Herrera}, Cinthya N. and {Kim}, Jaeyeon and {Leroy}, Adam K. and {Pety}, J{\'e}r{\^o}me and {Razza}, Alessandro and {Rosolowsky}, Erik and {Schinnerer}, Eva and {Schruba}, Andreas and {Barnes}, Ashley T. and {Bigiel}, Frank and {Blanc}, Guillermo A. and {Dale}, Daniel A. and {Emsellem}, Eric and {Faesi}, Christopher M. and {Grasha}, Kathryn and {Klessen}, Ralf S. and {Kreckel}, Kathryn and {Liu}, Daizhong and {Longmore}, Steven N. and {Meidt}, Sharon E. and {Querejeta}, Miguel and {Saito}, Toshiki and {Sun}, Jiayi and {Usero}, Antonio},
        title = "{Pre-supernova feedback mechanisms drive the destruction of molecular clouds in nearby star-forming disc galaxies}",
      journal = {\mnras},
         year = 2022,
        month = jan,
       volume = {509},
       number = {1},
        pages = {272-288},
          doi = {10.1093/mnras/stab2938},
archivePrefix = {arXiv},
       eprint = {2010.13788},
 primaryClass = {astro-ph.GA},
       adsurl = {https://ui.adsabs.harvard.edu/abs/2022MNRAS.509..272C}
}

@ARTICLE{Elmegreen1997,
       author = {{Elmegreen}, Bruce G. and {Efremov}, Yuri N.},
        title = "{A Universal Formation Mechanism for Open and Globular Clusters in Turbulent Gas}",
      journal = {\apj},
         year = 1997,
        month = may,
       volume = {480},
       number = {1},
        pages = {235-245},
          doi = {10.1086/303966},
       adsurl = {https://ui.adsabs.harvard.edu/abs/1997ApJ...480..235E}
}

@ARTICLE{Lada2003,
       author = {{Lada}, Charles J. and {Lada}, Elizabeth A.},
        title = "{Embedded Clusters in Molecular Clouds}",
      journal = {\araa},
         year = 2003,
        month = jan,
       volume = {41},
        pages = {57-115},
          doi = {10.1146/annurev.astro.41.011802.094844},
archivePrefix = {arXiv},
       eprint = {astro-ph/0301540},
 primaryClass = {astro-ph},
       adsurl = {https://ui.adsabs.harvard.edu/abs/2003ARA&A..41...57L}
}

@ARTICLE{Vogelsberger2020,
       author = {{Vogelsberger}, Mark and {Marinacci}, Federico and {Torrey}, Paul and {Puchwein}, Ewald},
        title = "{Cosmological simulations of galaxy formation}",
      journal = {Nature Reviews Physics},
         year = 2020,
        month = jan,
       volume = {2},
       number = {1},
        pages = {42-66},
          doi = {10.1038/s42254-019-0127-2},
archivePrefix = {arXiv},
       eprint = {1909.07976},
 primaryClass = {astro-ph.GA},
       adsurl = {https://ui.adsabs.harvard.edu/abs/2020NatRP...2...42V}
}

@ARTICLE{Girichidis2020,
       author = {{Girichidis}, Philipp and {Offner}, Stella S.~R. and {Kritsuk}, Alexei G. and {Klessen}, Ralf S. and {Hennebelle}, Patrick and {Kruijssen}, J.~M. Diederik and {Krause}, Martin G.~H. and {Glover}, Simon C.~O. and {Padovani}, Marco},
        title = "{Physical Processes in Star Formation}",
      journal = {\ssr},
         year = 2020,
        month = jun,
       volume = {216},
       number = {4},
          eid = {68},
        pages = {68},
          doi = {10.1007/s11214-020-00693-8},
archivePrefix = {arXiv},
       eprint = {2005.06472},
 primaryClass = {astro-ph.GA},
       adsurl = {https://ui.adsabs.harvard.edu/abs/2020SSRv..216...68G}
}

@ARTICLE{Wechsler2018,
       author = {{Wechsler}, Risa H. and {Tinker}, Jeremy L.},
        title = "{The Connection Between Galaxies and Their Dark Matter Halos}",
      journal = {\araa},
         year = 2018,
        month = sep,
       volume = {56},
        pages = {435-487},
          doi = {10.1146/annurev-astro-081817-051756},
archivePrefix = {arXiv},
       eprint = {1804.03097},
 primaryClass = {astro-ph.GA},
       adsurl = {https://ui.adsabs.harvard.edu/abs/2018ARA&A..56..435W}
}

@ARTICLE{White1991,
       author = {{White}, Simon D.~M. and {Frenk}, Carlos S.},
        title = "{Galaxy Formation through Hierarchical Clustering}",
      journal = {\apj},
         year = 1991,
        month = sep,
       volume = {379},
        pages = {52},
          doi = {10.1086/170483},
       adsurl = {https://ui.adsabs.harvard.edu/abs/1991ApJ...379...52W}
}

@ARTICLE{Gentry2017,
       author = {{Gentry}, Eric S. and {Krumholz}, Mark R. and {Dekel}, Avishai and {Madau}, Piero},
        title = "{Enhanced momentum feedback from clustered supernovae}",
      journal = {\mnras},
         year = 2017,
        month = feb,
       volume = {465},
       number = {2},
        pages = {2471-2488},
          doi = {10.1093/mnras/stw2746},
archivePrefix = {arXiv},
       eprint = {1606.01242},
 primaryClass = {astro-ph.GA},
       adsurl = {https://ui.adsabs.harvard.edu/abs/2017MNRAS.465.2471G}
}

@ARTICLE{KimCG2017,
       author = {{Kim}, Chang-Goo and {Ostriker}, Eve C. and {Raileanu}, Roberta},
        title = "{Superbubbles in the Multiphase ISM and the Loading of Galactic Winds}",
      journal = {\apj},
         year = 2017,
        month = jan,
       volume = {834},
       number = {1},
          eid = {25},
        pages = {25},
          doi = {10.3847/1538-4357/834/1/25},
archivePrefix = {arXiv},
       eprint = {1610.03092},
 primaryClass = {astro-ph.GA},
       adsurl = {https://ui.adsabs.harvard.edu/abs/2017ApJ...834...25K}
}

@ARTICLE{Wise2012,
       author = {{Wise}, John H. and {Turk}, Matthew J. and {Norman}, Michael L. and {Abel}, Tom},
        title = "{The Birth of a Galaxy: Primordial Metal Enrichment and Stellar Populations}",
      journal = {\apj},
         year = 2012,
        month = jan,
       volume = {745},
       number = {1},
          eid = {50},
        pages = {50},
          doi = {10.1088/0004-637X/745/1/50},
archivePrefix = {arXiv},
       eprint = {1011.2632},
 primaryClass = {astro-ph.CO},
       adsurl = {https://ui.adsabs.harvard.edu/abs/2012ApJ...745...50W}
}

@ARTICLE{Balogh2001,
       author = {{Balogh}, Michael L. and {Pearce}, Frazer R. and {Bower}, Richard G. and {Kay}, Scott T.},
        title = "{Revisiting the cosmic cooling crisis}",
      journal = {\mnras},
         year = 2001,
        month = oct,
       volume = {326},
       number = {4},
        pages = {1228-1234},
          doi = {10.1111/j.1365-2966.2001.04667.x},
archivePrefix = {arXiv},
       eprint = {astro-ph/0104041},
 primaryClass = {astro-ph},
       adsurl = {https://ui.adsabs.harvard.edu/abs/2001MNRAS.326.1228B}
}

@ARTICLE{Salpeter1955,
       author = {{Salpeter}, Edwin E.},
        title = "{The Luminosity Function and Stellar Evolution.}",
      journal = {\apj},
         year = 1955,
        month = jan,
       volume = {121},
        pages = {161},
          doi = {10.1086/145971},
       adsurl = {https://ui.adsabs.harvard.edu/abs/1955ApJ...121..161S}
}

@ARTICLE{Kimm2022,
       author = {{Kimm}, Taysun and {Bieri}, Rebekka and {Geen}, Sam and {Rosdahl}, Joakim and {Blaizot}, J{\'e}r{\'e}my and {Michel-Dansac}, L{\'e}o and {Garel}, Thibault},
        title = "{A Systematic Study of the Escape of LyC and Ly{\ensuremath{\alpha}} Photons from Star-forming, Magnetized Turbulent Clouds}",
      journal = {\apjs},
         year = 2022,
        month = mar,
       volume = {259},
       number = {1},
          eid = {21},
        pages = {21},
          doi = {10.3847/1538-4365/ac426d},
archivePrefix = {arXiv},
       eprint = {2110.02975},
 primaryClass = {astro-ph.GA},
       adsurl = {https://ui.adsabs.harvard.edu/abs/2022ApJS..259...21K}
}

@ARTICLE{KimCG2015,
       author = {{Kim}, Chang-Goo and {Ostriker}, Eve C.},
        title = "{Momentum Injection by Supernovae in the Interstellar Medium}",
      journal = {\apj},
         year = 2015,
        month = apr,
       volume = {802},
       number = {2},
          eid = {99},
        pages = {99},
          doi = {10.1088/0004-637X/802/2/99},
archivePrefix = {arXiv},
       eprint = {1410.1537},
 primaryClass = {astro-ph.GA},
       adsurl = {https://ui.adsabs.harvard.edu/abs/2015ApJ...802...99K}
}

@ARTICLE{Burkhart2018,
       author = {{Burkhart}, Blakesley},
        title = "{The Star Formation Rate in the Gravoturbulent Interstellar Medium}",
      journal = {\apj},
         year = 2018,
        month = aug,
       volume = {863},
       number = {2},
          eid = {118},
        pages = {118},
          doi = {10.3847/1538-4357/aad002},
archivePrefix = {arXiv},
       eprint = {1801.05428},
 primaryClass = {astro-ph.GA},
       adsurl = {https://ui.adsabs.harvard.edu/abs/2018ApJ...863..118B}
}

@ARTICLE{Ballesteros-Paredes2011,
       author = {{Ballesteros-Paredes}, Javier and {V{\'a}zquez-Semadeni}, Enrique and {Gazol}, Adriana and {Hartmann}, Lee W. and {Heitsch}, Fabian and {Col{\'\i}n}, Pedro},
        title = "{Gravity or turbulence? - II. Evolving column density probability distribution functions in molecular clouds}",
      journal = {\mnras},
         year = 2011,
        month = sep,
       volume = {416},
       number = {2},
        pages = {1436-1442},
          doi = {10.1111/j.1365-2966.2011.19141.x},
archivePrefix = {arXiv},
       eprint = {1105.5411},
 primaryClass = {astro-ph.GA},
       adsurl = {https://ui.adsabs.harvard.edu/abs/2011MNRAS.416.1436B}
}

@ARTICLE{Kainulainen2009,
       author = {{Kainulainen}, J. and {Beuther}, H. and {Henning}, T. and {Plume}, R.},
        title = "{Probing the evolution of molecular cloud structure. From quiescence to birth}",
      journal = {\aap},
         year = 2009,
        month = dec,
       volume = {508},
       number = {3},
        pages = {L35-L38},
          doi = {10.1051/0004-6361/200913605},
archivePrefix = {arXiv},
       eprint = {0911.5648},
 primaryClass = {astro-ph.GA},
       adsurl = {https://ui.adsabs.harvard.edu/abs/2009A&A...508L..35K}
}

@ARTICLE{HanS2026,
       author = {{Han}, S. and {Yi}, S.~K. and {Dubois}, Y. and {Rhee}, J. and {Jeon}, S. and {Jang}, J.~K. and {Byun}, G.-H. and {Cadiou}, C. and {Kim}, J. and {Kimm}, T. and {Pichon}, C.},
        title = "{Introducing NewCluster: First half of the history of a high-resolution cluster simulation}",
      journal = {\aap},
         year = 2026,
        month = jan,
       volume = {705},
          eid = {A169},
        pages = {A169},
          doi = {10.1051/0004-6361/202556291},
archivePrefix = {arXiv},
       eprint = {2507.06301},
 primaryClass = {astro-ph.GA},
       adsurl = {https://ui.adsabs.harvard.edu/abs/2026A&A...705A.169H}
}

@ARTICLE{Yi2024,
       author = {{Yi}, Sukyoung K. and {Jang}, J.~K. and {Devriendt}, Julien and {Dubois}, Yohan and {Han}, San and {Kimm}, Taysun and {Kraljic}, Katarina and {Park}, Minjung and {Peirani}, Sebastien and {Pichon}, Christophe and {Rhee}, Jinsu},
        title = "{On the Significance of the Thick Disks of Disk Galaxies}",
      journal = {\apjs},
         year = 2024,
        month = mar,
       volume = {271},
       number = {1},
          eid = {1},
        pages = {1},
          doi = {10.3847/1538-4365/ad0e71},
archivePrefix = {arXiv},
       eprint = {2308.03566},
 primaryClass = {astro-ph.GA},
       adsurl = {https://ui.adsabs.harvard.edu/abs/2024ApJS..271....1Y}
}

@ARTICLE{Larson1969,
       author = {{Larson}, Richard B.},
        title = "{Numerical calculations of the dynamics of collapsing proto-star}",
      journal = {\mnras},
         year = 1969,
        month = jan,
       volume = {145},
        pages = {271},
          doi = {10.1093/mnras/145.3.271},
       adsurl = {https://ui.adsabs.harvard.edu/abs/1969MNRAS.145..271L}
}

@ARTICLE{Penston1969,
       author = {{Penston}, M.~V.},
        title = "{Dynamics of self-gravitating gaseous spheres-III. Analytical results in the free-fall of isothermal cases}",
      journal = {\mnras},
         year = 1969,
        month = jan,
       volume = {144},
        pages = {425},
          doi = {10.1093/mnras/144.4.425},
       adsurl = {https://ui.adsabs.harvard.edu/abs/1969MNRAS.144..425P}
}

@ARTICLE{Bondi1952,
       author = {{Bondi}, H.},
        title = "{On spherically symmetrical accretion}",
      journal = {\mnras},
         year = 1952,
        month = jan,
       volume = {112},
        pages = {195},
          doi = {10.1093/mnras/112.2.195},
       adsurl = {https://ui.adsabs.harvard.edu/abs/1952MNRAS.112..195B}
}

@ARTICLE{Hoyle1939,
       author = {{Hoyle}, F. and {Lyttleton}, R.~A.},
        title = "{The effect of interstellar matter on climatic variation}",
      journal = {Proceedings of the Cambridge Philosophical Society},
         year = 1939,
        month = jan,
       volume = {35},
       number = {3},
        pages = {405},
          doi = {10.1017/S0305004100021150},
       adsurl = {https://ui.adsabs.harvard.edu/abs/1939PCPS...35..405H}
}

@ARTICLE{Kimm2011,
       author = {{Kimm}, Taysun and {Devriendt}, Julien and {Slyz}, Adrianne and {Pichon}, Christophe and {Kassin}, Susan A. and {Dubois}, Yohan},
        title = "{The angular momentum of baryons and dark matter halos revisited}",
      journal = {arXiv e-prints},
         year = 2011,
        month = jun,
          eid = {arXiv:1106.0538},
        pages = {arXiv:1106.0538},
          doi = {10.48550/arXiv.1106.0538},
archivePrefix = {arXiv},
       eprint = {1106.0538},
 primaryClass = {astro-ph.CO},
       adsurl = {https://ui.adsabs.harvard.edu/abs/2011arXiv1106.0538K}
}

@ARTICLE{Tremonti2004,
       author = {{Tremonti}, Christy A. and {Heckman}, Timothy M. and {Kauffmann}, Guinevere and {Brinchmann}, Jarle and {Charlot}, St{\'e}phane and {White}, Simon D.~M. and {Seibert}, Mark and {Peng}, Eric W. and {Schlegel}, David J. and {Uomoto}, Alan and {Fukugita}, Masataka and {Brinkmann}, Jon},
        title = "{The Origin of the Mass-Metallicity Relation: Insights from 53,000 Star-forming Galaxies in the Sloan Digital Sky Survey}",
      journal = {\apj},
         year = 2004,
        month = oct,
       volume = {613},
       number = {2},
        pages = {898-913},
          doi = {10.1086/423264},
archivePrefix = {arXiv},
       eprint = {astro-ph/0405537},
 primaryClass = {astro-ph},
       adsurl = {https://ui.adsabs.harvard.edu/abs/2004ApJ...613..898T}
}

@ARTICLE{Dave2011,
       author = {{Dav{\'e}}, Romeel and {Finlator}, Kristian and {Oppenheimer}, Benjamin D.},
        title = "{Galaxy evolution in cosmological simulations with outflows - II. Metallicities and gas fractions}",
      journal = {\mnras},
         year = 2011,
        month = sep,
       volume = {416},
       number = {2},
        pages = {1354-1376},
          doi = {10.1111/j.1365-2966.2011.19132.x},
archivePrefix = {arXiv},
       eprint = {1104.3156},
 primaryClass = {astro-ph.CO},
       adsurl = {https://ui.adsabs.harvard.edu/abs/2011MNRAS.416.1354D}
}

@ARTICLE{Danovich2015,
       author = {{Danovich}, Mark and {Dekel}, Avishai and {Hahn}, Oliver and {Ceverino}, Daniel and {Primack}, Joel},
        title = "{Four phases of angular-momentum buildup in high-z galaxies: from cosmic-web streams through an extended ring to disc and bulge}",
      journal = {\mnras},
         year = 2015,
        month = may,
       volume = {449},
       number = {2},
        pages = {2087-2111},
          doi = {10.1093/mnras/stv270},
archivePrefix = {arXiv},
       eprint = {1407.7129},
 primaryClass = {astro-ph.GA},
       adsurl = {https://ui.adsabs.harvard.edu/abs/2015MNRAS.449.2087D}
}

@ARTICLE{Semenov2018,
       author = {{Semenov}, Vadim A. and {Kravtsov}, Andrey V. and {Gnedin}, Nickolay Y.},
        title = "{How Galaxies Form Stars: The Connection between Local and Global Star Formation in Galaxy Simulations}",
      journal = {\apj},
         year = 2018,
        month = jul,
       volume = {861},
       number = {1},
          eid = {4},
        pages = {4},
          doi = {10.3847/1538-4357/aac6eb},
archivePrefix = {arXiv},
       eprint = {1803.00007},
 primaryClass = {astro-ph.GA},
       adsurl = {https://ui.adsabs.harvard.edu/abs/2018ApJ...861....4S}
}

@ARTICLE{Menon2025,
       author = {{Menon}, Shyam H. and {Burkhart}, Blakesley and {Somerville}, Rachel S. and {Thompson}, Todd A. and {Sternberg}, Amiel},
        title = "{Bursts of Star Formation and Radiation-driven Outflows Produce Efficient LyC Leakage from Dense Compact Star Clusters}",
      journal = {\apj},
         year = 2025,
        month = jul,
       volume = {987},
       number = {1},
          eid = {12},
        pages = {12},
          doi = {10.3847/1538-4357/add2f9},
archivePrefix = {arXiv},
       eprint = {2408.14591},
 primaryClass = {astro-ph.GA},
       adsurl = {https://ui.adsabs.harvard.edu/abs/2025ApJ...987...12M}
}

@ARTICLE{KimJG2018,
       author = {{Kim}, Jeong-Gyu and {Kim}, Woong-Tae and {Ostriker}, Eve C.},
        title = "{Modeling UV Radiation Feedback from Massive Stars. II. Dispersal of Star-forming Giant Molecular Clouds by Photoionization and Radiation Pressure}",
      journal = {\apj},
         year = 2018,
        month = may,
       volume = {859},
       number = {1},
          eid = {68},
        pages = {68},
          doi = {10.3847/1538-4357/aabe27},
archivePrefix = {arXiv},
       eprint = {1804.04664},
 primaryClass = {astro-ph.GA},
       adsurl = {https://ui.adsabs.harvard.edu/abs/2018ApJ...859...68K}
}

@ARTICLE{Cunningham2011,
       author = {{Cunningham}, Andrew J. and {Klein}, Richard I. and {Krumholz}, Mark R. and {McKee}, Christopher F.},
        title = "{Radiation-hydrodynamic Simulations of Massive Star Formation with Protostellar Outflows}",
      journal = {\apj},
         year = 2011,
        month = oct,
       volume = {740},
       number = {2},
          eid = {107},
        pages = {107},
          doi = {10.1088/0004-637X/740/2/107},
archivePrefix = {arXiv},
       eprint = {1104.1218},
 primaryClass = {astro-ph.SR},
       adsurl = {https://ui.adsabs.harvard.edu/abs/2011ApJ...740..107C}
}

@ARTICLE{Machida2012,
       author = {{Machida}, Masahiro N. and {Matsumoto}, Tomoaki},
        title = "{Impact of protostellar outflow on star formation: effects of the initial cloud mass}",
      journal = {\mnras},
         year = 2012,
        month = mar,
       volume = {421},
       number = {1},
        pages = {588-607},
          doi = {10.1111/j.1365-2966.2011.20336.x},
archivePrefix = {arXiv},
       eprint = {1108.3564},
 primaryClass = {astro-ph.SR},
       adsurl = {https://ui.adsabs.harvard.edu/abs/2012MNRAS.421..588M}
}

@ARTICLE{Orr2018,
       author = {{Orr}, Matthew E. and {Hayward}, Christopher C. and {Hopkins}, Philip F. and {Chan}, T.~K. and {Faucher-Gigu{\`e}re}, Claude-Andr{\'e} and {Feldmann}, Robert and {Kere{\v{s}}}, Du{\v{s}}an and {Murray}, Norman and {Quataert}, Eliot},
        title = "{What FIREs up star formation: the emergence of the Kennicutt-Schmidt law from feedback}",
      journal = {\mnras},
         year = 2018,
        month = aug,
       volume = {478},
       number = {3},
        pages = {3653-3673},
          doi = {10.1093/mnras/sty1241},
archivePrefix = {arXiv},
       eprint = {1701.01788},
 primaryClass = {astro-ph.GA},
       adsurl = {https://ui.adsabs.harvard.edu/abs/2018MNRAS.478.3653O}
}

@ARTICLE{Nunez-Castineyra2021,
       author = {{Nu{\~n}ez-Casti{\~n}eyra}, A. and {Nezri}, E. and {Devriendt}, J. and {Teyssier}, R.},
        title = "{Cosmological simulations of the same spiral galaxy: the impact of baryonic physics}",
      journal = {\mnras},
         year = 2021,
        month = jan,
       volume = {501},
       number = {1},
        pages = {62-77},
          doi = {10.1093/mnras/staa3233},
archivePrefix = {arXiv},
       eprint = {2004.06008},
 primaryClass = {astro-ph.GA},
       adsurl = {https://ui.adsabs.harvard.edu/abs/2021MNRAS.501...62N}
}

@ARTICLE{Wang2010,
       author = {{Wang}, Peng and {Li}, Zhi-Yun and {Abel}, Tom and {Nakamura}, Fumitaka},
        title = "{Outflow Feedback Regulated Massive Star Formation in Parsec-Scale Cluster-Forming Clumps}",
      journal = {\apj},
         year = 2010,
        month = jan,
       volume = {709},
       number = {1},
        pages = {27-41},
          doi = {10.1088/0004-637X/709/1/27},
archivePrefix = {arXiv},
       eprint = {0908.4129},
 primaryClass = {astro-ph.SR},
       adsurl = {https://ui.adsabs.harvard.edu/abs/2010ApJ...709...27W}
}

@ARTICLE{Federrath2015,
       author = {{Federrath}, Christoph},
        title = "{Inefficient star formation through turbulence, magnetic fields and feedback}",
      journal = {\mnras},
         year = 2015,
        month = jul,
       volume = {450},
       number = {4},
        pages = {4035-4042},
          doi = {10.1093/mnras/stv941},
archivePrefix = {arXiv},
       eprint = {1504.03690},
 primaryClass = {astro-ph.SR},
       adsurl = {https://ui.adsabs.harvard.edu/abs/2015MNRAS.450.4035F}
}

@ARTICLE{Appel2022,
       author = {{Appel}, Sabrina M. and {Burkhart}, Blakesley and {Semenov}, Vadim A. and {Federrath}, Christoph and {Rosen}, Anna L.},
        title = "{The Effects of Magnetic Fields and Outflow Feedback on the Shape and Evolution of the Density Probability Distribution Function in Turbulent Star-forming Clouds}",
      journal = {\apj},
         year = 2022,
        month = mar,
       volume = {927},
       number = {1},
          eid = {75},
        pages = {75},
          doi = {10.3847/1538-4357/ac4be3},
archivePrefix = {arXiv},
       eprint = {2109.13271},
 primaryClass = {astro-ph.GA},
       adsurl = {https://ui.adsabs.harvard.edu/abs/2022ApJ...927...75A}
}

@ARTICLE{Hopkins2013SF,
       author = {{Hopkins}, Philip F. and {Narayanan}, Desika and {Murray}, Norman},
        title = "{The meaning and consequences of star formation criteria in galaxy models with resolved stellar feedback}",
      journal = {\mnras},
         year = 2013,
        month = jul,
       volume = {432},
       number = {4},
        pages = {2647-2653},
          doi = {10.1093/mnras/stt723},
archivePrefix = {arXiv},
       eprint = {1303.0285},
 primaryClass = {astro-ph.CO},
       adsurl = {https://ui.adsabs.harvard.edu/abs/2013MNRAS.432.2647H}
}

@ARTICLE{Nobels2024,
       author = {{Nobels}, Folkert S.~J. and {Schaye}, Joop and {Schaller}, Matthieu and {Ploeckinger}, Sylvia and {Chaikin}, Evgenii and {Richings}, Alexander J.},
        title = "{Tests of subgrid models for star formation using simulations of isolated disc galaxies}",
      journal = {\mnras},
         year = 2024,
        month = aug,
       volume = {532},
       number = {3},
        pages = {3299-3321},
          doi = {10.1093/mnras/stae1390},
archivePrefix = {arXiv},
       eprint = {2309.13750},
 primaryClass = {astro-ph.GA},
       adsurl = {https://ui.adsabs.harvard.edu/abs/2024MNRAS.532.3299N}
}

@ARTICLE{Katz2022,
       author = {{Katz}, Harley and {Liu}, Shenghua and {Kimm}, Taysun and {Rey}, Martin P. and {Andersson}, Eric P. and {Cameron}, Alex J. and {Rodriguez-Montero}, Francisco and {Agertz}, Oscar and {Devriendt}, Julien and {Slyz}, Adrianne},
        title = "{PRISM: A Non-Equilibrium, Multiphase Interstellar Medium Model for Radiation Hydrodynamics Simulations of Galaxies}",
      journal = {arXiv e-prints},
         year = 2022,
        month = nov,
          eid = {arXiv:2211.04626},
        pages = {arXiv:2211.04626},
          doi = {10.48550/arXiv.2211.04626},
archivePrefix = {arXiv},
       eprint = {2211.04626},
 primaryClass = {astro-ph.GA},
       adsurl = {https://ui.adsabs.harvard.edu/abs/2022arXiv221104626K}
}

@ARTICLE{Semenov2025turb,
       author = {{Semenov}, Vadim A.},
        title = "{Capturing Turbulence with Numerical Dissipation: A Simple Dynamical Model for Unresolved Turbulence in Hydrodynamic Simulations}",
      journal = {\apjs},
         year = 2025,
        month = dec,
       volume = {281},
       number = {2},
          eid = {37},
        pages = {37},
          doi = {10.3847/1538-4365/ae0cc6},
archivePrefix = {arXiv},
       eprint = {2410.23339},
 primaryClass = {astro-ph.GA},
       adsurl = {https://ui.adsabs.harvard.edu/abs/2025ApJS..281...37S}
}

@ARTICLE{Crain2023,
       author = {{Crain}, Robert A. and {van de Voort}, Freeke},
        title = "{Hydrodynamical Simulations of the Galaxy Population: Enduring Successes and Outstanding Challenges}",
      journal = {\araa},
         year = 2023,
        month = aug,
       volume = {61},
        pages = {473-515},
          doi = {10.1146/annurev-astro-041923-043618},
archivePrefix = {arXiv},
       eprint = {2309.17075},
 primaryClass = {astro-ph.GA},
       adsurl = {https://ui.adsabs.harvard.edu/abs/2023ARA&A..61..473C}
}

@ARTICLE{HanD2026,
       author = {{Han}, Daniel and {Kimm}, Taysun and {Kang}, Cheonsu and {Lee}, Jaehyun and {Katz}, Harley and {Rosdahl}, Joki},
        title = "{From short-lived to long-lived clouds: impact of star formation models on giant molecular cloud evolution in simulations of an NGC 300-like galaxy}",
      journal = {arXiv e-prints},
         year = 2026,
        month = apr,
          eid = {arXiv:2604.25911},
        pages = {arXiv:2604.25911},
archivePrefix = {arXiv},
       eprint = {2604.25911},
 primaryClass = {astro-ph.GA},
       adsurl = {https://ui.adsabs.harvard.edu/abs/2026arXiv260425911H}
}

@ARTICLE{Hopkins2013PDF,
       author = {{Hopkins}, Philip F.},
        title = "{A model for (non-lognormal) density distributions in isothermal turbulence}",
      journal = {\mnras},
         year = 2013,
        month = apr,
       volume = {430},
       number = {3},
        pages = {1880-1891},
          doi = {10.1093/mnras/stt010},
archivePrefix = {arXiv},
       eprint = {1211.3119},
 primaryClass = {astro-ph.GA},
       adsurl = {https://ui.adsabs.harvard.edu/abs/2013MNRAS.430.1880H}
}

@ARTICLE{Castaing1996,
       author = {{Castaing}, B.},
        title = "{The Temperature of Turbulent Flows}",
      journal = {Journal de Physique II},
         year = 1996,
        month = jan,
       volume = {6},
       number = {1},
        pages = {105-114},
          doi = {10.1051/jp2:1996172},
       adsurl = {https://ui.adsabs.harvard.edu/abs/1996JPhy2...6..105C}
}

@ARTICLE{Hu2019,
       author = {{Hu}, Chia-Yu},
        title = "{Supernova-driven winds in simulated dwarf galaxies}",
      journal = {\mnras},
         year = 2019,
        month = mar,
       volume = {483},
       number = {3},
        pages = {3363-3381},
          doi = {10.1093/mnras/sty3252},
archivePrefix = {arXiv},
       eprint = {1805.06614},
 primaryClass = {astro-ph.GA},
       adsurl = {https://ui.adsabs.harvard.edu/abs/2019MNRAS.483.3363H}
}

@ARTICLE{Hu2023,
       author = {{Hu}, Chia-Yu and {Smith}, Matthew C. and {Teyssier}, Romain and {Bryan}, Greg L. and {Verbeke}, Robbert and {Emerick}, Andrew and {Somerville}, Rachel S. and {Burkhart}, Blakesley and {Li}, Yuan and {Forbes}, John C. and {Starkenburg}, Tjitske},
        title = "{Code Comparison in Galaxy-scale Simulations with Resolved Supernova Feedback: Lagrangian versus Eulerian Methods}",
      journal = {\apj},
         year = 2023,
        month = jun,
       volume = {950},
       number = {2},
          eid = {132},
        pages = {132},
          doi = {10.3847/1538-4357/accf9e},
archivePrefix = {arXiv},
       eprint = {2208.10528},
 primaryClass = {astro-ph.GA},
       adsurl = {https://ui.adsabs.harvard.edu/abs/2023ApJ...950..132H}
}

@ARTICLE{Pontzen2012,
       author = {{Pontzen}, Andrew and {Governato}, Fabio},
        title = "{How supernova feedback turns dark matter cusps into cores}",
      journal = {\mnras},
         year = 2012,
        month = apr,
       volume = {421},
       number = {4},
        pages = {3464-3471},
          doi = {10.1111/j.1365-2966.2012.20571.x},
archivePrefix = {arXiv},
       eprint = {1106.0499},
 primaryClass = {astro-ph.CO},
       adsurl = {https://ui.adsabs.harvard.edu/abs/2012MNRAS.421.3464P}
}

@ARTICLE{Chan2015,
       author = {{Chan}, T.~K. and {Kere{\v{s}}}, D. and {O{\~n}orbe}, J. and {Hopkins}, P.~F. and {Muratov}, A.~L. and {Faucher-Gigu{\`e}re}, C.-A. and {Quataert}, E.},
        title = "{The impact of baryonic physics on the structure of dark matter haloes: the view from the FIRE cosmological simulations}",
      journal = {\mnras},
         year = 2015,
        month = dec,
       volume = {454},
       number = {3},
        pages = {2981-3001},
          doi = {10.1093/mnras/stv2165},
archivePrefix = {arXiv},
       eprint = {1507.02282},
 primaryClass = {astro-ph.GA},
       adsurl = {https://ui.adsabs.harvard.edu/abs/2015MNRAS.454.2981C}
}

@ARTICLE{Keller2014,
       author = {{Keller}, B.~W. and {Wadsley}, J. and {Benincasa}, S.~M. and {Couchman}, H.~M.~P.},
        title = "{A superbubble feedback model for galaxy simulations}",
      journal = {\mnras},
         year = 2014,
        month = aug,
       volume = {442},
       number = {4},
        pages = {3013-3025},
          doi = {10.1093/mnras/stu1058},
archivePrefix = {arXiv},
       eprint = {1405.2625},
 primaryClass = {astro-ph.GA},
       adsurl = {https://ui.adsabs.harvard.edu/abs/2014MNRAS.442.3013K}
}

@ARTICLE{Smith2019,
       author = {{Smith}, Matthew C. and {Sijacki}, Debora and {Shen}, Sijing},
        title = "{Cosmological simulations of dwarfs: the need for ISM physics beyond SN feedback alone}",
      journal = {\mnras},
         year = 2019,
        month = may,
       volume = {485},
       number = {3},
        pages = {3317-3333},
          doi = {10.1093/mnras/stz599},
archivePrefix = {arXiv},
       eprint = {1807.04288},
 primaryClass = {astro-ph.GA},
       adsurl = {https://ui.adsabs.harvard.edu/abs/2019MNRAS.485.3317S}
}

@ARTICLE{Smith2021,
       author = {{Smith}, Matthew C. and {Bryan}, Greg L. and {Somerville}, Rachel S. and {Hu}, Chia-Yu and {Teyssier}, Romain and {Burkhart}, Blakesley and {Hernquist}, Lars},
        title = "{Efficient early stellar feedback can suppress galactic outflows by reducing supernova clustering}",
      journal = {\mnras},
         year = 2021,
        month = sep,
       volume = {506},
       number = {3},
        pages = {3882-3915},
          doi = {10.1093/mnras/stab1896},
archivePrefix = {arXiv},
       eprint = {2009.11309},
 primaryClass = {astro-ph.GA},
       adsurl = {https://ui.adsabs.harvard.edu/abs/2021MNRAS.506.3882S}
}

@ARTICLE{Fielding2018,
       author = {{Fielding}, Drummond and {Quataert}, Eliot and {Martizzi}, Davide},
        title = "{Clustered supernovae drive powerful galactic winds after superbubble breakout}",
      journal = {\mnras},
         year = 2018,
        month = dec,
       volume = {481},
       number = {3},
        pages = {3325-3347},
          doi = {10.1093/mnras/sty2466},
archivePrefix = {arXiv},
       eprint = {1807.08758},
 primaryClass = {astro-ph.GA},
       adsurl = {https://ui.adsabs.harvard.edu/abs/2018MNRAS.481.3325F}
}

@ARTICLE{Michel-Dansac2020,
       author = {{Michel-Dansac}, L. and {Blaizot}, J. and {Garel}, T. and {Verhamme}, A. and {Kimm}, T. and {Trebitsch}, M.},
        title = "{RASCAS: RAdiation SCattering in Astrophysical Simulations}",
      journal = {\aap},
         year = 2020,
        month = mar,
       volume = {635},
          eid = {A154},
        pages = {A154},
          doi = {10.1051/0004-6361/201834961},
archivePrefix = {arXiv},
       eprint = {2001.11252},
 primaryClass = {astro-ph.GA},
       adsurl = {https://ui.adsabs.harvard.edu/abs/2020A&A...635A.154M}
}

@ARTICLE{Semenov2025bursty,
       author = {{Semenov}, Vadim A. and {Conroy}, Charlie and {Hernquist}, Lars},
        title = "{From UV-bright Galaxies to Early Disks: The Importance of Turbulent Star Formation in the Early Universe}",
      journal = {\apj},
         year = 2025,
        month = aug,
       volume = {989},
       number = {2},
          eid = {219},
        pages = {219},
          doi = {10.3847/1538-4357/ade22d},
archivePrefix = {arXiv},
       eprint = {2410.09205},
 primaryClass = {astro-ph.GA},
       adsurl = {https://ui.adsabs.harvard.edu/abs/2025ApJ...989..219S}
}

@ARTICLE{Shen2023,
       author = {{Shen}, Xuejian and {Vogelsberger}, Mark and {Boylan-Kolchin}, Michael and {Tacchella}, Sandro and {Kannan}, Rahul},
        title = "{The impact of UV variability on the abundance of bright galaxies at z {\ensuremath{\geq}} 9}",
      journal = {\mnras},
         year = 2023,
        month = nov,
       volume = {525},
       number = {3},
        pages = {3254-3261},
          doi = {10.1093/mnras/stad2508},
archivePrefix = {arXiv},
       eprint = {2305.05679},
 primaryClass = {astro-ph.GA},
       adsurl = {https://ui.adsabs.harvard.edu/abs/2023MNRAS.525.3254S}
}

@ARTICLE{Kravtsov2024,
       author = {{Kravtsov}, Andrey and {Belokurov}, Vasily},
        title = "{Stochastic star formation and the abundance of $z>10$ UV-bright galaxies}",
      journal = {arXiv e-prints},
         year = 2024,
        month = may,
          eid = {arXiv:2405.04578},
        pages = {arXiv:2405.04578},
          doi = {10.48550/arXiv.2405.04578},
archivePrefix = {arXiv},
       eprint = {2405.04578},
 primaryClass = {astro-ph.GA},
       adsurl = {https://ui.adsabs.harvard.edu/abs/2024arXiv240504578K}
}

@ARTICLE{Naidu2022,
       author = {{Naidu}, Rohan P. and {Oesch}, Pascal A. and {van Dokkum}, Pieter and {Nelson}, Erica J. and {Suess}, Katherine A. and {Brammer}, Gabriel and {Whitaker}, Katherine E. and {Illingworth}, Garth and {Bouwens}, Rychard and {Tacchella}, Sandro and {Matthee}, Jorryt and {Allen}, Natalie and {Bezanson}, Rachel and {Conroy}, Charlie and {Labbe}, Ivo and {Leja}, Joel and {Leonova}, Ecaterina and {Magee}, Dan and {Price}, Sedona H. and {Setton}, David J. and {Strait}, Victoria and {Stefanon}, Mauro and {Toft}, Sune and {Weaver}, John R. and {Weibel}, Andrea},
        title = "{Two Remarkably Luminous Galaxy Candidates at z {\ensuremath{\approx}} 10-12 Revealed by JWST}",
      journal = {\apjl},
         year = 2022,
        month = nov,
       volume = {940},
       number = {1},
          eid = {L14},
        pages = {L14},
          doi = {10.3847/2041-8213/ac9b22},
archivePrefix = {arXiv},
       eprint = {2207.09434},
 primaryClass = {astro-ph.GA},
       adsurl = {https://ui.adsabs.harvard.edu/abs/2022ApJ...940L..14N}
}

@ARTICLE{Finkelstein2023,
       author = {{Finkelstein}, Steven L. and {Bagley}, Micaela B. and {Ferguson}, Henry C. and {Wilkins}, Stephen M. and {Kartaltepe}, Jeyhan S. and {Papovich}, Casey and {Yung}, L.~Y. Aaron and {Arrabal Haro}, Pablo and {Behroozi}, Peter and {Dickinson}, Mark and {Kocevski}, Dale D. and {Koekemoer}, Anton M. and {Larson}, Rebecca L. and {Le Bail}, Aur{\'e}lien and {Morales}, Alexa M. and {P{\'e}rez-Gonz{\'a}lez}, Pablo G. and {Burgarella}, Denis and {Dav{\'e}}, Romeel and {Hirschmann}, Michaela and {Somerville}, Rachel S. and {Wuyts}, Stijn and {Bromm}, Volker and {Casey}, Caitlin M. and {Fontana}, Adriano and {Fujimoto}, Seiji and {Gardner}, Jonathan P. and {Giavalisco}, Mauro and {Grazian}, Andrea and {Grogin}, Norman A. and {Hathi}, Nimish P. and {Hutchison}, Taylor A. and {Jha}, Saurabh W. and {Jogee}, Shardha and {Kewley}, Lisa J. and {Kirkpatrick}, Allison and {Long}, Arianna S. and {Lotz}, Jennifer M. and {Pentericci}, Laura and {Pierel}, Justin D.~R. and {Pirzkal}, Nor and {Ravindranath}, Swara and {Ryan}, Russell E. and {Trump}, Jonathan R. and {Yang}, Guang and {Bhatawdekar}, Rachana and {Bisigello}, Laura and {Buat}, V{\'e}ronique and {Calabr{\`o}}, Antonello and {Castellano}, Marco and {Cleri}, Nikko J. and {Cooper}, M.~C. and {Croton}, Darren and {Daddi}, Emanuele and {Dekel}, Avishai and {Elbaz}, David and {Franco}, Maximilien and {Gawiser}, Eric and {Holwerda}, Benne W. and {Huertas-Company}, Marc and {Jaskot}, Anne E. and {Leung}, Gene C.~K. and {Lucas}, Ray A. and {Mobasher}, Bahram and {Pandya}, Viraj and {Tacchella}, Sandro and {Weiner}, Benjamin J. and {Zavala}, Jorge A.},
        title = "{CEERS Key Paper. I. An Early Look into the First 500 Myr of Galaxy Formation with JWST}",
      journal = {\apjl},
         year = 2023,
        month = mar,
       volume = {946},
       number = {1},
          eid = {L13},
        pages = {L13},
          doi = {10.3847/2041-8213/acade4},
archivePrefix = {arXiv},
       eprint = {2211.05792},
 primaryClass = {astro-ph.GA},
       adsurl = {https://ui.adsabs.harvard.edu/abs/2023ApJ...946L..13F}
}

@ARTICLE{Harikane2024,
       author = {{Harikane}, Yuichi and {Nakajima}, Kimihiko and {Ouchi}, Masami and {Umeda}, Hiroya and {Isobe}, Yuki and {Ono}, Yoshiaki and {Xu}, Yi and {Zhang}, Yechi},
        title = "{Pure Spectroscopic Constraints on UV Luminosity Functions and Cosmic Star Formation History from 25 Galaxies at z $_{spec}$ = 8.61-13.20 Confirmed with JWST/NIRSpec}",
      journal = {\apj},
         year = 2024,
        month = jan,
       volume = {960},
       number = {1},
          eid = {56},
        pages = {56},
          doi = {10.3847/1538-4357/ad0b7e},
archivePrefix = {arXiv},
       eprint = {2304.06658},
 primaryClass = {astro-ph.GA},
       adsurl = {https://ui.adsabs.harvard.edu/abs/2024ApJ...960...56H}
}

@ARTICLE{Sun2023,
       author = {{Sun}, Guochao and {Faucher-Gigu{\`e}re}, Claude-Andr{\'e} and {Hayward}, Christopher C. and {Shen}, Xuejian and {Wetzel}, Andrew and {Cochrane}, Rachel K.},
        title = "{Bursty Star Formation Naturally Explains the Abundance of Bright Galaxies at Cosmic Dawn}",
      journal = {\apjl},
         year = 2023,
        month = oct,
       volume = {955},
       number = {2},
          eid = {L35},
        pages = {L35},
          doi = {10.3847/2041-8213/acf85a},
archivePrefix = {arXiv},
       eprint = {2307.15305},
 primaryClass = {astro-ph.GA},
       adsurl = {https://ui.adsabs.harvard.edu/abs/2023ApJ...955L..35S}
}

@ARTICLE{Ren2019,
       author = {{Ren}, Keven and {Trenti}, Michele and {Mason}, Charlotte A.},
        title = "{The Brightest Galaxies at Cosmic Dawn from Scatter in the Galaxy Luminosity versus Halo Mass Relation}",
      journal = {\apj},
         year = 2019,
        month = jun,
       volume = {878},
       number = {2},
          eid = {114},
        pages = {114},
          doi = {10.3847/1538-4357/ab2117},
archivePrefix = {arXiv},
       eprint = {1905.04848},
 primaryClass = {astro-ph.GA},
       adsurl = {https://ui.adsabs.harvard.edu/abs/2019ApJ...878..114R}
}

\begin{appendix}
\nolinenumbers
\section{Impact of dust attenuation on the measured galaxy size}
\label{sec:appendix_rascas}

Dust attenuation is not included when measuring the half-light radii in Fig.~\ref{fig:r50}, which could affect the inferred galaxy sizes. To assess whether our conclusion based on the intrinsic luminosities remains valid in the presence of dust, we perform radiative transfer calculations using RASCAS \citep{Michel-Dansac2020} on the five snapshots of the \gtt\ and the \sink\ runs shown in Fig.~\ref{fig:r50}. For the dust physics, we adopt the same methods as those used to measure $f^\mathrm{LyC}_\mathrm{esc}$ in Fig.~\ref{fig:fesc}. We generate three $512^2$ flux maps for the central $(0.5\Rvir)^3$ volume by projecting along the three principal axes and measure the two-dimensional half-light radii independently along each projection axis.

\begin{figure}
    \includegraphics[width=\linewidth]{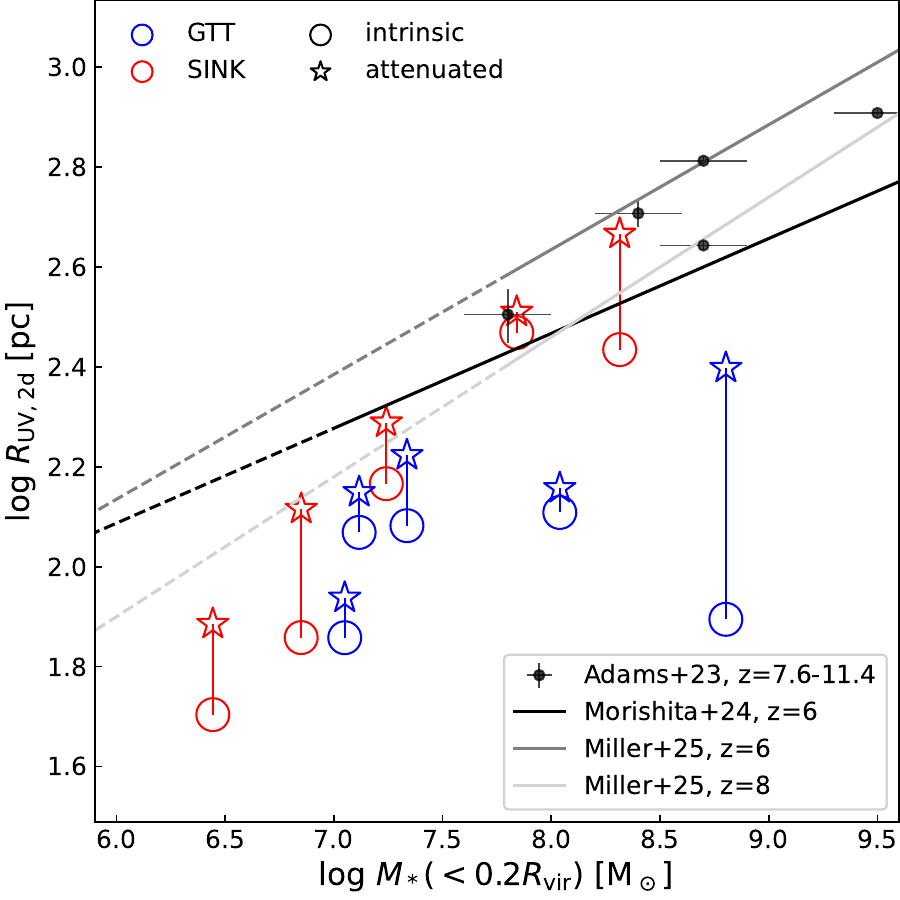}
    \vspace{-5mm}
    \caption{Comparison between the intrinsic (circles) and dust-attenuated (stars) half-light radii in the UV band for the \gtt\ and the \sink\ runs at five different redshifts, as in Fig.~\ref{fig:r50}. Each point represents the mean of the logarithmic two-dimensional radii measured along three orthogonal projection axes.}
    \label{fig:r50_rascas}
\end{figure}

Fig.~\ref{fig:r50_rascas} shows the average intrinsic and dust-attenuated two-dimensional half-light radii in the UV band. Dust attenuation increases the inferred half-light radii in all cases, indicating that the central part of the galaxy is more strongly obscured than the outer region. Although the magnitude of this increase varies with the redshift and between models, the overall trend shown in Fig.~\ref{fig:r50} remains largely unchanged. In particular, the \gtt\ run remains relatively compact even after the dust attenuation is included, whereas the \sink\ run continues to show good agreement with the observed mass-size relation \citep{Adams2023, Morishita2024, Miller2025}. This suggests that our conclusion based on the intrinsic UV luminosities is not significantly altered by dust attenuation.

\section{Long-lived clumps and inefficient feedback in the \gtt\ model}
\label{sec:appendix_overcooling}

\begin{figure*}
    \includegraphics[width=\textwidth]{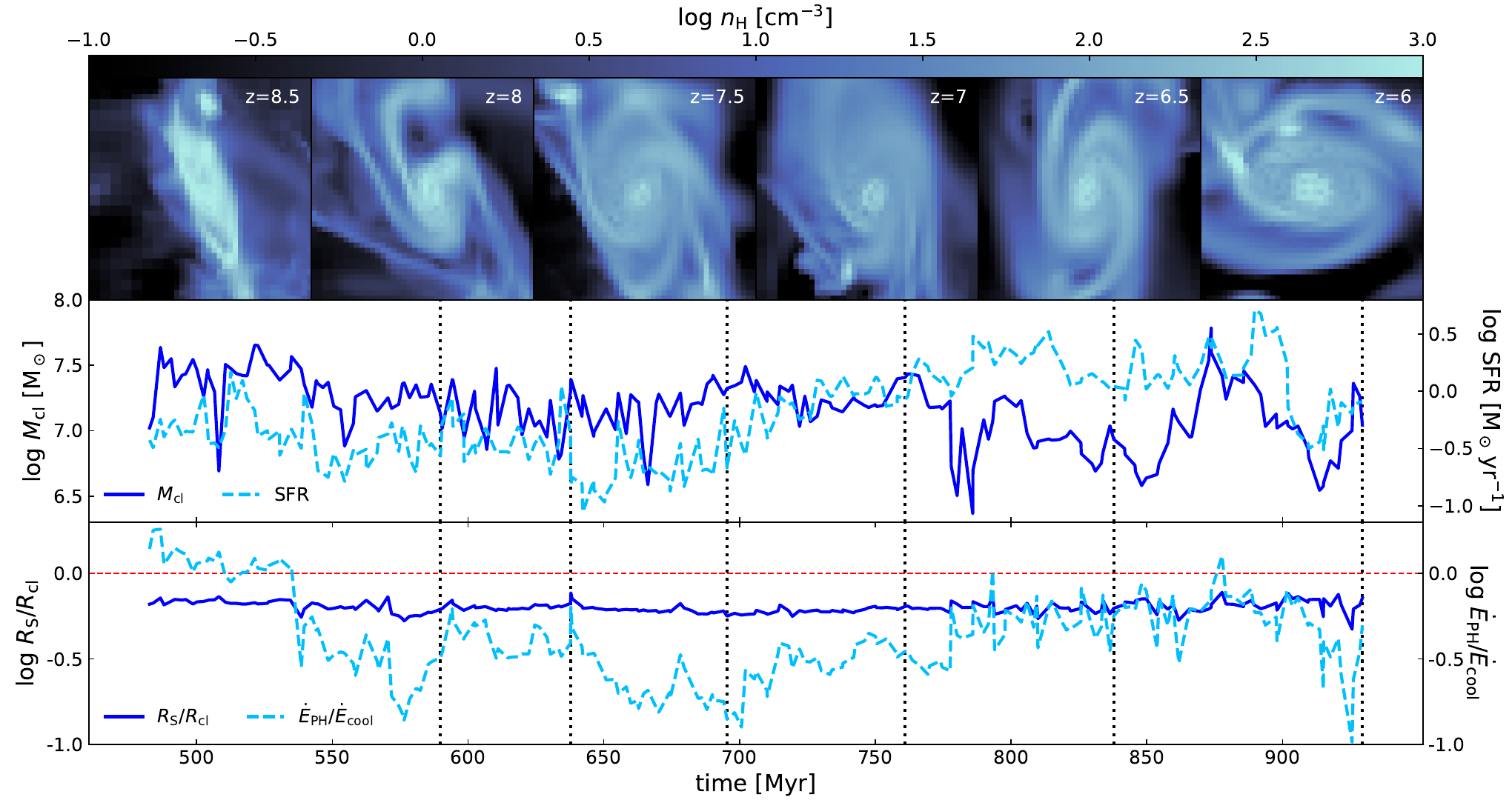}
    \vspace{-5mm}
    \caption{Evolution of the longest-living clump in the \gtt\ run. (a) Hydrogen number density around the clump at six different epochs labeled on the top right corner of each panel. Each panel is centered on the clump's density peak, and the side length is $\sim2400\,\mathrm{cpc}$ ($\sim340\,\mathrm{pc}$ at $z=6$). Middle: Time evolution of the clump mass ($M_\mathrm{cl}$, solid line) and the SFR within the clump radius (dashed line). The six vertical lines correspond to the epochs shown in the top panels. Bottom: Ratios between $R_\mathrm{S}$ and the clump radius ($R_\mathrm{cl}$, solid line), and between the photoionization heating rate ($\dot E_\mathrm{PH}$) and net cooling rate ($\dot E_\mathrm{cool}$, dashed line). Four quantities are measured using star particles and gas cells within the clump radius. The red horizontal line marks where the ratio is equal to 1.}
    \label{fig:clump_zoom}
\end{figure*}

In the main text, we show that star formation in the \gtt\ model tends to be inefficient, resulting in prolonged star formation that contributes to the overcooling problem. In this section, we provide a detailed example of how inefficient feedback enables such structures to survive and sustain star formation over extended periods.

Among star-forming clumps at $z=6$ in the \gtt\ run, we track the evolution of the longest-lived clump shown in Fig.~\ref{fig:clump_zoom}.
This clump initially forms within a satellite galaxy (bottom right side of the $z=7$ panel in Fig.~\ref{fig:map}) and survives for more than 400 Myr, during which the host satellite travels over $\sim20\, \mathrm{kpc}$. Although the clump undergoes a rapid mass decline at $t\sim780\,\mathrm{Myr}$ during its closest approach to the halo center, it is not fully disrupted, and the SFR remains comparable. By $z=6$, the clump settles at the center of the main galaxy after completing a full orbit and accounts for $\sim 50\%$ of the total SFR within $0.2\Rvir$. The total stellar mass formed within this single clump over its lifetime reaches $1.1\times10^8\,\Msun$, corresponding to 16\% of the total stellar mass of the \gtt\ run at $z=6$.

The bottom panel of Fig.~\ref{fig:clump_zoom} shows two quantities that characterize the efficiency of star formation regulation: the ratio of the Str{\"o}mgren radius to the clump radius ($R_\mathrm{S}/R_\mathrm{cl}$) and the ratio of the photoionization heating rate to the net cooling rate ($\dot E_\mathrm{PH}/\dot E_\mathrm{cool}$). As shown in Fig.~\ref{fig:cluster}, the ionizing photon budget is insufficient to ionize the clump or enable significant photon escape. Consequently, the clump maintains $R_\mathrm{S}/R_\mathrm{cl}<1$ throughout its evolution, with a time-averaged value of $\sim0.64$. $\dot E_\mathrm{PH}/\dot E_\mathrm{cool}$ also remains below unity over most of the clump lifetime, except at very early times and during a brief interval at $t\sim880\,\mathrm{Myr}$. These diagnostics clearly show that star formation within this clump is too inefficient to heat the gas, thereby leading to an unusually long clump lifetime.



\end{appendix}

\end{document}